%% file: arXiv.tex
\documentclass[12pt]{aastex63}
\usepackage{booktabs, url, amsmath}
\usepackage{IEEEtrantools}
\def\lea{\mathrel{<\kern-1.0em\lower0.9ex\hbox{$\sim$}}}
\def\gea{\mathrel{>\kern-1.0em\lower0.9ex\hbox{$\sim$}}}

\graphicspath{{./}{./Figures/}{figures/}}
\received{}
\revised{}
\accepted{}
\submitjournal{ApJ}

\shorttitle{Embedded Clusters in the Antennae}
\shortauthors{Chandar et al.}

\begin{document}


\title{Nowhere Left to Hide: Uncovering All of the Massive Young Embedded Star Clusters in the Antennae with JWST\footnote{This work is based on observations made with the NASA/ESA/CSA James Webb Space Telescope. The data were obtained from the Mikulski Archive for Space Telescopes at the Space Telescope Science Institute, which is operated by the Association of Universities for Research in Astronomy, Inc., under NASA contract NAS 5-03127 for JWST. These observations are associated with program \#2581.}}

\correspondingauthor{Rupali Chandar}
\email{rupali.chandar@utoledo.edu}

\author[0000-0003-0085-4623]{Rupali Chandar}
\affiliation{Ritter Astrophysical  Research Center, University of Toledo, Toledo, OH 43606, USA} 
\author[0000-0002-2957-3924]{Miranda Caputo}
\affiliation{Ritter Astrophysical  Research Center, University of Toledo, Toledo, OH 43606, USA}
\author[0000-0002-5728-1427]{Paul Goudfrooij}
\affiliation{Space Telescope Science Institute, 3700 San Martin Drive, Baltimore, MD 21218, USA}
\author[0000-0002-1000-6081]{Sean T. Linden}
\affiliation{Steward Observatory, University of Arizona, 933 N Cherry Avenue, Tucson, AZ 85721, USA}
\author[0000-0001-7413-7534]{Angus Mok} \affiliation{OCAD University, Toronto, Ontario, M5T 1W1, Canada}
\author[0000-0003-2093-4452]{Cory Whitcomb}
\affiliation{Ritter Astrophysical  Research Center, University of Toledo, Toledo, OH 43606, USA} 
\author[0009-0001-6065-0414]{Grant Donnelly}
\affiliation{Ritter Astrophysical  Research Center, University of Toledo, Toledo, OH 43606, USA} 
\author[0000-0001-5073-2267]{Florent Renaud}
\affiliation{Observatoire Astronomique de Strasbourg, Universit\'e de Strasbourg, CNRS UMR 7550, F-67000 Strasbourg, France}
\affiliation{University of Strasbourg Institute for Advanced Study, 5 all\'ee du G\'en\'eral Rouvillois, F-67083 Strasbourg, France}
\author[0000-0003-1545-5078]{John-David T. Smith}
\affiliation{Ritter Astrophysical  Research Center, University of Toledo, Toledo, OH 43606, USA} 
\author[0000-0002-5480-5686]{Alberto Bolatto}
\affiliation{Department of Astronomy, University of Maryland, College Park, MD 20742, USA}
\affiliation{Joint Space-Science Institute, University of Maryland, College Park, MD 20742, USA}
\author[0000-0002-5782-9093]{Danny A. Dale}
\affiliation{Department of Physics and Astronomy, University of Wyoming, Laramie, WY 82071, USA}
\author[0000-0003-0014-0508]{Sara Duval}
\affiliation{Ritter Astrophysical  Research Center, University of Toledo, Toledo, OH 43606, USA} 
\author[0009-0005-0750-2956]{Lindsey Hands}\affiliation{Department of Astronomy \& Astrophysics, University of California, San Diego, 9500 Gilman Drive, San Diego, CA 92093, USA}
\author[0000-0002-0560-3172]{Ralf Klessen}
\affiliation{Universit\"{a}t Heidelberg, Zentrum f\"{u}r Astronomie, Institut f\"{u}r Theoretische Astrophysik, Albert-Ueberle-Str 2, D-69120 Heidelberg, Germany}
\author[0009-0001-3989-1731]{Caroline Kuczek}
\affiliation{Ritter Astrophysical  Research Center, University of Toledo, Toledo, OH 43606, USA} 
\author[0000-0002-4378-8534]{Karin Sandstrom}
\affiliation{Department of Astronomy \& Astrophysics, University of California, San Diego, 9500 Gilman Drive, San Diego, CA 92093, USA}
\author[0000-0002-3933-7677]{Eva Schinnerer}
\affiliation{Max-Planck-Institut f\"{u}r Astronomie, K\"{o}nigstuhl 17, D-69117, Heidelberg, Germany}
\author[0000-0003-4793-7880]{Fabian Walter}
\affiliation{Max-Planck-Institut f\"{u}r Astronomie, K\"{o}nigstuhl 17, D-69117, Heidelberg, Germany}


\begin{abstract}
The Antennae galaxies merger produces the brightest infrared emission of any galaxy within $\approx20$~Mpc, mostly from intense star formation taking place in supergiant molecular cloud complexes in the overlap region.
Here, we present new, high-resolution NIRCam and MIRI images of the Antennae galaxies taken with the F150W, F187N, F335M, F360M, F410M, and F770W filters on JWST to search for the predicted but as-yet-undiscovered population of deeply embedded, optically obscured star clusters.
We identify a population of 45 sources, 
40 previously unknown,
with high Br$\alpha/\mbox{H}\alpha$ and Pa$\alpha/\mbox{H}\alpha$ flux ratios which are likely very young clusters still embedded or just emerging from their natal cocoons, and estimate their age, extinction (A$_V$), and mass.
We find that all are extremely young ($\lea 2.5$~Myr), have A$_V$ between 2 and 10~mag, and masses between $\approx 10^4$ and several$\times10^6~M_{\odot}$. 
We believe we have now uncovered all clusters with $M\gea3\times10^4~M_{\odot}$ and A$_V \gea2$~mag in the Antennae.
While our sample represents a small fraction($\approx15$\%) of clusters younger than 3~Myr by number, it dominates the ionizing photon luminosity across the galaxy pair ($\approx60$\%).
We find elevated H$_2/$PAH ratios of the ISM surrounding the most massive pair of embedded clusters, 
supporting the idea that merger-induced shock-heated gas play an important role in the formation of extremely massive clusters.

\end{abstract}

\section{Introduction}\label{sec:intro}

The likely progenitors of globular star clusters (GCs) forming just $\sim0.5-1$~Gyr after the Big Bang have recently been discovered in James Webb Space Telescope (JWST) images of high redshift galaxies \citep[e.g.,][]{Mowla22,Vanzella22,Claeyssens23,Vanzella23,Senchyna24,Adamo24,Messa24}.  This direct new evidence of massive cluster formation so early in cosmic history underscores the importance of understanding the physical processes that drive massive cluster formation and early evolution. 
In several systems, the estimated cluster masses are extremely high, reaching $\sim10^6 - 10^7~M_{\odot}$ \citep[][]{Vanzella22,Vanzella23,Adamo24}. 
Several of the parent galaxies show tidal features and morphological evidence of interactions, indications that processes commonly found in galaxy collisions almost certainly drive massive cluster formation \citep[e.g.,][]{Whitaker25}.
Because observations are currently limited to the high-mass end, the initial mass function of globular clusters has not yet been measured directly.

The shape of the initial cluster mass function is a fundamental diagnostic of the star formation process on galaxy scales.
Globular cluster masses reach  $\approx10^6~M_{\odot}$ in Milky-Way mass galaxies and have a steepening or cutoff at the high end \citep[e.g.,][]{Goudfrooij03,Peng10,Goudfrooij16,DeLucia24}, which could either result from billions of years of evolution or have been imprinted at birth \citep[e.g.,][]{Fall77, Capriotti96, Kavelaars97, Fall01, Elmegreen10a}.  If imprinted at birth, such a cutoff would indicate a physical limit to the mass with which clusters can form. 
Unfortunately, the intense star formation needed to create such massive clusters is mostly taking place in luminous infrared galaxies that are dust-obscured, distant, or both \citep[e.g.,][]{Adamo20, Linden23}. As the closest system forming clusters with masses $>10^6~M_{\odot}$, the merging Antennae galaxies are ideal to assess the shape of the initial cluster mass function.



\begin{figure}[!ht]
	\centering
\includegraphics[width=6.0in]{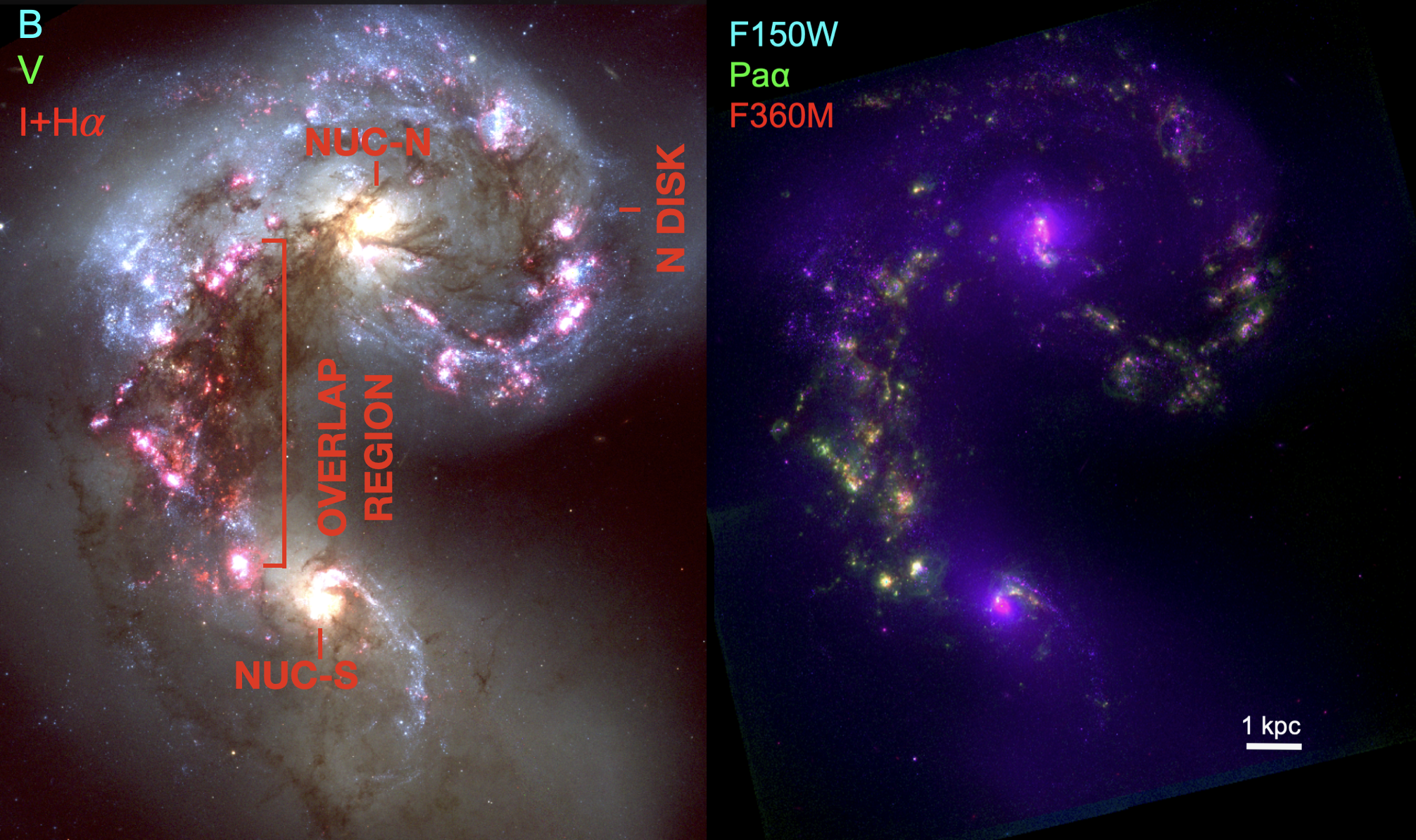}
	\caption{High resolution color images of the merging Antennae galaxies are shown in the optical (left) from HST and in the infrared (right) from JWST in the indicated filters.  The optical (B, V, I$+$H$\alpha$) filter combination gives a good view of dark dust lanes and warm ionized gas (pinkish red color) found throughout the Antennae and particularly in the overlap region, which is the region between and to the east of the two nuclei. The infrared (F150W, Pa$\alpha$, F360M) filter combination peers through this dust, and allows us to identify optically obscured, very young star clusters which are still embedded in their natal ISM. North is up and East to the left, and the 1~kpc covers $10.75\arcsec$ in each panel.} \label{fig:colorimages} 
\end{figure}

The overlap region in the Antennae, where the two progenitor spiral galaxies are merging together and recent star formation is driving strong infrared emission (the strongest observed in any galaxy within $\sim20$~Mpc), is the closest system where extremely massive ($\approx 10^6 - 10^7~M_{\odot}$) clusters are forming.
This region has intense star formation, a vast reservoir of cold gas organized into super giant molecular cloud complexes \citep[][]{Gao01,Wilson03}, and many radio sources \citep{Neff00}.  
Velocity maps from ALMA/CO indicate that cloud-cloud collisions may play an important role in triggering the formation of massive clusters \citep[][]{Tsuge21b}.
A color image of the Antennae constructed from BVI$+$H$\alpha$ filters taken by the Hubble Space Telescope (HST) is shown in the left panel of Figure~\ref{fig:colorimages}.  A number of very young, recently formed clusters are observed by their H$\alpha$ emission (which makes them look pink) in the overlap region, but most of the area is dark because it is obscured by large amounts of dust.
The high dust obscuration in the overlap region \citep{Mirabel98} and in other LIRGs suggests that optically-based studies may be largely incomplete, and may miss as much as $\approx60$\% of the recently formed ($\lea 6$~Myr) cluster population \citep[e.g.,][]{Linden23}.
While extremely massive clusters are forming in other dusty, merger-driven (U)LIRGs, most of these systems are much further away and the massive clusters tend to be significantly older (e.g., Arp 220, \citealt{Chandar23}; NGC 7252, \citealt{Miller97}; NGC 34 \citealt{Schweizer07}). 

Our current understanding of the recent star and cluster formation histories in the Antennae relies mostly on optical studies with the Hubble Space Telescope \citep[e.g.,][]{Whitmore99,Fall01,Fall05,Whitmore10}, and
predictions from high resolution (parsec-scale) hydrodynamical simulations of galaxy mergers \citep[][]{Renaud15,Li17}. 
A number of works have concluded that most of the active star formation in the Antennae is embedded in thick layers of dust with high extinction $A_V\sim 30-70$~mag, and is therefore invisible at optical (and possibly) near-infrared wavelengths \citep[e.g.,][]{Kunze96, Mirabel98, Neff00}.  
If this is confirmed, current conclusions are based on an incomplete census of star clusters and may need to be revised.
On the other hand, \citet{Whitmore02} concluded that most ($\approx85$\%) recently formed massive clusters have extinction values A$_V \lea 10$~mag and should be detected at optical wavelengths, at least in the I band.

The source WS80 is the best-known example of the embedded clusters we are searching for.
It is the most massive ($>10^6~M_{\odot}$), recently formed, highly extinguished cluster known in the Antennae \citep[][]{Herrera17}.
WS80 is the brightest source of thermal bremsstrahlung radiation at radio wavelengths in the Antennae, but is barely visible in a deep V-band HST image.  
If the mass function of very young ($<3$~Myr) clusters in the Antennae follows a power law with index $\beta \approx -2$ as found for the cluster population younger than 10~Myr \citep{Fall01}, there should be on the order of $\approx 50-100$ WS80-like clusters with high extinction and masses greater than few$\times10^5M_{\odot}$ forming in the overlap region, although very few such clusters have been discovered to date.
The main goal of this paper is to use new infrared images taken with the JWST of the merging Antennae galaxies to search for the predicted but as-yet-undiscovered population of deeply embedded/optically obscured massive ($\approx10^4- 10^6~M_{\odot}$) clusters and to estimate their basic properties, i.e. age, extinction (A$_V$), and mass.  In a future work, newly discovered embedded sources will be combined with optically-selected ones to create a complete census of obscured$+$unobscured clusters.
We adopt a distance of 21.5~Mpc (distance modulus of $m-M$ = 31.66; \citealt{Riess+11}) to the Antennae, which gives a physical scale of $\approx104$~pc per arsecond.

The rest of this paper is organized as follows.
The new JWST observations and basic reduction are presented in Section~2, along with color images of the Antennae created from different optical-infrared filter combinations.
In Section~3 we use maps of Br$\alpha/$H$\alpha$ and Pa$\alpha/$H$\alpha$ flux ratios to select a new catalog of embedded cluster candidates in the Antennae, and in Section~4 we estimate the age, extinction, and mass for each cluster.
Section~5 estimates the reddening and mass ranges of clusters that might still be missed in the JWST observations, determines the distribution of embedded cluster masses, and examines the local environment of the two most massive clusters to constrain formation mechanisms.
We summarize the main results and discuss future plans in Section~6.  The Appendix investigates the infrared emission from bright 6~cm sources from the \citet{Neff00} catalog.


\section{Observations of the Antennae}
\label{sec:obs}

\subsection{New JWST Images}
\begin{table*}
\begin{center}
\caption{HST $+$ JWST Observations of the Antennae\label{tab:obs}}
\begin{tabular}{lccccccccccc}
\hline\hline
\colhead{Instrument$+$} & \colhead{Filter}  & \colhead{\# exposures} & \colhead{Total Exposure} & \colhead{Proposal} & \colhead{VEGAMAG} \\
\colhead{Camera} & \colhead{Name}  & \colhead{ } & \colhead{Time (sec)} & \colhead{ID} & \colhead{zeropoint$^1$} \\ \hline \hline 
HST/WFC3 & F275W (NUV) & 3 & 2610 &  14593 & 22.64~mag \\
HST/WFC3 & F336W (U)  & 5 & 5534 & 11577 & 23.526~mag  \\
HST/ACS  & F435W (B) &  3 &  1320 &  10188 & 25.766~mag \\
HST/WFC3 & F555W (V) & 2 & 1032 & 11577 & 25.843~mag \\
HST/ACS & F658N (H$\alpha$) &   4 & 2300 & 10188 & 22.390~mag \\
HST/WFC3 & F814W (I) &  2 & 1032 & 11577  & 25.520~mag \\
JWST/NIRCam & F150W &   24 & 7472 &  2581 & 1192.64 Jy \\
JWST/NIRCam & F187N (Pa$\alpha$)   & 12 & 3736 & 2581  & 779.34 Jy \\
JWST/NIRCam & F335M (3.3$\mu$m PAH)   & 12 & 3736 & 2581  & 300.24 Jy \\
JWST/NIRCam & F360M  & 12 & 3736 & 2581  & 262.24 Jy  \\
JWST/NIRCam & F410M (Br$\alpha$)   & 12 & 3736 &  2581  & 208.45 Jy  \\
JWST/MIRI & F770W (7.7$\mu$m PAH)   & 4 & 244 & 2581 & ...  \\
\hline 
\hline
\end{tabular} 
\end{center}
\tablecomments{$^1$Zeropoints are from instrument documentation available at www.stsci.edu.}

\end{table*}

Images of the Antennae galaxies were taken by JWST as part of program GO-2581 (PI: R. Chandar) in June 2023 using the F150W, F187N, F335M, F360M, F410M filters on NIRCam and the F770M filter on MIRI.  These filters cover stellar and dust continuum emission (F150W, F360M), hydrogen emission lines (Pa$\alpha$ at 1.87$\mu$m and Br$\alpha$ at 4.1$\mu$m), and PAH features (3.3$\mu$m and 7.7 $\mu$m).  Basic information about the observations, including the number of dithered exposures and total photon collection time for each filter, are compiled in Table~1. All photometry in this work is in the VEGAMAG system.

With NIRCam, we use Module B, the FULL subarray, and an INTRAMODULEBOX primary dither pattern with 4 positions. At each primary dither position, we take three exposures using a ``small-grid'' sub-pixel dither pattern for a total of 12 exposures and 3736~seconds of photon collection time in the F187N, F335M, F360M, and F410M filters.  The F150W filter was observed twice in this configuration so has 24 exposures and a total photon collection time of 7472~seconds.  The SHALLOW4 readout pattern with one integration per exposure was used for all NIRCam observations.
For our MIRI F770W observations, we use a 4-point dither and the FASTR1 readout pattern yielding a total photon collection time of 244~sec. 

We start with the uncalibrated raw (“uncal”) files obtained from MAST and use Detector1Pipeline, which converts from “ramps” to “slopes” and applies basic detector-level corrections to the data, including flagging saturated pixels and correcting for chip persistence. We ran the Detector1Pipeline with the default parameters for most steps, except for the {\tt ramp\_fitting} step for which we used {\tt suppress\_one\_group = False} to avoid hard saturation for the brightest objects in the field.
Stage 2 of the pipeline produces calibrated individual exposures.  During this stage, the pixel coordinates are translated into WCS coordinates, which includes a correction for geometric distortion. This stage also includes flat-fielding and background subtraction (the latter was only performed for the MIRI data).  

Before combining the individual exposures, the individual, flatfielded NIRCam images are run through an algorithm which minimizes the effect of 
electronic 1/$f$ noise. This algorithm, which is publicly available and described in \href{https://jwst-docs.stsci.edu/known-issues-with-jwst-data/niriss-known-issues/niriss-1-f-noise-removal}{JWST documentation}, masks pixels that have intensities $\ga 3\sigma$ above the background, takes the median value of non-masked pixels in
each line in the slow readout direction and adjusts the level of each of those lines by subtracting the difference between those line medians and the overall median level of non-masked pixels. 
Stage 3 combines individual tiles to produce a final, geometrically rectified mosaic of the Antennae in units of surface brightness (MJy/sr). 
The final redrizzling step produces mosaics oriented with north pointing up and east to the left.

The standard mosaics produced by the Level-3 pipeline have good relative alignment between NIRCam tiles, but the alignment of the JWST images relative to other data with well-established absolute astrometry tends to be poorer, with offsets as large as $\sim1\arcsec$ in some cases. Furthermore, there were non-negligible offsets in pixel space between the Level-3 products of the different filters. 
For our final Stage-3 processing, we therefore performed additional corrections to improve the relative and absolute astrometry of the NIRCam and MIRI images using the {\tt tweakreg} step of the pipeline.
These corrections included: \emph{(i)} creating a single reference star catalog to align all individual exposures, \emph{(ii)} measuring ``true'' sky levels from custom sky regions; this was only important for the short-wavelength F150W and F187N filters where the 4 individual detectors sample different regions within the full field-of-view. For this purpose, we use relative sky levels measured in the long-wavelength F360M image as a reference; \emph{(iii)} matching the sky coordinates in the catalog to the Gaia DR3 reference frame, \emph{(iv)} drizzling all Stage-3 products to a common celestial position for the reference pixel, and \emph{(v)} adopting a common pixel scale and image size across all Stage-3 products. The final pixel scale of the images is 0\farcs031 pix$^{-1}$.

\subsection{Archival HST Images}

The Antennae galaxies have been imaged by HST through several different programs over the years.
In this work, we include archival observations in both broad- and narrow-band filters covering the near-ultraviolet and optical. 
We use observations taken with the WFC3/UVIS camera in the F275W/NUV (GO-14593, PI: N. Bastian), F336W/U, F555W/V, and F814W/I (GO-11577, PI: B. Whitmore) filters,
and with the ACS/WFC camera in the F435W/B, 
and F658N/H$\alpha$ filters (GO-10188; PI: B. Whitmore).
Details of the HST observations, including the Proposal ID, number of exposures and total exposure time, are compiled in Table~1.

Individual exposures are processed through the standard 
CALACS or CALWFC3 software, which performs initial data quality flagging, bias subtraction, gain correction, bias stripe removal, correction for CTE losses, dark current subtraction, flat-fielding and photometric calibration. These individual files are aligned and drizzled to the common grid to create a single image per filter in counts~sec$^{-1}$ using the DRIZZLEPAC software package. The WFC3 (ACS) images have a pixel scale of 0\farcs04 pix$^{-1}$ (0\farcs05 pix$^{-1}$) and are used for producing spectral energy distributions.

\subsection{Flux-Calibrated, Continuum-Subtracted Maps}

The narrow-band images contain both starlight and recombination line emission from ionized gas, and trace the locations of very recently formed massive stars and clusters.
Because the optical light from the very youngest of these clusters is heavily extinguished because they are still embedded in their natal gas and dust,  hydrogen recombination lines in the infrared (such as Pa$\alpha$ and Br$\alpha$) are well-suited to identify these young dusty objects.
The $4.05\mu$m Br$\alpha$ line in particular is near the minimum in the Galactic extinction curve \citep{Gordon2023}.

We produce maps of hydrogen recombination lines by subtracting the stellar continuum from the narrow- and medium-band filters: F658N (H$\alpha$), F187N (Pa$\alpha$), 
and F410M (Br$\alpha$). For Pa$\alpha$, we estimate the continuum level by interpolating between the F150W and F360M images (using a first degree polynomial fit) to the central wavelength of the F187N filter.  We then use several isolated, non-saturated, stars and intermediate-age clusters in the Antennae to determine the scale factor ($\sim 1.3$) between our interpolated continuum image and the F187N line$+$continuum image.
These sources do not have nebular or dust emission, so should give a good measure of the stellar continuum in all filters.
For the F410M filter we adopt the nearby F360M image as our ‘continuum’ and scale it based on the same line-free sources described above.  This filter is the closest one in wavelength that does not contain significant contributions from dust or recombination line emission.  To create H$\alpha$ line maps we adopt a scaled (by a factor of 0.64) F814W filter as the continuum and subtract this from the F658N image.
All continuum-subtracted images are in physical flux density units of erg~s$^{-1}~\mbox{cm}^{-2}~\mbox{Hz}^{-1}$.
These continuum-subtracted H$\alpha$, Pa$\alpha$, and Br$\alpha$ maps will be used to search for recently formed, embedded clusters.


\subsection{Spitzer Mid-Infrared Spectroscopy}

Most of the Antennae, including the interaction region, was observed by the Infrared Spectrograph (IRS) on Spitzer as part of program 21 (PI: J.R. Houck).  The observations span the wavelength range 5.2 - 14.7$\mu$m with a scale of $\sim1.8\arcsec~\mbox{pix}^{-1}$ and from 14.2 - $38.4\mu$m at $\sim5.1\arcsec~\mbox{pix}^{-1}$. 
Previously, \citet{Brandl09} extracted and analyzed the spectra for 6 bright infrared peaks and the two galactic nuclei from these observations.

Spitzer/IRS data cubes are created as part of a uniform reduction of all low spectral resolution IRS mapping programs, which includes hundreds of galaxies, by the Spitzer/IRS Mapping Legacy Archive (SIMLA) project \citep{Donnelly25}.
The SIMLA pipeline leveraged off-source observations from the entire IRS archive to create custom backgrounds for each spectral map, resulting in significantly reduced noise in the extracted spectra. In this work, we measure the H$_2/$PAH ratio from the Spitzer spectra in a few special locations in the Antennae, and leave a more detailed investigation of the full region and rich spectral features to a future work.

\subsection{An Infrared View of Optically-Dark Regions}

\begin{figure}[ht]
	\centering
\includegraphics[width=6.0in]{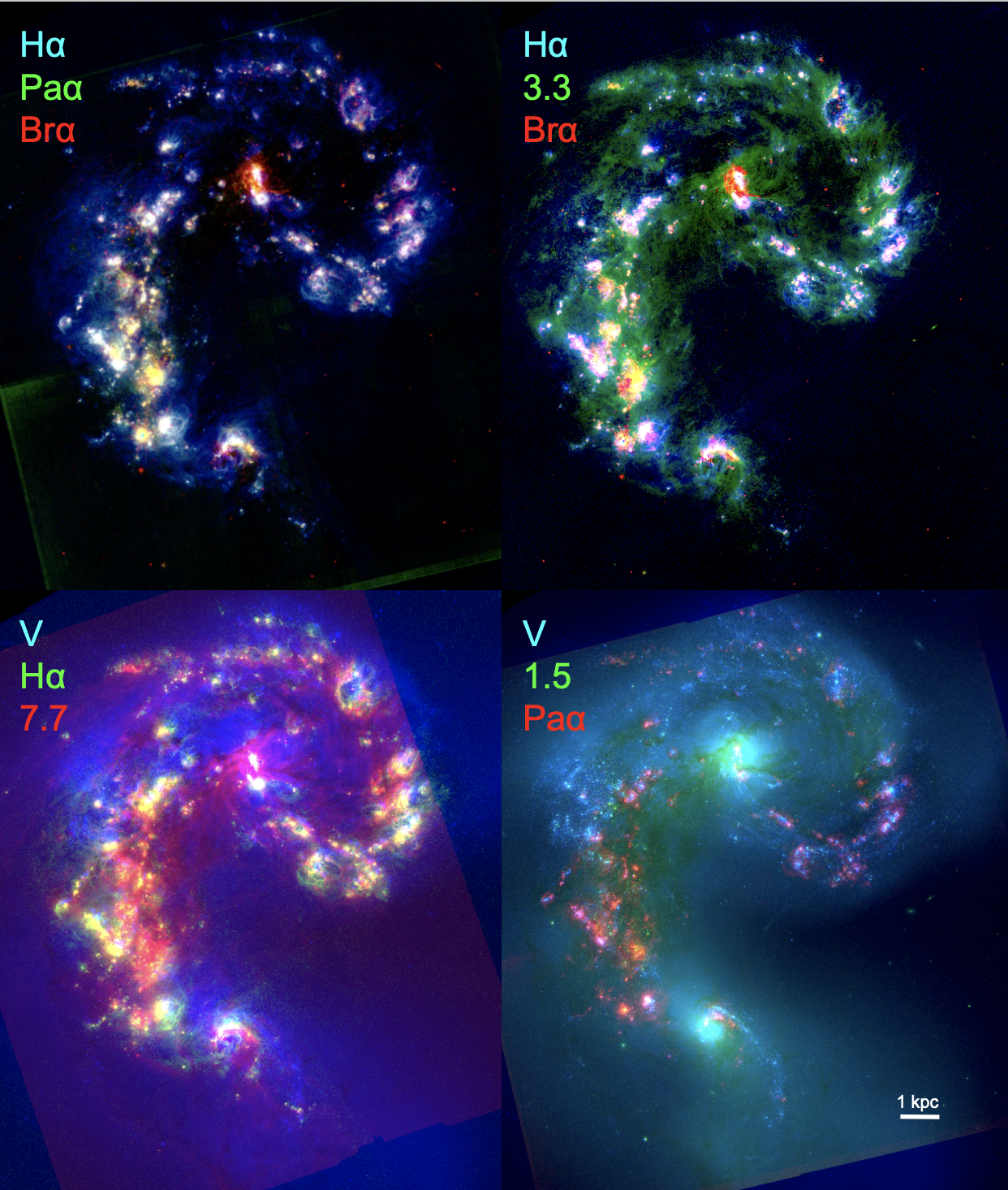}
	\caption{Color images of the Antennae are shown in four different filter combinations, as indicated in each panel.  Maps which include H$\alpha$, Pa$\alpha$, Br$\alpha$, and 3.3$\mu$m emission have all been continuum subtracted as described in Section~2.3. } \label{fig:colorimages2}
\end{figure}

Figure~\ref{fig:colorimages} contrasts the Antennae in optical  (left panel) and infrared light (right panel).
The north and south nuclei, partially intact northern disk, and dusty overlap region (the region to the east between the two galaxy nuclei) are all clearly seen in the optical filter combination of B, V, I$+$H$\alpha$ displayed in the left panel.  This filter combination highlights the dark dust lanes found throughout the Antennae, particularly in the overlap region, where some areas appear fully dark with little H$\alpha$ emission from recently formed star-clusters. 
The infrared (F150W, Pa$\alpha$, F360M) filter combination in the right panel minimizes this obscuration and reveals sites where star formation is (and is not) taking place.
With the new JWST images {\em we can essentially see all of massive clusters forming in the Antennae.}

In Figure~\ref{fig:colorimages2} we show different combinations of JWST$+$HST filters to highlight different aspects of this iconic system.

\begin{itemize}

\item {\bf Top-left:} This image combines the continuum-subtracted H$\alpha$ (blue), Pa$\alpha$ (green), and Br$\alpha$ (red) maps to highlight warm ionized gas. 
For the most part, this emission traces natal gas ionized by massive stars within recently formed star clusters.
Warm ionized gas is observed throughout the Antennae, from the northern disk, throughout the overlap region and near the southern nucleus.  The gas shows a range of colors, indicating a range of extinction, with the most highly extinguished regions found mostly in the overlap region. The latter appear yellow since they are dominated by Pa$\alpha$ emission and their inferred extinctions lie between 4.5 and 15~mag.

\item {\bf Top-right:} An image of 3.3\,$\mu$m emitting dust grains replaces the
continuum-subtracted Pa$\alpha$ image from the top-left panel (in green).
Most of the green in this image, where the 3.3$\mu$m PAH emission dominates over H$\alpha$ and Br$\alpha$, is diffuse and extends away from regions of ionized gas.  Strong 3.3$\mu$m emission is also co-spatial with many (but not all) shells and bubbles around HII regions, as also observed in a sample of PHANGS galaxies \citep{Rodriguez25}.
We will study the association between PAH emission and young stellar clusters in the Antennae in an upcoming paper.

\item {\bf Bottom-left:} This color composite is created from the V (blue), H$\alpha$ (green), and 7.7$\mu$m PAH band images.  Like the previous image, there is a significant amount of diffusely distributed PAH emission, now seen in red.  This image also reveals an interesting 'bubble' in the northern disk, where the 7.7$\mu$m emission appears to fall just outside of an H$\alpha$ shell.  
This bubble is reminiscent of the 'Phantom Void', a large, feedback-driven $\sim$kpc-scale bubble observed in PAH emission in the spiral galaxy NGC~628 \citep{Barnes23}. 

\item {\bf Bottom-right:} The combination of the V band (blue), F150W (green), and continuum-subtracted Pa$\alpha$ (red) filters highlights differences in the warm ionized gas properties in the northern disk, which are a dark pink color, with those in the overlap region which are more of an orange-red due to their higher extinction. Many blue, slightly older (by a few Myr) clusters are also observed in the northern disk.  

\end{itemize}

\section{A New Catalog of Very Young, Embedded Clusters in the Antennae}

\subsection{Cluster Selection Based on Hydrogen Recombination Line Ratios}

Recently formed star clusters are typically still surrounded by their parental cloud of gas and dust, which makes their emission difficult to detect at optical wavelengths but easier at longer wavelengths.
Because the intrinsic flux ratios for H$\alpha$/Br$\alpha$ and H$\alpha$/Pa$\alpha$ are 36.679 and 8.584, respectively \citep{Hummer87}, H$\alpha$ will be stronger than Pa$\alpha$ and Br$\alpha$ in \ion{H}{2} regions with nebular extinction $A_V \lea 4.5$~mag, Pa$\alpha$ will be strongest for sources with $A_V \approx4.5$ to $\approx15$~mag, and Br$\alpha$ will be strongest for sources with $A_V \gea 15$~mag.\footnote{These estimates assume Case B recombination for a a typical electron temperature of 10,000~K and electron density $1\times10^4$ and a Milky Way-like attenuation curve with $R_V=3.1$.}
Note that the \cite{Calzetti00} law finds that the nebular reddening is $\approx2.27$ times higher than the reddening experienced by stellar emission.
In this section, we use our high resolution maps of hydrogen recombination line emission in H$\alpha$, Pa$\alpha$, and Br$\alpha$ to identify embedded star clusters in the Antennae.

\begin{figure}[!ht]
\includegraphics[width=\textwidth]{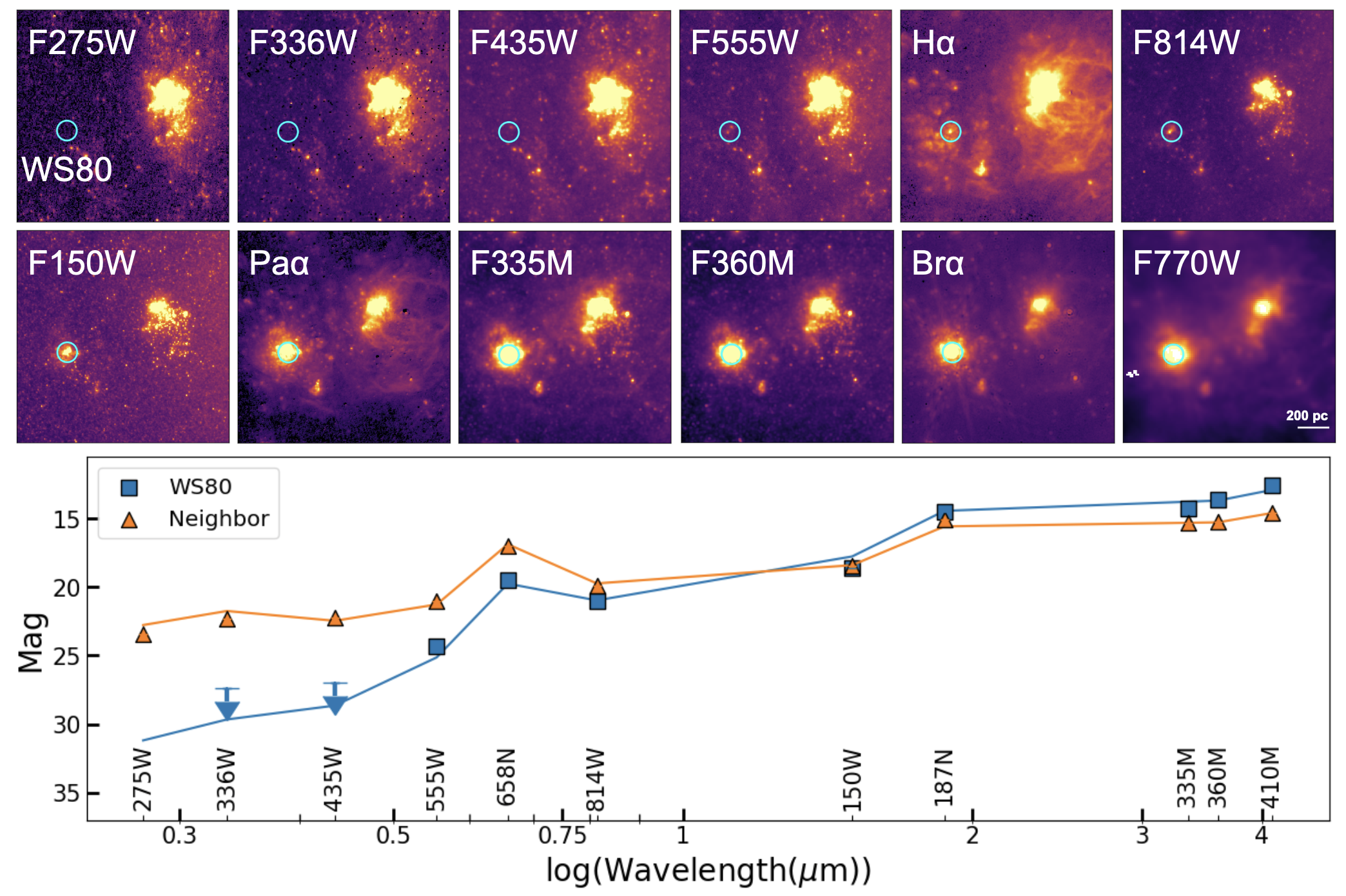}
	\caption{The embedded, extremely massive star cluster WS80 (identified by the cyan {\bf circle}) and its nearby companion (also a massive young cluster but with more moderate extinction) are shown in 12 HST$+$JWST filters, starting from the near-ultraviolet in the upper left through the JWST/F770W filter in the lower-right. Each filter shows a 12.5$\times$12.5$\arcsec$ region around the clusters. WS80 is not seen in the NUV, U, or B filters but can be observed as a faint source in the V, H$\alpha$ and I filters through the cyan cross.  WS80 gets brighter and brighter towards longer wavelengths, and by 3.3$\mu$m is the brightest source in the Antennae.  By contrast, its companion cluster is bright in all 12 filters.  The lower panel shows the measured photometry for the two clusters, with the best-fit SED models overplotted (see Section~4.1).
    }\label{fig:WS80}
\end{figure}

\begin{figure}[!ht]
\includegraphics[width=\textwidth]{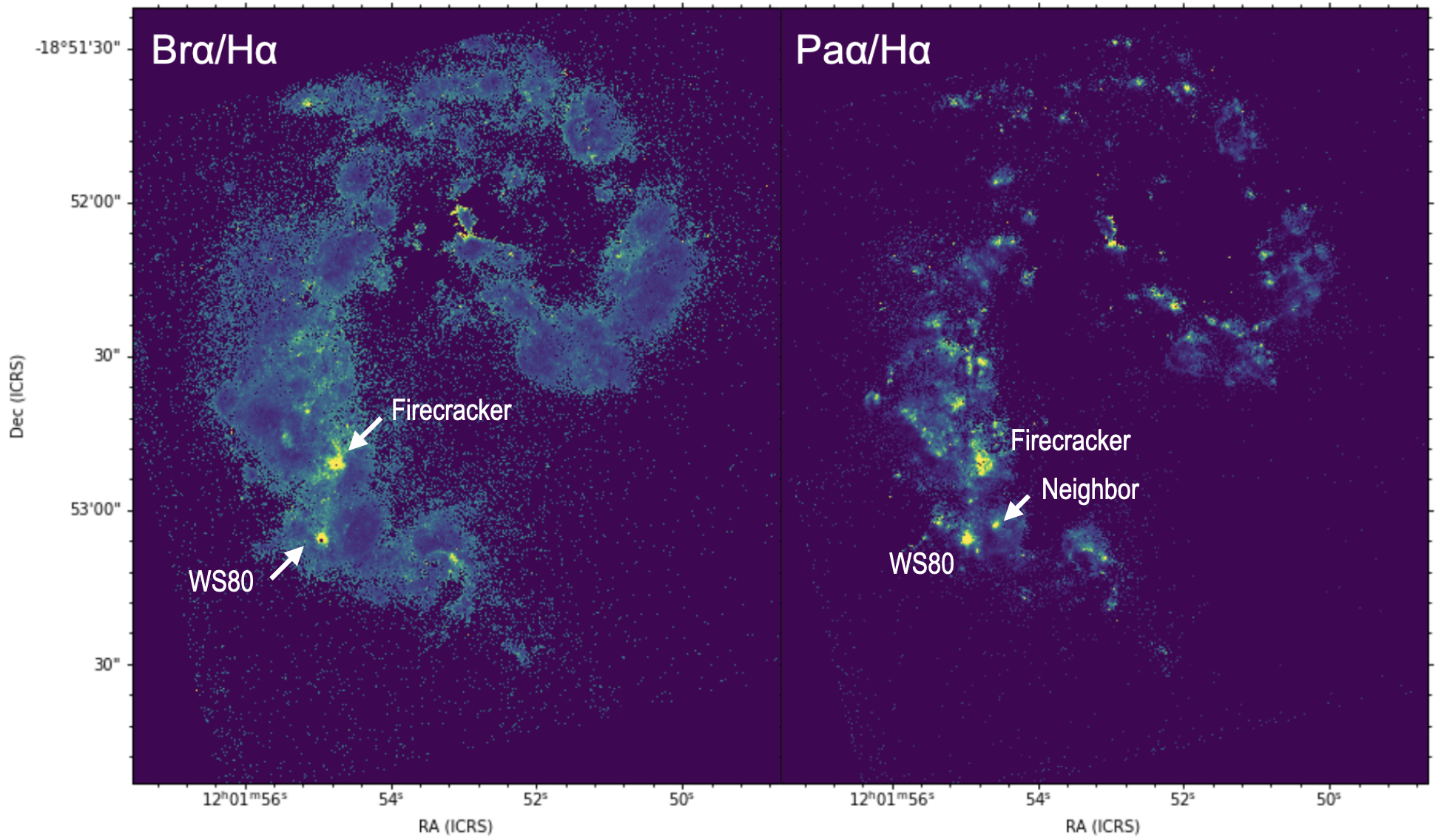}
	\caption{Br$\alpha / \mbox{H}\alpha$ (left) and Pa$\alpha / \mbox{H}\alpha$ flux ratio maps used to identify embedded clusters from their high extinction.  WS80 is the brightest source in both images, while its neighbor, which is known to have lower extinction is only bright in Pa$\alpha$/H$\alpha$. The Firecracker is bright in both ratio maps but star formation has yet to take place in this region (it may in the near future).} \label{fig:detection}
\end{figure}

Maps of Br$\alpha$/H$\alpha$ and Pa$\alpha$/H$\alpha$ flux ratios are ideal for identifying sources with high nebular extinction \citep[e.g.,][]{Brandl12,Messa21,Graham25}.
As a guide to the specific flux ratios to use from our maps, we examine the measured ratios for WS80 and its companion, the latter being another very young, massive cluster with moderate reddening which was first identified by \citet{Rubin70}. 
Images of this cluster pair are shown in Figure~\ref{fig:WS80} from $0.275\,\mu$m through $7.7\,\mu$m, with WS80 identified by the cyan cross.  WS80 is not detected in the NUV, U, or B filters, is very faint in V, H$\alpha$, and I, becomes progressively brighter towards longer wavelengths and by $\approx1.5\,\mu$m is the brightest source in the Antennae.
By contrast, its neighbor is the brightest source (excluding the galactic nuclei) in the near-ultraviolet (F275W) and U (F336W) bands, and remains bright over all plotted wavelengths, including Pa$\alpha$ and Br$\alpha$.
WS80 and its companion are among the brightest sources in the Pa$\alpha/$H$\alpha$ line ratio map shown in the right panel of Figure~\ref{fig:detection}, and WS80 is the brightest source in both maps.

Based on the observed ratios for WS80 and its neighbor, we require a candidate embedded cluster to have (at least) one pixel with Pa$\alpha$/H$\alpha \geq5$ or
Br$\alpha$/H$\alpha \geq1$, or to have at least
3 contiguous pixels with Pa$\alpha$/H$\alpha \geq3$ or
Br$\alpha$/H$\alpha \geq0.6$.
An initial list of embedded cluster candidates which satisfied these criteria are selected automatically from the flux ratio maps, followed by careful visual inspection of each candidate in both maps. 
Two additional criteria were imposed on sources that satisfied these line ratios:
(i) the source must have similar morphology in the Pa$\alpha$/H$\alpha$ and Br$\alpha$/H$\alpha$ ratio maps (this step eliminated a number of point sources with imperfect continuum subtraction); and
(ii) the source must be broader than the point spread function (PSF) at 3.6\,$\mu$m (more below).

We identify 45 candidate embedded clusters which satisfy our selection criteria and list their coordinates in Table~\ref{tab:sample}. 
All have Pa$\alpha$ and Br$\alpha$ emission; we did not discover any sources with Br$\alpha$ but no Pa$\alpha$ emission.
Not surprisingly, most of the embedded clusters (28/45 or $\approx60$\%) are found in the overlap region between the two galaxies.
Approximately 15\% (6/45) are in the southern galaxy (all but one in the optically-dark star-forming arc around the nucleus) while $\approx 25$\% (11/45) are forming in the disk of the northern galaxy.
We do not identify any embedded cluster candidate in the `Firecracker' \citep{Whitmore14}, despite high flux ratios (Br$\alpha$/H$\alpha>1$) throughout the region (see Figure~\ref{fig:detection}).
The Firecracker is a well-studied, infrared-bright region located north and slightly west of WS80 (see Figure~\ref{fig:detection}) with no known star or cluster formation (\citealt{Whitmore14, Johnson15, Finn19}, although see \citealt{Herrera17}) for a different view), lending confidence to our selection criteria.

\input{catalog}

\begin{figure}[!ht]
	\centering
\includegraphics[width=4.0in]{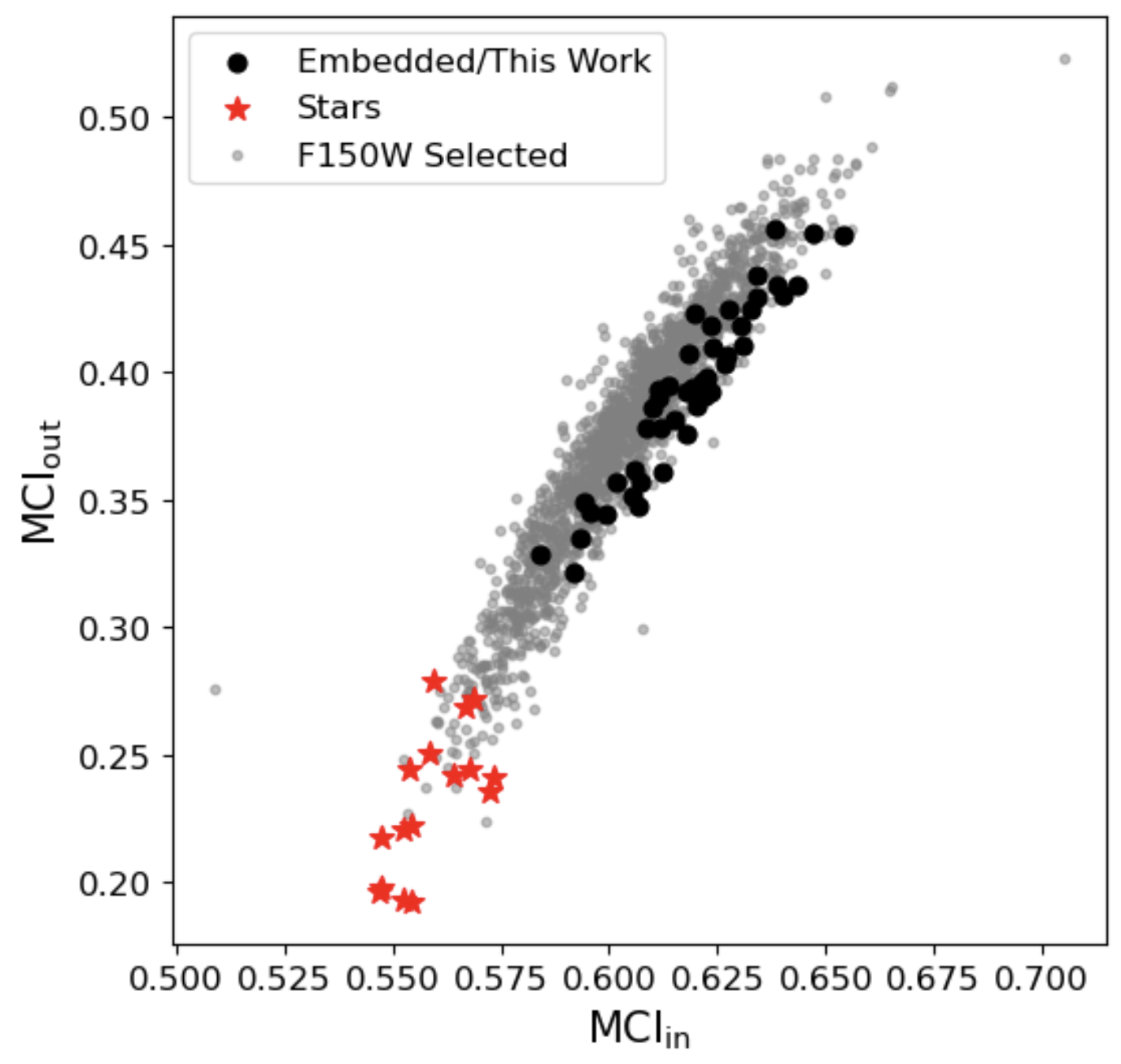}
	\caption{Structural properties measured in the F360W filter for the embedded cluster candidates detected in this work (black circles) are compared with a training set of stars (red symbols) and a new catalog of F150W-selected clusters (gray points).  The embedded sources overlap with the cluster sample.  Multiple concentration indices (MCI) measure the magnitude differences between different apertures, and are a simple method to determine object size.  The definitions of MCI$_{\rm in}$ and MCI$_{\rm out}$ are given in Section~3.1. } \label{fig:CI}
\end{figure}
  
As mentioned earlier, embedded cluster candidates are selected to be broader than the PSF in the F360M filter, since many are faint at $1.5\mu$m but brighter at 3.6$\mu$m.
Following \citet{Thilker22}, we measure a set of multiple concentration indices (MCI) to determine if a source is broader than the PSF, where a concentration index is simply the difference in magnitude measured in two apertures.
MCI is defined for pairs of radii $(a,b)$, $(b,c)$, and $(c,d)$ as the average:   
$\mbox{MCI} = \overline{[{\rm NCI}(a,b),{\rm NCI}(b,c),{\rm NCI}(c,d)]}$, where NCI is the concentration index 'normalized' relative to a fiducial cluster.
We calculate an inner and outer value using radii of 1.0, 1.5, 2.0, and 2.5~pixels (2.5, 3.0, 4.0, 5.0~pixels) as $a, b, c, d$ for MCI$_{\rm in}$ (MCI$_{\rm out}$).
The results in Figure~\ref{fig:CI} show that the embedded cluster candidates have higher MCI values than 
a hand-selected training set of stars, indicating they are more extended than the PSF. These overlap with the structural parameters measured for a new catalog of F150W-selected clusters, although the embedded clusters tend to have somewhat less extended outskirts than older clusters (i.e., they have slightly lower MCI$_{\rm out}$ for a given MCI$_{\rm in}$), consistent with expectations that these very young clusters will expand as they age; \citealt{Chandar16}).

%
\begin{figure}[!ht]
	\centering
\includegraphics[width=4.0in]{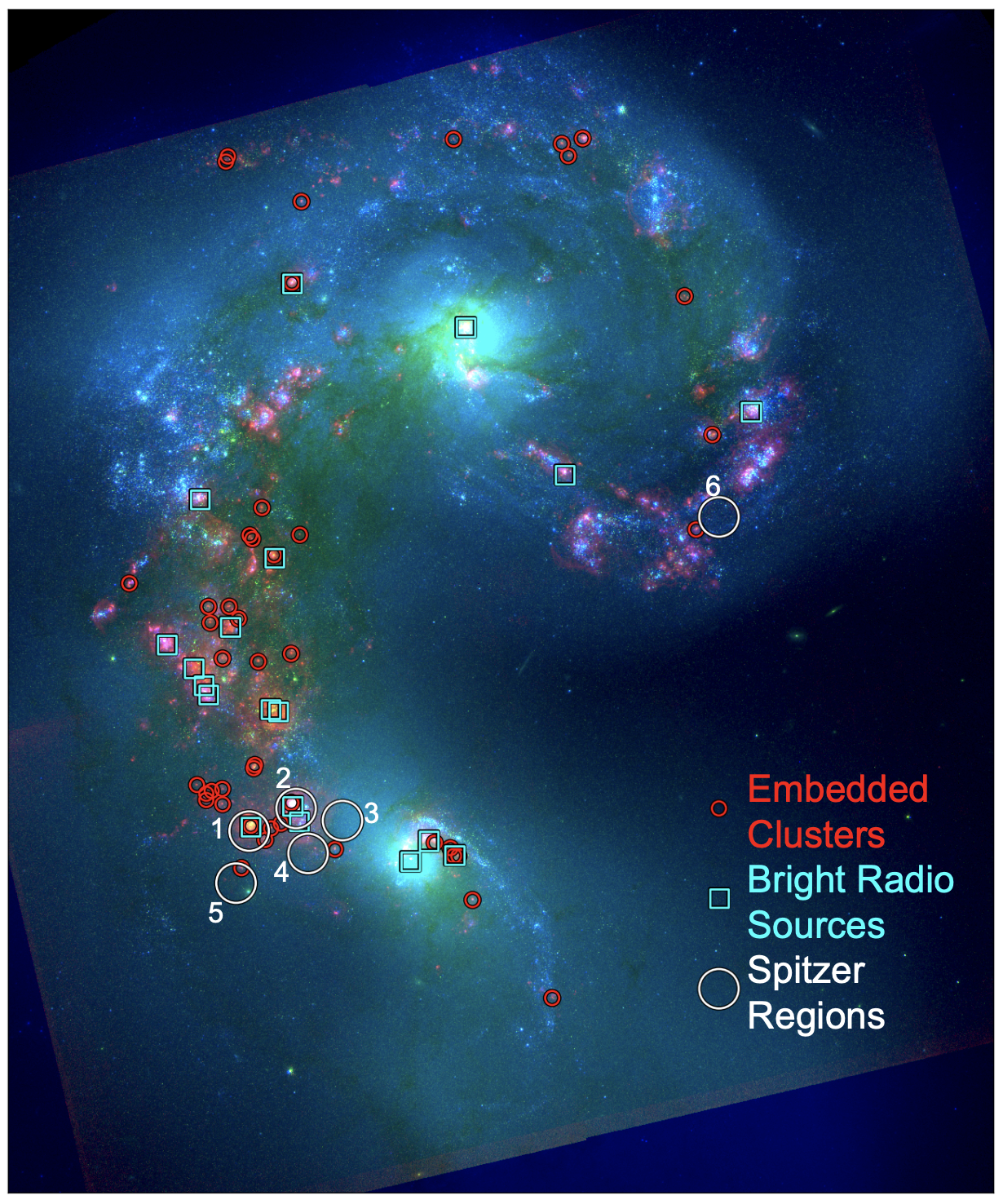}
	\caption{Locations of the embedded clusters identified in this work are identified by the red circles. All 19 `bright' radio sources (6~cm flux $\gea 200$~mJy; see Appendix) are shown as cyan squares, and the six apertures (circles with a $2.55\arcsec$ radius) used to extract Spitzer/IRS spectra are shown as the white circles. The background color image is a combination of V-F150W-Pa$\alpha$ filters.  }
\label{fig:radiolocation}
\end{figure}

We emphasize that our 45 embedded cluster candidates are not the full sample of clusters in the Antennae, just some of the youngest and most extinguished ones.
In an upcoming work (M. Caputo et al.), we will present a new catalog of star clusters selected to be broader than the PSF in the F150W JWST image.  This catalog includes clusters which are both obscured and unobscured in V band images taken with HST.
Figure~\ref{fig:CI} shows that the embedded clusters selected here (black circles) have similar structural parameters to the 1.5\,$\mu$m selected clusters (gray circles).

\subsection{An Assessment of Different Infrared Tracers for Identifying Embedded Clusters}

A number of recent works have used different combinations of high-resolution Pa$\alpha$, Br$\alpha$, and 3.3\,$\mu$m JWST imaging to identify embedded star clusters in nearby galaxies \citep[e.g.,][]{Rodriguez23,Rodriguez25,Gregg24,Knutas25,Graham25}.
The FEAST collaboration has defined embedded candidates (designated as emerging young star clusters or 'eYSCI') as sources with a Br$\alpha$ and 3.3$\mu$m PAH peak within 4~pixels ($\sim0.16\farcs$) of a Pa$\alpha$ emission peak.
Their selection results in 737 sources in the spiral galaxy NGC~628 that have an average reddening $E(B\!-\!V) \sim 0.5$~mag or extinction A$_V\sim 1.5$~mag \citep{Gregg24}, and in 946 sources in M83 with a median $E(B\!-\!V) \sim 0.8$~mag or A$_V\sim 2.5$~mag \citep{Knutas25} for a Milky Way-like extinction curve.
They suggest that clusters with peaked Pa$\alpha$ and Br$\alpha$ emission but diffuse (and weaker) 3.3$\mu$m PAH emission (designated 'eYSCII') are somewhat older (by $\approx1-2$~Myr) than the eYSCI sources.

As part of the PHANGs collaboration, \citet{Rodriguez23, Rodriguez25} selected embedded cluster candidates which may be missed in optically-selected samples in spiral galaxies through their 3.3$\mu$m PAH emission.
While they did not estimate extinction, they found that $\sim40$\% of the 3.3$\mu$m sources detected in NGC~7469 were not selected as clusters at optical wavelengths \citep{Rodriguez23}.

While identifying embedded cluster candidates in the Antennae based on the selection criteria used by the FEAST and PHANGS collaborations is beyond the scope of this paper, we offer a qualitative assessment of the source catalogs that would result from applying these selection criteria to our Antennae data.
We can easily count a few hundred sources with overlapping  Pa$\alpha$, Br$\alpha$, and 3.3$\mu$m emission, including our embedded clusters. These sources have significantly lower Br$\alpha$/H$\alpha$ and Pa$\alpha$/H$\alpha$ ratios than the thresholds adopted here, and therefore almost certainly experience less extinction.
Most of these sources are identifiable in the HST V-band image, supporting the idea that they are young, line-emitting clusters with low extinction A$_V$.  
We conclude that combinations of Pa$\alpha$ (Br$\alpha$), and 3.3$\mu$m emission provide an excellent way to identify very young clusters regardless of the amount of extinction they experience.
The method used here of identifying clusters based on their high Pa$\alpha$/H$\alpha$ and Br$\alpha$/H$\alpha$ line ratios preferentially selects the subset of very young clusters with higher extinction, most of which are not identifiable as clusters from their broad-band optical emission.


\section{Deriving the Properties of Embedded Clusters}

\subsection{SED Fitting}

In this section we estimate the age, reddening, and mass of each embedded cluster by fitting its observed SED with predictions from cluster evolution models that extend to infrared wavelengths and include emission from both stars and ISM.

In preparation for SED fitting, we perform aperture photometry using the {\sc photutils} package \citep{photutils} using a measurement radius of 0\farcs155 uniformly across all HST and JWST/NIRCam images. The sky background is measured in an annulus with radii between 0\farcs50 and 0\farcs62 from the source centers. 
The shape of the SED is affected by the age of the cluster stars and the amount and condition of the ISM, and its normalization is tied to cluster mass.
We estimate the amount of light missing in the aperture by measuring the magnitude difference of each source between 0\farcs155 and 0\farcs465 in the F360W filter (difference between 5 and 15~pixel radii).  The median aperture correction is $-1.29$~mag, and the values measured for each cluster are listed in column~7 of Table~\ref{tab:sample}.
In a future paper we will present size-dependent aperture corrections for a full catalog of optically obscured$+$unobscured star clusters in the Antennae, but curves of growth for the embedded clusters suggest that the simple approach adopted here gives results within $\approx25$\% of the total flux.

For the SED fitting, we adopt the flexible stellar population synthesis (FSPS) code \citep{Conroy+09, Conroy+10} to determine the best fit combination of age and reddening.  The models include dust extinction and emission, nebular line and continuum emission, and circumstellar emission around AGB stars. Full details of the fitting procedure will be provided along with the total sample of clusters at all ages in an upcoming work (M. Caputo et al., in preparation); here we summarize the basic assumptions and method. 
We use the Python package \textsc{python-FSPS} \citep{python-fsps} to create integrated-light spectra of single-age stellar populations (SSPs) with solar metallicity and a \citet{Kroupa01} initial mass function (IMF). The synthetic cluster spectra are placed at the redshift of the Antennae ($z$ = 0.0055, cf.\ \href{https://ned.ipac.caltech.edu}{NED}).
For the age range, we choose $5.5 \leq {\rm log\,(age/yr)} \leq 10.2$ in steps of 0.1 dex.  
For extinction, we allow the range $0.0 \leq A_V \leq 80.0$~mag in steps of 0.1 mag, assuming a Milky Way-like value for $R_V$ of 3.1, but find very similar age-A$_V$ results if we adopt the \citet{Calzetti10} extinction law instead.
We select models from the PARSEC isochrone family \citep{Bressan+12, Nguyen+22} and the MILES spectral library \citep{MILES1, Vazdekis+10, Vazdekis+16}.

To create predicted cluster photometry, the spectra are integrated over the HST/ACS, HST/WFC3, and JWST/NIRCam passbands by incorporating throughput tables downloaded from the STScI website into the FSPS package.
Predicted magnitudes are derived based on an assumed distance to the Antennae of $m-M$ = 31.66 \citep{Riess+11}. 

The best-fit age and reddening for each cluster are determined as follows. For each cluster, we first normalize the model magnitudes by subtracting the weighted mean offset between the model grid and observed cluster magnitudes, using inverse variance weighting. After this normalization, we determine $\chi^2$ values for each predicted combination of age and reddening:
\begin{IEEEeqnarray}{rCl}
  \chi^2 & = &  \sum_{i = 1}^{N} \; \left(\frac{m_{{\rm model},\,i} - m_{{\rm obs,}\,i}}{\sigma(m_{{\rm obs},\,i})}\right)^2 \IEEEyesnumber
    \label{eq:chi2}
\end{IEEEeqnarray}
where $N$ is the number of filter passbands being fitted, and $\sigma (m_{{\rm obs},\,i})$ represents the observational uncertainty (in magnitudes) for the filter. 

Uncertainties for log\,(age/yr) and $A_V$ are determined as the values
within $\pm$ 34 percentile of the lowest $\chi^{-2}$ value, equivalent to $\pm 1 \sigma$ for a normal distribution. The log\,(age/yr) vs.\ $A_V$ grid is oversampled by a factor 10 in this step, using linear interpolation. 
The mass 
and associated uncertainty is calculated from the extinction-corrected F360W magnitude and best fit, 
plus uncertainties for log\,(age) and $A_V$.
The best-fit age, A$_V$, mass and associated uncertainties are listed for each cluster in Columns~4-6 of Table~\ref{tab:sample}. 

\subsection{\texorpdfstring{Age, Extinction, Mass Results}{Age and Av Results}}

\begin{figure}[!ht]
	\centering
\includegraphics[width=5.0in]{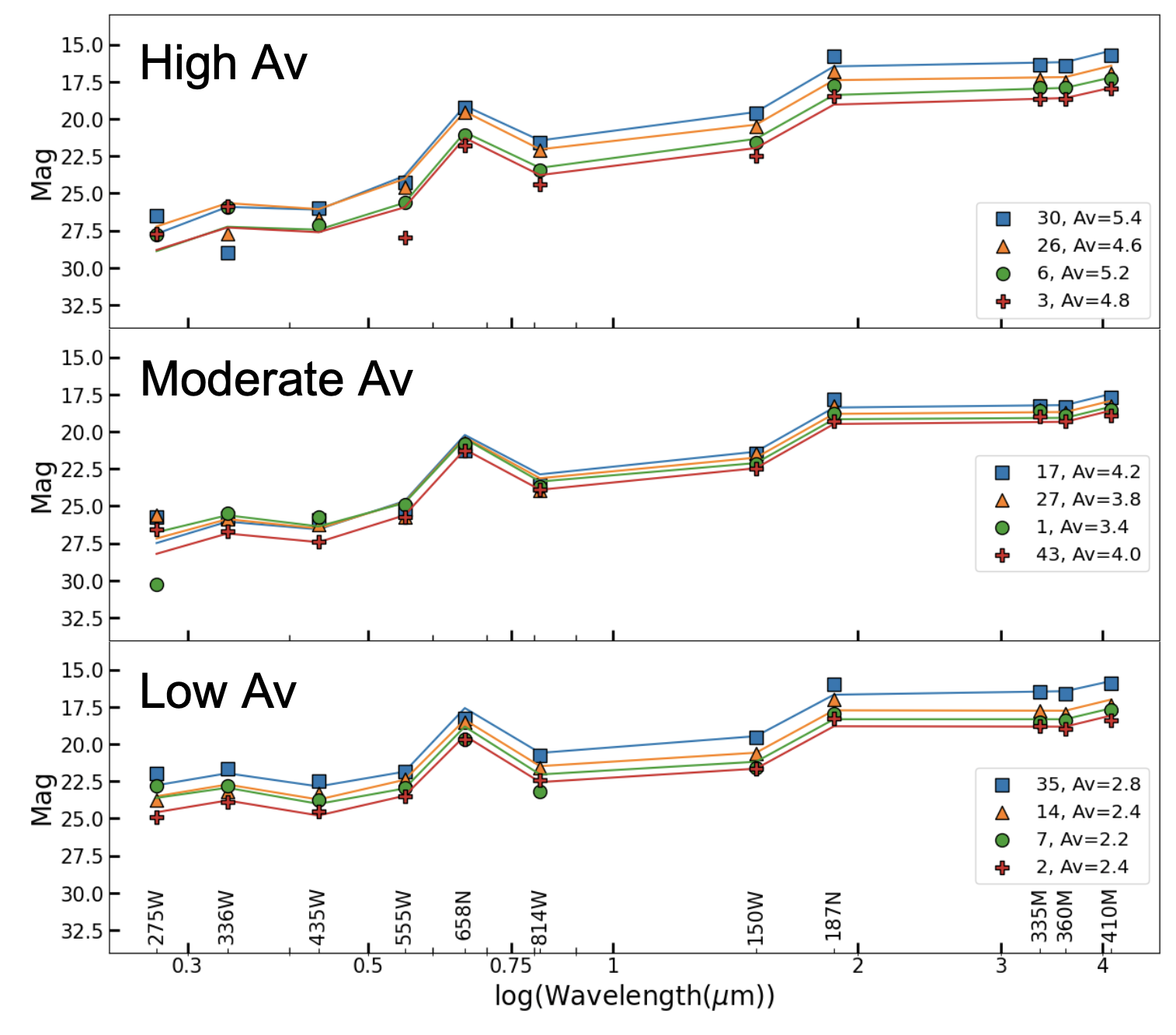}
	\caption{Observed photometry and best fit spectral energy distributions for a subset of clusters with high (top), moderate (middle), and low (bottom) extinction A$_V$.  Photometric uncertainties are included, but are mostly smaller than the symbol size.
    All of the embedded clusters are very young with best fit ages $\lea 2.5$~Myr so their spectral energy distributions (SEDs) are driven mostly by extinction. Differences in the shapes of the SED can be seen, with the high A$_V$ clusters becoming quite faint at short wavelengths (many are not detected in the NUV and U filters),
    the low A$_V$ clusters have blue SEDs, while those with moderate A$_V$ have intermediate slopes at short wavelengths.  Note that H$\alpha$ emission is detected in most embedded clusters, even those with the highest A$_V$. } \label{fig:SEDfits}
\end{figure}

\begin{figure}[!ht]
	\centering
\includegraphics[width=3.5in]{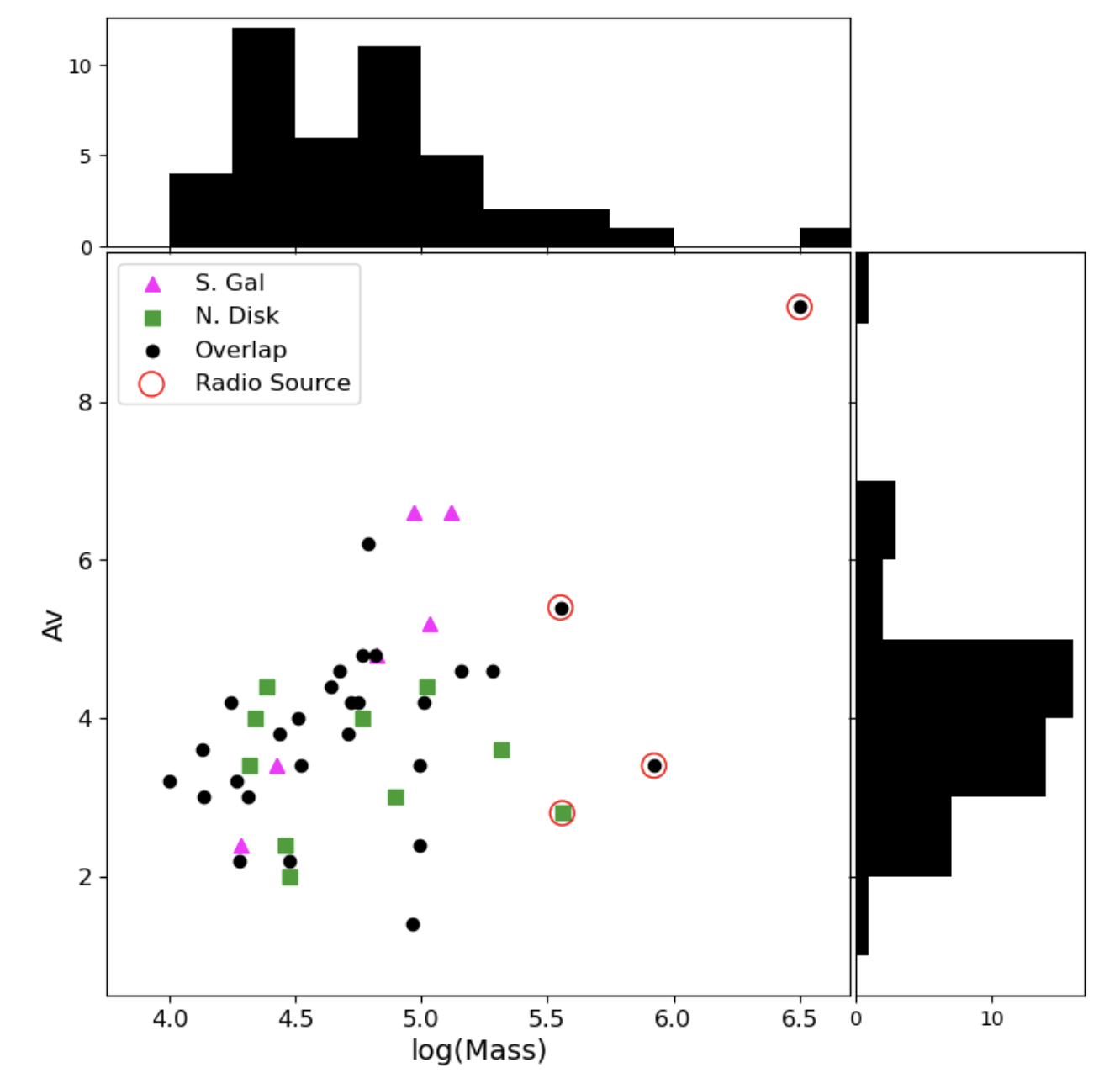}
	\caption{Best fit mass and extinction A$_V$ for the embedded cluster sample. Clusters in the southern galaxy are plotted in purple, overlap region in black, and northern disk in green. All clusters have estimated ages of 2.5~Myr or younger.  The A$_V$ values range from $\approx2$~mag up to $\approx10$~mag with half the sample (22 of 45) having a best fit A$_V$ of 4~mag or higher. The red circles show that clusters also detected at radio wavelengths are the most massive in our sample (see \S5.4).} \label{fig:massAv}
\end{figure}

In Figure~\ref{fig:SEDfits} we show results from the SED fitting described above for 12 clusters from our sample.  
We show the photometry measured in all 11 HST$+$JWST filters (data points) and the best fit results (solid line) based on the procedure described above.
Photometric uncertainties are included, but are mostly smaller than the data points.
The panels show examples of similar-age clusters with high A$_V$ ($\approx6-7$~mag, top panel), moderate A$_V$ ($\approx4$~mag, middle panel), and low A$_V$ ($\approx2.5$~mag, bottom panel) from our sample.

Overall, the fits look quite good.
As expected, all clusters in the sample are very young with a best-fit age of $\approx2.5$~Myr or younger,
and the shape of the SED is mainly driven by extinction.
We see that none of the Antennae clusters with the highest extinction (including WS80) are detected in the F275W/NUV or F336W/U filters, and that their measured and best-fit continuum slopes drop sharply towards short (blue) wavelengths.
Despite experiencing high extinction, H$\alpha$ emission is detected from massive clusters with $A_V \gea 6$~mag.
Clusters with moderate extinction have shallower slopes (decrease less) towards short wavelengths and have strong emission lines, and those with low extinction have SEDs that are flat or increase towards the near-ultraviolet.


In Figure~\ref{fig:massAv} (central panel) we plot the best fit mass vs. A$_V$ for each embedded cluster in our sample.
The clusters have a range of extinction A$_V$ between $\approx2-10$~mag.
The histogram along the right side shows a predominant value of A$_V \approx4$~mag.
The embedded clusters have estimated masses (listed in Column~6 of Table~\ref{tab:sample}) that range from $\approx3\times 10^4~M_{\odot}$ to $\approx3\times 10^6~M_{\odot}$.
Overall, we see there are fewer high-mass clusters (the distribution of embedded cluster masses will be discussed in Section~5.3) and fewer clusters with high extinction. 
  
\subsection{Comparison with Published Cluster Properties}

Our estimates of the age, extinction, and mass for the most massive embedded cluster in the Antennae, WS80, agree well with many previously published results.
We find this cluster has an age log~$(\tau/\mbox{yr})=6.3^{+0.63}_{-0.80}$, A$_V=9.2^{+1.36}_{-1.16}$~mag 
(the highest in our sample) and an estimated mass of at least log~(M$/M_{\odot})=6.5$ (for this cluster and its neighbor, the aperture corrections are lower-limits).
Our age compares well with published estimates of log~$(\tau/\mbox{yr})\approx 6.0$ \citep[e.g.,][]{Whitmore10}.
Published values of extinction include A$_V=7.3$~mag \citep{Whitmore10} and $\approx9-10$~mag \citep{Gilbert00}, and the published range of masses is between $10^6$ and  $10^7~M_{\odot}$ \citep[][]{Whitmore95,Whitmore10,Gilbert07,Herrera17}.

The companion to WS80 is also quite young with a best fit age of log~$(\tau/\mbox{yr})=6.4$, A$_V=3.4\pm0.2$~mag, and log~$(M/M_{\odot})=5.9$. 
Previously, \citet{Whitmore10} estimated log~$(\tau/\mbox{yr})=6.4$, A$_V\approx 3$~mag, and log~$(M/M_{\odot})=6.6$ from optical photometry alone.
Our embedded cluster \#\,30 has best-fit values of log~$(\tau/\mbox{yr}) < 6.0$, A$_V=5.4\pm1.2$~mag, and log~$(M/M_{\odot})=5.5$, in good agreement with the \citet{Whitmore10} results of log~$(\tau/\mbox{yr}) \approx 6.0$, A$_V \approx 4$~mag, and log~$(M/M_{\odot})\approx6.0$.

\citet{He22} estimated the optical extinction $A_V$ towards several clusters including WS80 and its neighbor, from high resolution continuum observations at 100~GHz and 345~GHz taken with ALMA.
They derived extinction values from estimates of the gas surface density, and 
found A$_V \approx 2200$~mag and $\approx 450$~mag for WS80 and its neighbor, respectively, significantly higher than the estimates presented here, although their calculations rely on assuming the shape of the thermal dust and free-free spectral indices.
It is possible that the frequency coverage of their radio observations are insufficient to constrain cluster extinction.


\section{Discussion}

\subsection{Nowhere Left to Hide}


Some of the most intense star formation in the nearby universe is happening in merging galaxies.  Since star formation takes place in dense, dusty molecular clouds, the earliest stages of star and cluster formation are obscured at optical wavelengths.
However it still is not known whether most sites of cluster formation are optically obscured and embedded in thick layers of dust with A$_V\approx40-70$~mag \citep[e.g.,][]{Stanford90,Kunze96,Mirabel98,Klaas10},
or if most (massive) clusters have A$_V \lea 10$~mag and have already been detected at optical wavelengths \citep{Whitmore02}.
With our deep, high resolution infrared JWST images we are now in a position to directly answer this question by uncovering the deeply embedded, recently formed cluster population in the Antennae.

We believe there are no massive clusters with A$_V \gea 10$~mag in the Antennae, and that we have uncovered essentially all clusters with $M \gea 3\times10^4~M_{\odot}$ and extinction A$_V\gea2$~mag.
One key clue is that we did not find any sources that have Br$\alpha$ but no Pa$\alpha$ emission; Pa$\alpha$ emission generally indicates that a source must have extinction A$_V$ less than 20~mag.
More quantitatively, we can estimate the range of detectable A$_V$ for a given mass by scaling the observed parameters of WS80, which has a mass $M\approx3\times 10^6~M_{\odot}$, extinction A$_V \approx10$~mag, and is $\approx 7$~mag brighter than the detection limit in the Pa$\alpha$ filter.
Since A$_{Pa\alpha} \approx 0.1\times A_V$, if WS80 experienced no extinction it would be $\approx8$~mag brighter than the detection limit.
Simple scaling arguments indicate that our Pa$\alpha$ observations are sufficiently deep to detect $3\times 10^5~M_{\odot}$ ($3\times10^4~M_{\odot}$) clusters with extinction A$_V$ up to $\approx55$~mag ($\approx 30$~mag).
All but 13 of the embedded clusters have masses $M \gea 3\times10^4~M_{\odot}$, so the Antennae has formed plenty of massive clusters, we just do not detect any with extinction higher than A$_V \approx10$~mag.

We estimate the fraction of very young clusters missing from optical surveys by combining our new embedded cluster list with the HST-based cluster catalog published by \citet{Whitmore10}.  We adopt a mass limit of $3\times10^4~M_{\odot}$, the approximate completeness limit for the embedded cluster catalog (see Section~5.3). Our new catalog has 38 clusters above this mass limit, but four are included in the optical catalog, so there are 34 unique new clusters above the completeness limit.
The \citet{Whitmore10} catalog has 156 clusters younger than $\tau \leq 3$~Myr and more massive than $3\times10^4~M_{\odot}$, suggesting their optical study is missing  $\approx18$\% ($34/(156+34)$) by number of these very young clusters in the Antennae.  
If we consider the 306 (788) optically-selected clusters above this mass that have estimated ages $\leq10$~Myr ($\leq100$~Myr) this percentage drops to $\approx10$\% ($\approx4$\%), which
represents a small fraction of the total {\em number} of clusters in the Antennae.
 While the embedded population is just a few percent of the total cluster population, it emits the majority ($\approx60$\%) of the total ionizing photon luminosity in the Antennae, with 
WS80 alone emitting $\approx27$\% of the ionizing photons based on estimates from the ages and masses of the embedded and optically-detected clusters. 
This is consistent with the 15\% of the total 12.5-18$\mu$m luminosity measured for WS80 \citep{Mirabel98}.

\begin{figure}
    \centering   \includegraphics[width=0.90\textwidth]{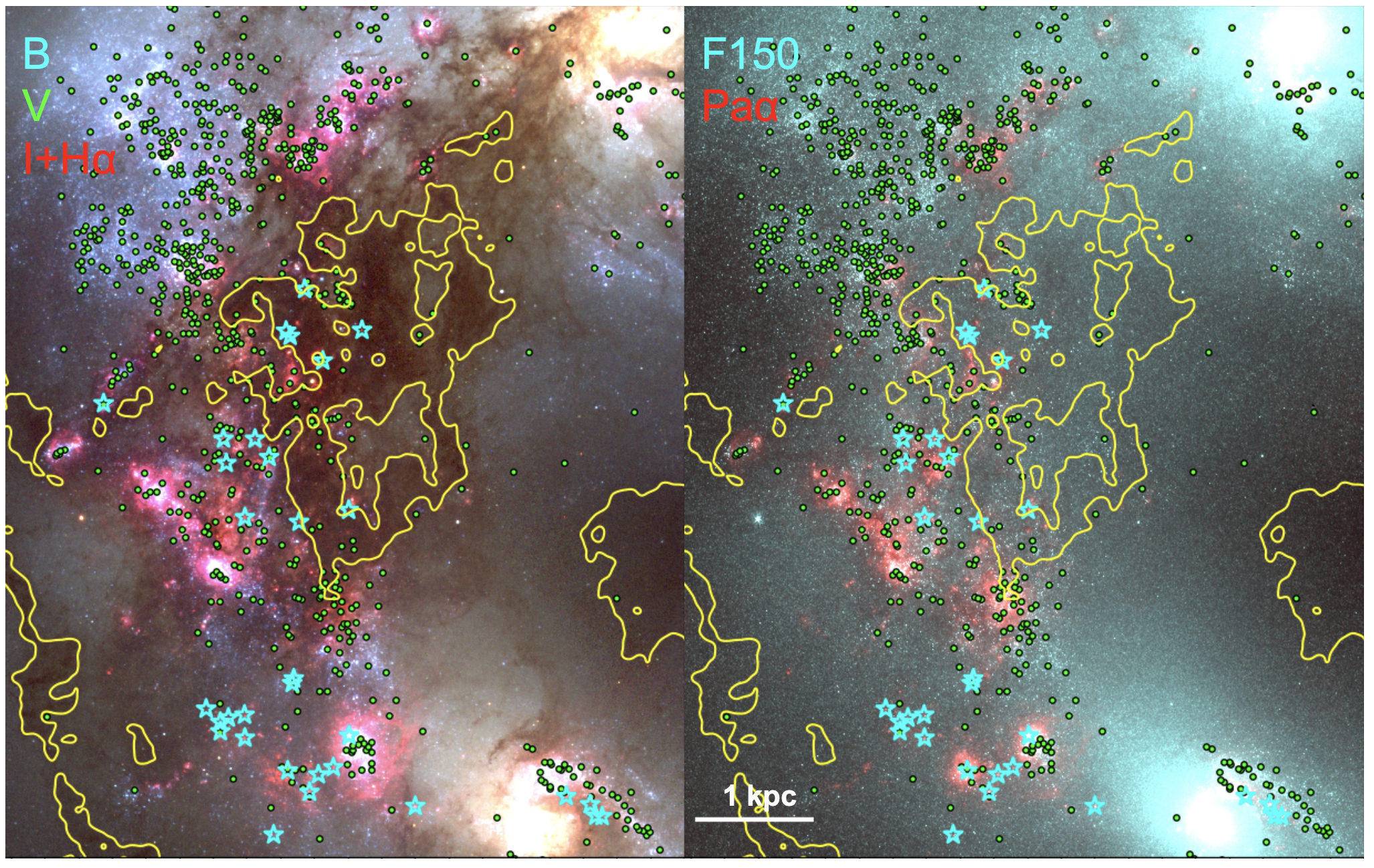}
    \caption{Few clusters (of any age) are forming in the prominent, dark dusty regions seen in optical images of the overlap region in the Antennae. This indicates the ISM in these regions is not currently participating in star formation and is likely in the foreground.  
    Different combinations of optical and infrared filters are shown in each panel (as indicated).  Contours identify dark regions in the B-band image, small green circles show the locations of the F150W-detected clusters and the cyan stars show the locations and amount of ionization from very young clusters. 
    }
    \label{fig:darkareas}
\end{figure}

Figure~\ref{fig:darkareas} illustrates that star formation, traced in pink by H$\alpha$ (left) and Pa$\alpha$ (right), is on-going throughout much of the overlap region, with no obvious pattern from east to west.
However, little star-formation is taking place in a number of optically dark, dusty regions.
The yellow contours highlight dark regions in the B band image.  It is clear that there are very few embedded clusters forming in these dusty regions (cyan stars), or clusters of any age, since there are almost no small green circles (candidate star clusters selected in the JWST/F150W image) within the contours.
The most likely explanation for the dust lanes and clusters having different spatial distributions is that the dust is cold and in the foreground of any optically emitting material.
In fact, the brightest clusters in our Br$\alpha$ and 3.6$\mu$m
infrared images do not coincide with the most prominent dark, dust-affected areas in the optical HST images, indicating that much of the dust/ISM in the overlap region is not hiding any significant amount of star formation.

\subsection{The Embedded Cluster Population in the Antennae}

\begin{figure}
    \centering
    \includegraphics[width=0.60\textwidth]{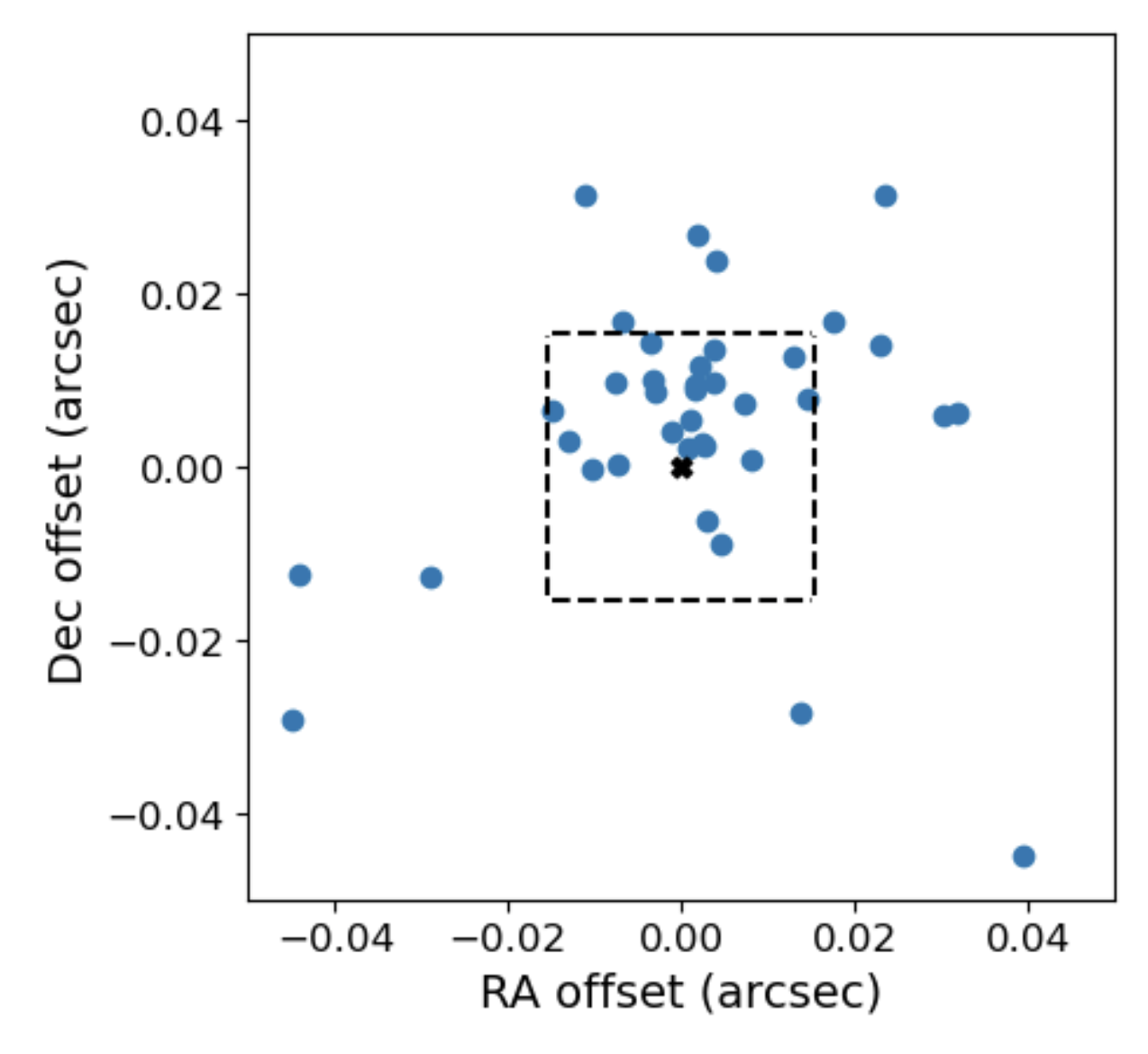}
    \caption{Offsets (in arcseconds) between the measured centers of the embedded clusters in the F187N (Pa$\alpha$) and F150W (stellar continuum)
    filters, where $1\arcsec$ subtends 104~pc at the distance of the Antennae. The dashed box shows the size of a NIRCAM short wavelength pixel.  Despite Pa$\alpha$ and 1.5$\mu$m originating from different processes, we see that the sources peak in similar locations in both filters.  Some of the embedded clusters are quite faint at 1.5$\mu$m, while others are somewhat extended in Pa$\alpha$, which likely contributes some scatter to the measurements.  Overall, the emission from recombination lines and stellar continuum are consistent with zero spatial offset for these very young, embedded clusters.  
    }
    \label{fig:offset}
\end{figure}

Massive stars formed in very young $\lea 3$~Myr clusters ionize the gas in their immediate vicinity but live inside a larger, surrounding cloud of neutral and molecular gas. 
Pre-supernova stellar feedback (stellar winds, photoionization, and radiation pressure) can carve channels in the ISM through which the starlight and ionizing radiation can escape \citep[e.g.,][]{Whelan11,Howard17}, although the timescale for this process is not well constrained.

We assess the multi-wavelength contributions of ionized gas, stellar continuum, and dust emission from the embedded clusters based on the JWST$+$HST images and their measured photometry.  
All embedded clusters have compact Pa$\alpha$ and Br$\alpha$ emission, but nine of them (\#2, 3, 4, 7, 10, 24, 31, 36, 37) have no detectable or very weak H$\alpha$ emission.
All have detected stellar continuum emission at 1.5$\mu$m
and all have strong 3.6$\mu$m emission.
We also detect compact 3.3$\mu$m PAH emission from all of the embedded clusters except for \#4, which is the closest source to the southern nucleus and has a strange SED; 3.3$\mu$m PAH emission has been shown to decrease towards the nucleus of at least one nearby Seyfert galaxy (NGC~7469; \citealt{Lai23}).

\citet{He22} suggested that the near-infrared and radio emission from embedded clusters might originate from slightly different parts of the dense molecular cloud, as predicted by the 'blister' model, as a possible explanation for the high extinction values they derived (Section~4.3).
In Figure~\ref{fig:offset} we present the spatial offsets in arcseconds measured between Pa$\alpha$ (ionized gas) and 1.5$\mu$m (stellar continuum)
emission where each cluster was centered individually in each image.
The dashed square represents the size of a NIRCAM pixel.
The figure shows that the stellar continuum and ionized gas emission peak at similar locations, as expected for massive clusters with A$_V \lea 10$~mag.  Quantitatively, we find that the embedded cluster locations in Pa$\alpha$ and 1.5$\mu$m are consistent with zero mean offset, with $\Delta$RA$=-0.01 \pm 0.09^{\prime\prime}$, and $\Delta$Dec$=0.01\pm 0.03^{\prime\prime}$.
All of the embedded clusters in our new catalog have estimated ages of 2.5~Myr and younger, with the majority being dated to ages of just 1~Myr or less.
The fact that we detect starlight (even if faint) at 1.5$\mu$m from the embedded clusters from these very young clusters indicates that pre-supernova feedback carves channels in the dense ISM within $\approx1$~Myr, allowing recombination line emission and starlight to escape the parent molecular cloud \citep[e.g.,][]{Kim21,Ramambason25}, and essentially rules out the blister model for the sources studied here.  
And it is not just the most massive clusters that have such short emergence timescales.  Starlight (1.5$\mu$m emission) is detected from clusters over the full observed mass range,
suggesting this timescale does not have a strong dependence on cluster mass in the Antennae.

We checked for differences in cluster properties based on location within the Antennae, considering clusters in the south galaxy (mostly near the nucleus, sources \# 1-6), the overlap region between the two galaxies (\# 7-34), and the north galaxy disk (\# 35-45).
The estimated mass vs. A$_V$ for clusters in the three locations are shown in Figure~\ref{fig:massAv}.  Clusters with masses higher than a few$\times10^4~M_{\odot}$ are forming throughout the Antennae, but the most massive clusters only form in the overlap region.
Clusters with the highest extinction (A$_V \gea 5$~mag) are in the overlap region (black circles) and in the star-forming arc around the southern nucleus (purple triangles), while embedded clusters in the disk of the north galaxy (green squares) have systematically lower extinction, all with A$_V \lea 4$~mag. 
There is no obvious variation in the age estimates with location.

\subsection{Embedded (Initial) Cluster Mass Function}

\begin{figure}[!ht]
	\centering
\includegraphics[width=5.in]{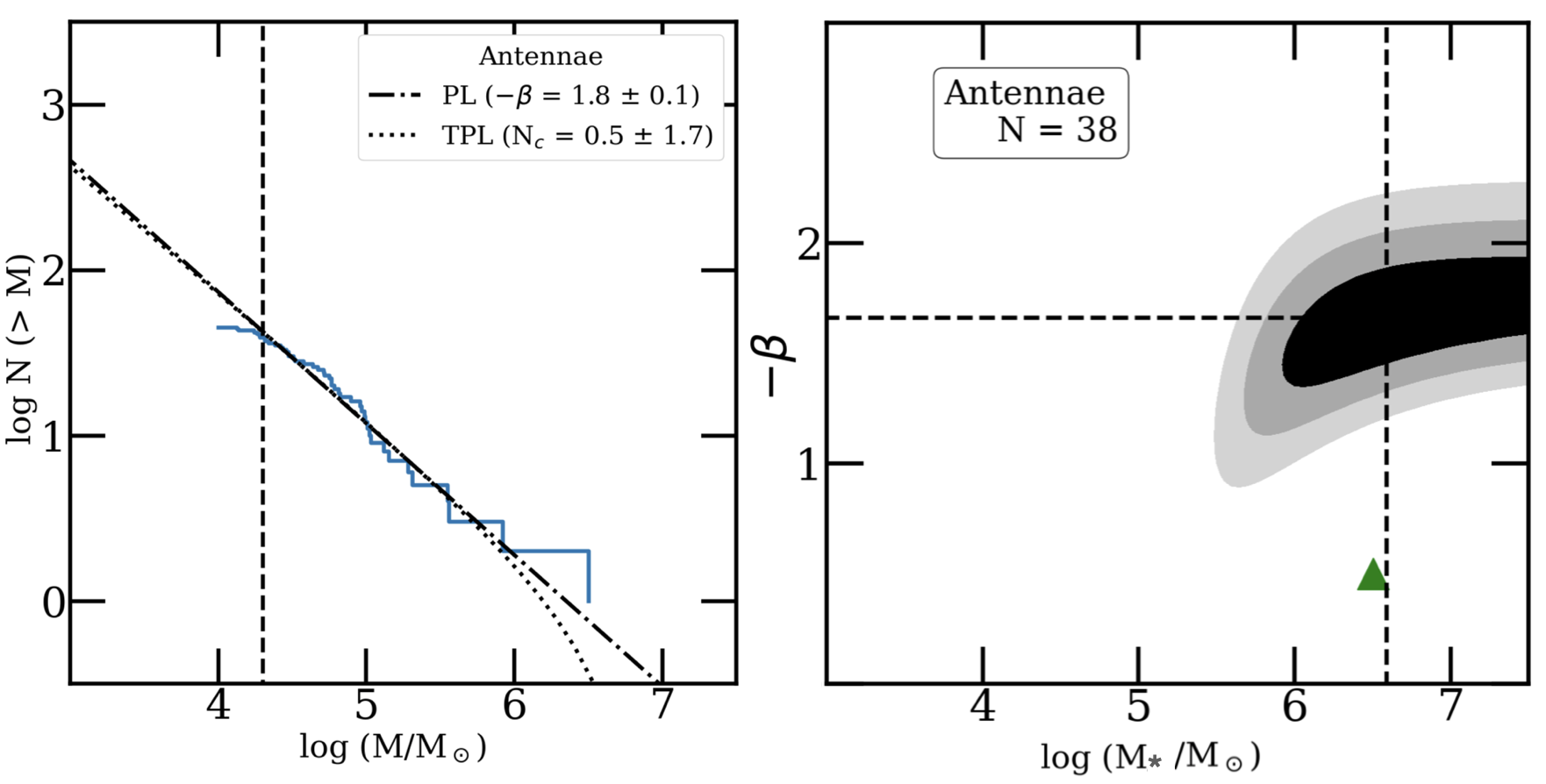}
	\caption{{\bf Left:} Cumulative mass function of the embedded cluster sample.  The completeness limit is shown as the dashed vertical line, and determined to be at the mass where the distribution flattens from a power law. The best power law fit is shown as the dotted–dashed line, and has an index of $\beta=-1.8\pm0.1$.  The best truncated power law fit is shown as the dotted line, and the statistic N$_C$, which gives the number of embedded clusters above the truncation mass, is not statistically significant suggesting that no truncation mass is detected at the upper end of the distribution.
{\bf Right:} The contours show $1\sigma$ (black), $2\sigma$ (gray), and $3\sigma$ (light gray) confidence intervals for the power-law index $-\beta$ and upper mass cutoff M$_*$ for the maximum likelihood fit to a Schechter function for the embedded clusters above the mass limit.
The most massive cluster is plotted as the green triangle.  The upper mass cutoff M$_*$ does not converge for the embedded cluster sample presented here, indicating there is no preference for a Schechter-like cutoff over a power-law. }\label{fig:mf}
\end{figure}

The recent formation of massive clusters in the Antennae make it an ideal system to assess the shape of the initial cluster mass function.
 The cumulative mass distribution of the embedded clusters in the Antennae clearly extends above $10^6~M_{\odot}$, as shown in Figure~\ref{fig:mf} (left panel).  A lower mass limit of log($M/M_{\odot})=4.2$ (dashed vertical line) is determined from the deviation (flattening) of the distribution from a power-law.
A power law, $dN/dM \propto M^{\beta}$, is a good representation of the data, and we find a best-fit value of $\beta=-1.8\pm0.1$, shown by the dash-dotted line in the figure.  
This result is similar to the power-law index of $\beta\approx-2$ found for the mass function of young clusters in many other nearby star-forming galaxies \citep[e.g.,][]{Zhang99,Bik03,McCrady07}.
While a pc-scale simulation of an Antennae-like system suggests that the mass function of the gas flattens in regions of very high pressure \citep{Li20}, we do not find strong evidence for this.
It is worth noting that nearly all previous works include clusters with ages between 1 and 10~Myr in their fits, which may already have experienced some evolution, whereas our distribution includes only the very youngest ($\lea2.5$~Myr) clusters and is therefore a better representation of the 'initial' cluster mass function. 

To assess whether the high-mass end of the embedded cluster distribution shows any hint of a downturn or cutoff, we first fit a truncated power law (TPL) to the cumulative distribution using the MSPECFIT software \citep{Rosolowsky05}.  The results are shown by the dotted line in the left panel of Figure~\ref{fig:mf}.  The N$_C$ value of $0.5\pm1.7$,  the number of embedded clusters above the truncation mass, indicates there is no statistically significant truncation found in the distribution.
Because binned distributions can hide weak features at the ends of the distribution, we also fit a Schechter function, $dN/dM \propto M^{\beta} \,\mbox{exp}\,(-M/M_*)$ with power-law index $\beta$ and exponential cutoff $M_*$ using the maximum likelihood method developed in \citet{Mok19}.  
The resulting 1-, 2-, and 3-$\sigma$ contours for $\beta$ and $M_*$ are shown in the right panel of Figure~\ref{fig:mf}.  The green triangle indicates the most massive embedded cluster in the sample, and the dashed lines show the best formal fit to $\beta$ and $M_*$.
However, all of the contours remain open to the highest mass allowed in the fit, indicating this value of cutoff mass $M_*$ is essentially a lower limit.
This preliminary result suggests that the initial cluster mass function continues as a power law to masses above $10^6~M_{\odot}$ in galaxies with intense star formation, and that the exponential-like downturn observed in the high-end of the globular cluster mass function may result from evolution rather than be imprinted during formation \citep{Fall01}.

\subsection{No Infrared-Dark Thermal Radio Sources in the Antennae}

Radio observations are considered the gold standard for identifying deeply embedded star clusters, since they detect thermal free-free continuum emission from gas ionized by recently formed massive stars  \citep[][]{Kobulnicky99,Johnson01,Johnson04,Johnson09,Johnson03,Aversa11,Kepley14}
and are unaffected by high extinction caused by large amounts of dust \citep{Murphy11}.
In the Appendix we identify the infrared counterparts to 6~cm sources catalogued by \citet{Neff00} and a subset of seven sources identified by \citet{He22} at 100, 200, 345 GHz, and summarize the results of that investigation here.

We identify four and possibly five of the radio sources as embedded clusters from their high Pa$\alpha$/H$\alpha$ and Br$\alpha$/H$\alpha$ flux ratios.
These include WS80 (\#9), its neighbor (\#13), and two other sources in the overlap region (\#30, \#35).
One near the southern nucleus (\#4) may also have radio emission.
These sources have a range of extinction, with three having among the highest in our sample with A$_V \gea6$~mag and two with more moderate values of A$_V \approx3$~mag.
The clusters associated with these radio sources are the most massive in our sample  (red
circles in Figure~\ref{fig:massAv}), with $M \gea \mbox{few}\times10^5~M_{\odot}$.
Two of the sources (4-4 and 8-4 from \citealt{Neff00}) have bright Pa$\alpha$, optical, and near-infrared emission with low Pa$\alpha$/H$\alpha$ and Br$\alpha$/H$\alpha$ flux ratios, and appear to be optically bright star clusters with low extinction.
We did not find any published radio properties (e.g., spectral index or flux ratios) which appear to correlate with extinction, although there are only a few clusters and the published radio properties are incomplete.
The remaining 10 non-nuclear radio sources listed in the Appendix do not appear to be associated with clusters at all and may be non-thermal (e.g., supernova remnants). 
Some of these show a bubble-like Pa$\alpha$ morphology with no obvious counterpart in the broad-band JWST images, and include two sources in the Firecracker region (3-5A and 3-5B).

We find no thermal radio sources that are infrared-dark, since all are detected in the 1.5$\mu$m through 4.05$\mu$m JWST images.
The thermal radio sources are detected in all of our infrared filters, but 40 of the 45 embedded clusters discovered here have no radio counterpart.
These 'radio-dark' clusters mostly have estimated masses less than $\lea 10^5~M_{\odot}$.
Thermal radio sources in the Antennae include deeply embedded clusters ($A_V$ up to $\approx10$~mag), but also clusters that are optically bright and experiencing little extinction.
Regardless of the amount of extinction, only the most massive clusters ($M \gea \mbox{few}\times 10^5~M_{\odot}$) are identified in current radio surveys of the Antennae.

The estimated depth of the two radio surveys discussed above are consistent with the cluster mass limits we have found here. In particular, \citet{He22} find that 100~GHz continuum point sources detected with ALMA at a $S/N \geq 5$ correspond to star clusters with a derived stellar mass of several$\times 10^5~M_{\odot}$. Given the much lower number of 100 GHz detections (17) compared with expectations from existing optically surveys ($\sim 200$), 
\citet{He22} conclude they detect so few embedded clusters because of their high mass limit plus the short-lived phase when embedded clusters are bright at submillimeter wavelengths. 
The \citet{Neff00} radio survey has a $5\sigma$ point-source detection limit of 50~$\mu$Jy at 4 and 6~cm, which corresponds to a luminosity of $\sim3\times10^{25}~\mbox{erg~s}^{-1}~\mbox{Hz}^{-1}$.
We calculate an ionizing photon rate of $\approx10^{51}$ from this luminosity using Equation~10 in \citealt{Murphy11} (for an assumed thermal fraction of 50\% at 5~GHz), which is expected from very young clusters with masses $\approx10^5~M_{\odot}$ \citep{Emig20}. Free-free absorption could potentially suppress the 6~cm thermal flux from embedded clusters, which would effectively raise this mass threshold further.
Our infrared line-ratio based selection method therefore detects embedded clusters that are $\approx10\times$ less massive than current radio surveys in the Antennae.


\subsection{Constraints on the Formation of Very Massive Clusters}

\begin{figure}[!ht]
	\centering
\includegraphics[width=6.5in]{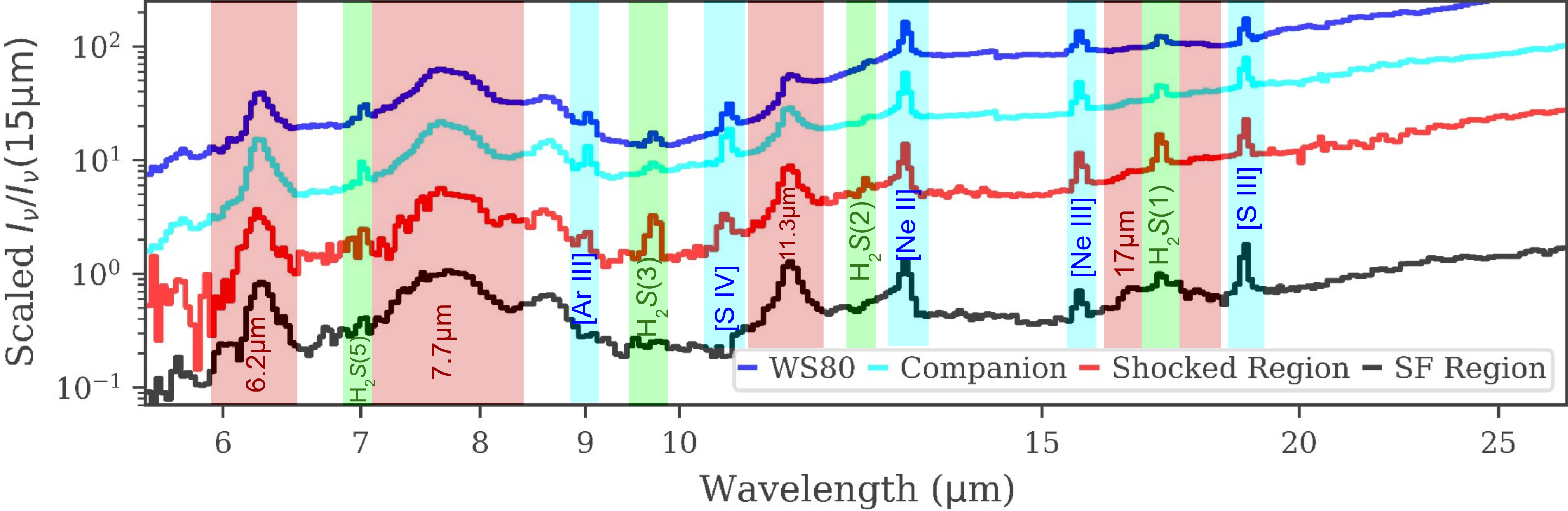}
	\caption{Extracted Spitzer/IRS spectra of WS80 (dark blue), it's companion (cyan), a region of shock-heated ISM just below the cluster pair (red, designated region~3 in Figure~\ref{fig:radiolocation}), and a random star-forming region in the north disk. Key diagnostic features are highlighted, including the main PAH bands, low \& high ionization metal, H$_{2}$ rotational, and H recombination lines. Note the strong H$_2$ lines in the shock-heated region.}\label{fig:spitzer}
\end{figure}

\begin{figure}[!ht]
	\centering
\includegraphics[width=4.0in]{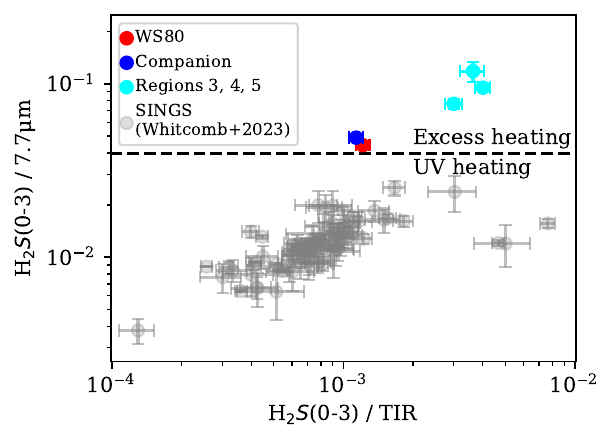}
	\caption{The measured ratios of the (summed) H$_2$ S(0-3) lines to $7.7\mu$m PAH feature are plotted for the massive embedded cluster pair WS80 and its companion, a random star-forming (SF) region in the north disk, and three regions of ISM just south of the clusters (their locations are shown in Figure~\ref{fig:radiolocation}).  The dashed horizontal line shows the typical value measured for H$_2/$PAH in photo-dissociation regions (PDRs) associated with star formation.  The ISM regions have significantly stronger ratios, indicating they are affected by excess heating, likely from shocks. }\label{fig:H2PAH}
\end{figure}

Now that we have identified all of the obscured clusters more massive than a few$\times10^4~M_{\odot}$, we are confident that WS80 and its companion, both with masses $\gea 10^6~M_{\odot}$, are the two most massive recently formed clusters in the Antennae.
This massive cluster pair have formed in a special location, at the edge of a super-giant molecular cloud very close to the southern edge where the two galaxies overlap, which may offer clues to the specific processes responsible for their formation.

Diffuse, soft x-ray emission is observed just to the south of the clusters in Chandra X-ray Observatory images {\bf \citep{Fabbiano01}}, suggesting there is shock-heated gas nearby.
However, previous infrared-based studies have reached different conclusions about the presence of shocked gas.
\citet{Haas05} concluded that strong H$_2$\,S(3) emission observed in an ISO spectrum covering most of the Antennae (including the overlap region and both nuclei), likely originated from large-scale shocks due to the collision between HI clouds.
Conversely, \citet{Brandl12} found from Spitzer spectra that the H$_2$ emission in the overlap region agrees well with active star formation and that far ultraviolet photons and local shocks from supernovae are likely responsible for exciting the molecular hydrogen rather than shocks from the large-scale interaction of the two galaxies. They did not, however, specifically extract spectra from regions surrounding WS80 and its companion.

Here, we revisit the H$_2$ emission near this remarkable pair of massive star clusters.
We use SIMLA cubes (see \citealt{Donnelly25}, Section~2.4) to extract Spitzer spectra for WS80, its companion, three regions just south of the massive cluster pair (overlapping the diffuse x-ray emission), and a control region in the northern disk using a 2\farcs5 aperture radius ($\sim$smallest reasonable aperture for the observations), although the main results are not sensitive to the exact aperture size or location.
Extracted spectra are shown in Figure~\ref{fig:spitzer}, and their apertures are indicated as the larger white circles in Figure~\ref{fig:radiolocation}.
Key diagnostic features in the spectra are highlighted in Figure~\ref{fig:spitzer}, including the main PAH bands (pink), low and high ionization metal lines (light blue), and H$_2$ rotational lines (light green).
The PAH bands at 6.2, 7.7, and 11.3$\mu$m are strong in the spectrum of both WS80 and its companion cluster, and WS80 in particular shows a significant amount of absorption.  
Figure~\ref{fig:spitzer} also shows a spectrum (in red) of region~3, located south of the clusters.
PAH features and nebular emission lines are observed in the spectrum.  
One key difference, however, is that the H$_2$ lines in region 3 are clearly stronger than that associated with the clusters themselves.
In an upcoming work we will present a detailed analysis of the Spitzer spectra for $\approx20$ of the brightest young clusters throughout the Antennae and tens of regions between them.

Observations and radiation modeling have shown that the H$_2$/PAH ratio can be used as a diagnostic to identify regions heated primarily by star-formation and those with additional heating from shocked gas \citep[e.g.,][]{Ogle10,Guillard12}.
PAH molecules are excited primarily by UV photons but H$_2$ molecules will experience additional heating if shocks are present, so the ratio of H$_2$ to PAH emission is higher in regions with additional mechanical heating than in normal star-forming regions.
We measure the strengths of the 7.7$\mu$m PAH feature and the four H$_2$ S(0), S(1), S(2), and S(3) lines, which are summed together to produce H$_2$ S(0-3). 
The strength of these features are determined from PAHFIT, which models the mid-infrared spectrum as a combination of blackbodies of fixed temperature, Gaussian emission lines, Drude profiles for PAH sub-feature emission, and a power law describing silicate extinction with features at 9.7 and $18\mu$m. PAHFIT is also used to combine sub-features and propagate uncertainties to derive the intensity of the $7.7\mu$m complex (see \citealt{Smith07} for more details).

The ratio measured for the six extracted regions is plotted in Figure~\ref{fig:H2PAH}.
The dashed line at 0.04 represents the near-constant ratio between the H$_2$ S(0-3) line and the $7.7\mu$m PAH feature determined for normal star-forming regions with a large range of infrared luminosities \citep{Ogle10}. 
The star-forming region in the northern disk and the companion to WS80 both have measured H$_2$/PAH ratios which are consistent with $\sim0.04$.
Meanwhile, the three regions just below the massive cluster pair (designated 3, 4, 5) have significantly higher H$_2$/PAH ratios, $\approx0.10-0.12$. 
These regions are unlikely to be excited by an AGN in the nucleus of NGC~4039, since the ratio is strongest for region 6 which is the furthest away.
The H$_2$/PAH ratio for WS80 is somewhat higher than those found for normal star-forming regions in nearby galaxies as part of the SINGS sample (gray points; \citealt{Whitcomb23}), but it is possible that some shocked-heated gas is included in the aperture.  
These results support a picture where shocks are responsible for the elevated H$_2$/PAH ratios measured in areas just surrounding WS80 and its companion.
Higher spatial resolution observations would help map out the strength and spatial distribution of the shocks in this region.

All of the clues discussed in this section for the massive cluster pair are consistent with the picture of massive cluster formation in merging galaxies suggested by \citet{Jog92}.
In their analytical models, they separately followed atomic and molecular hydrogen during a merger between two galaxies, and found that typically HI clouds collide but molecular clouds, which have smaller volume filling factors, do not.  The collisions between HI clouds result in radiative shock compression of the outer layers of nearby, pre-existing GMCs that then trigger the formation of massive clusters.
Similarly, the  fully hydrodynamic simulations of galaxy mergers performed by \citet{Maji17} found that massive star clusters (with masses $\sim 3 \times 10^5-3\times 10^7~M_{\odot}$) preferentially form in the highly-shocked regions found in galaxy interactions.
Consistent with these predictions, WS80 and its companion cluster have formed near the southern edge of a super-giant molecular cloud (WS80 is still embedded in weak CO emission while its companion is not; \citealt{He22}) which is adjacent to the shocked gas identified in regions 3, 4, 5 studied here.

\section{Summary \& Conclusions}

In this work we presented new images of the merging Antennae galaxies taken with JWST in the F150W, F187N (Pa$\alpha$), F335M ($3.3\mu$m PAH), F360M, and F410M (Br$\alpha$) NIRCam filters and the F770W ($7.7\mu$m PAH) MIRI filter. These observations, particularly when combined with archival images taken with HST in the F275W (near ultraviolet), F336W (U), F435W (B), F555W (V), F658N (H$\alpha$), and F814W (I) filters, give a new view of the star clusters and interstellar medium in this iconic merging system. The JWST images penetrate even the dustiest portions of the overlap region, allowing us to identify and study the most recently formed and deeply embedded star clusters.

We discovered a population of 45 embedded cluster candidates in the Antennae with high Br$\alpha/\mbox{H}\alpha$ and/or Pa$\alpha/\mbox{H}\alpha$ flux ratios which are broader than the point-spread function at $3.6\mu$m.
We did not identify any sources with Br$\alpha$ but no Pa$\alpha$ emission; if such a source was detected, it would have extremely high extinction in the V-band (A$_V > 20$~mag).
The embedded clusters are detected at all wavelengths between 1.5$\mu$m and $7.7\mu$m.
Most of the embedded cluster population (28/45 or $\approx60$\%) is located in the region where the two galaxies overlap and most current star formation is taking place, with 11/45 or $\approx25$\% forming in the northern disk and 6/45 or $\approx15$\% near the southern nucleus.

We estimated the age, V-band extinction, and mass of each embedded cluster by fitting its measured 0.275-4.1$\mu$m spectral energy distributions (SEDs) with predictions from the FSPS models of \citet{Conroy+09}.  These models include dust extinction and emission, nebular continuum and line emission and fit the observed SEDs well. All embedded clusters have best-fit ages of 2.5~Myr and younger, with nearly 80\% having best-fit ages consistent with 1~Myr or less. 
Estimates of extinction A$_V$ range between $\approx2$ and 10~mag, and mass between $\approx3\times 10^4 - 3\times10^6~M_{\odot}$.
The embedded cluster mass function can  be described by a single power law with a formal best-fit index of $\beta=-1.7\pm0.1$, and shows no evidence for a downturn at the high-mass end.


The observations are sufficiently deep that we believe we have detected all clusters in the Antennae galaxies more massive than $3\times 10^4~M_{\odot}$ with A$_V\gea2$~mag.
Previous suggestions that clusters deeply embedded in dusty cocoons with A$_V >30$~mag dominate the infrared emission from the Antennae likely resulted from poorer resolution observations.

We compared the locations of the embedded clusters with those of thermal radio sources identified by \citet{Neff00} and \citet{He22}, and with optically-selected clusters in the \citet{Whitmore10} catalog.  
The spatial analysis reveals:
\begin{itemize}
    \item Four and possibly five of the embedded clusters in our catalog are also thermal radio sources (\#8, 12, 29, 34, and possibly \#3), including the well-known massive cluster pair WS80 (\#8), which is highly embedded, and its nearby more moderately extinguished massive companion cluster (\#12).  The remaining 40 sources are newly identified star clusters.
    \item Three thermal radio sources (companion to WS80 plus 4-2A and 8-4 from \citealt{Neff00}) are bright at optical wavelengths and included in the optically selected HST-based cluster catalog of \citet{Whitmore10}.  
    \item We were unable to identify a cluster associated with three other potential thermal radio sources in the JWST or HST images.  There is Pa$\alpha$ emission from these sources, but it has a different morphology than from the embedded clusters, with shells rather than a compact structure.  These radio sources may be supernova remnants.   
    \item None of the currently known bright thermal radio sources are infrared-dark, and all are detected from 1.5 through 4.1$\mu$m, suggesting they have moderate extinction.
    \item Based on the derived properties, we found that current radio observations identify only the most massive clusters, including ones that are optically-bright and have low extinction. 
    It appears challenging to isolate the embedded phase of cluster evolution with only radio measurements.
    \item When added to optically selected catalogs, the embedded sources provide a complete census of obscured and unobscured clusters.   We estimated that optical studies are missing $\approx15$\% of very young ($\lea 3$~Myr) clusters in the Antennae. 
The newly discovered embedded cluster population is important for studying the earliest phases of cluster evolution, but represents a very small fraction ($\lea 3$\%) of the observable cluster population in the Antennae galaxies. 
\end{itemize}



We found that using a combination of H$\alpha$, Pa$\alpha$, and Br$\alpha$ line ratios is an excellent method of identifying the youngest, most embedded clusters which are missing from optically-based cluster catalogs, and reach clusters that are $\approx10$ times less massive than the limits of current radio survey of the Antennae.

\acknowledgments
R.C.\ and P.G.\ acknowledge support for program number JWST-GO-2581, which was provided through a grant from the Space Telescope Science Institute (STScI) under NASA contract NAS5-03127.
JDS, KS, SD, GD, CW, and LH acknowledge funding support from NASA/ADAP grant 80NSSC21K0851.
We thank the anonymous referee for suggestions that improved our manuscript.
The data presented in this paper were obtained from the Mikulski Archive for Space Telescopes (MAST) at the STScI. The specific observations analyzed can be accessed via
\dataset[https://doi.org/10.17909/k2y4-2z87]{https://doi.org/10.17909/k2y4-2z87}.

Software: Photutils \citep{photutils}, Matplotlib (Hunter 2007), NumPy (Oliphant 2006; Van Der Walt et al. 2011), Astropy (Astropy Collaboration et al. 2018), SciPy (Virtanen et al. 2020), SAOImage DS9 (Smithsonian Astrophysical Observatory 2000), APLpy \citep{Robitaille2012}.
\bibliography{master}
\appendix

\section{Identifying Infrared Counterparts to Radio Sources}

Radio sources with a thermal component are likely to be the youngest, most massive clusters forming in the Antennae.
In this Appendix, we investigate the infrared counterparts to radio sources in the high-resolution 6\,cm and 4\,cm catalog compiled by \citet{Neff00}, which is the largest, high-resolution catalog of radio sources in the Antennae currently available.
\citet{He22} published properties for a subset of these sources which are identified at 100, 200, and 345~GHz (from ALMA observations) from their cold dust emission.

Typically, objects with a spectral index $\alpha$, defined as $S_{\nu} \propto \nu^{\alpha}$, with $\alpha_{6-4cm} \gea -0.4$ are more likely to have a thermal component and hence to be an embedded star cluster, whereas non-thermal sources ($\alpha_{6-4cm} < -0.4$) are typically supernova remnants. 
In Table~\ref{tab:brightradio}, we compile the coordinates, 6~cm flux, and spectral index (if available) for the 19 sources brighter than 200~$\mu$Jy from \citet{Neff00}.
\citet{Whitmore02} previously found that $\sim85$\% (11/13) of the bright, thermal radio sources in the \citet{Neff00} catalog can be detected in the HST $I$-band image, but did not study the objects individually or in the infrared.


\begin{table*}
\begin{center}
\caption{Infrared Counterparts to Bright 6~cm Radio Sources\label{tab:brightradio}}
\begin{tabular}{lccccccccccc}
\hline\hline
\colhead{Region} & \colhead{RA} & \colhead{Dec} & \colhead{6~cm Flux} & \colhead{Spectral} & \colhead{Notes} \\
 \colhead{Name} & \colhead{(J2000)} &  \colhead{(J2000)} & \colhead{($\mu$Jy)} & \colhead{Index ($\alpha$)} & \colhead{} \\ \hline \hline 
1-2 & 12:01:53.52  & -18:53:10.5  &  $513\pm11$   & ... &   Nucleus-S \\
1-3 & 12:01:53.12  & -18:53:09.7  &  $303\pm11$  & $+0.38$ &  bright Pa$\alpha$ source, possible match with embedded \#4 \\
1-4 & 12:01:53.35  & -18:53:07.9  &  $365\pm21$ & $-0.20$ &  bright, Pa$\alpha$ source\\
2-1 & 12:01:54.96  & -18:53:06.1  &  $5161\pm21$   & $-0.26$ &  WS80, brightest Pa$\alpha$ source, embedded \#9, He22 \#1 \\
2-2 & 12:01:54.52  & -18:53:05.4  &  $241\pm31$  & $-0.51$ &  non-thermal, no Pa$\alpha$ source  \\
2-6 & 12:01:54.58  & -18:53:03.4  &  $2257\pm20$ & ... &  WS80 neighbor, bright Pa$\alpha$ source, embedded \#13, He22 \#2 \\
3-5A & 12:01:54.78  & -18:52:51.1  &  $1121\pm19$  & ...  & Firecracker region, Pa$\alpha$ shell, no counterpart  \\
3-5B & 12:01:54.71  & -18:52:51.4  &  $2291\pm49$  & ...&  Firecracker region, Pa$\alpha$ shell, no counterpart \\
4-2A & 12:01:55.34  & -18:52:49.2  &  $1215\pm25$  & ... &  bright Pa$\alpha$, optical-NIR continuum source \\
4-2B & 12:01:55.38  & -18:52:48.0  &  $493\pm17$  & ... &  Pa$\alpha$ shell, no obvious counterpart, He22 \#6 \\
4-4 & 12:01:55.47  & -18:52:45.9  &  $744\pm11$  & $-0.16$ &  Pa$\alpha$ source, no 3.3$\mu$m, He22 \#7 \\
4-5 & 12:01:55.71  & -18:52:42.8  &  $243\pm11$  & $-0.74$ &  non-thermal, Pa$\alpha$ shell, no counterpart \\
4-7 & 12:01:55:14 & -18:52:40.6 & $616\pm25$  & ... &  Pa$\alpha$ shell, no counterpart, He22 \#8 \\
4A-6 & 12:01:54.74 & -18:52:31.7 & $341\pm11$  & $-0.36$  &  bright Pa$\alpha$ source embedded \#30, He22 \#9 \\
4A-16 & 12:01:55.41  & -18:52:24.3  &  $262\pm27$  & ... &  Pa$\alpha$ shell, no counterpart \\ 
6-1 & 12:01:54.58  & -18:51:56.7  &  $493\pm23$  & $-0.65$ &  bright Pa$\alpha$ source, embedded \#35, He22 \#13 \\
7-8 & 12:01:53.02  & -18:52:02.3  &  $1354\pm23$  & ... &  Nucleus-N \\
11-2B & 12:01:50.45  & -18:52:13.1  &  $355\pm29$  & ... &  Pa$\alpha$ shell, no counterpart \\
8-4 & 12:01:52.13  & -18:52:21.1  &  $198\pm11$ & $0.18$ & bright Pa$\alpha$, optical-NIR continuum source \\
\hline
\hline
\end{tabular} 
\end{center}
\tablecomments{Region names and positions of radio sources with 6~cm flux measurements greater than $\approx200~\mu$Jy are from the catalog of \citet{Neff00}.}
\end{table*}

\begin{figure}[!ht]
\includegraphics[width=\textwidth]{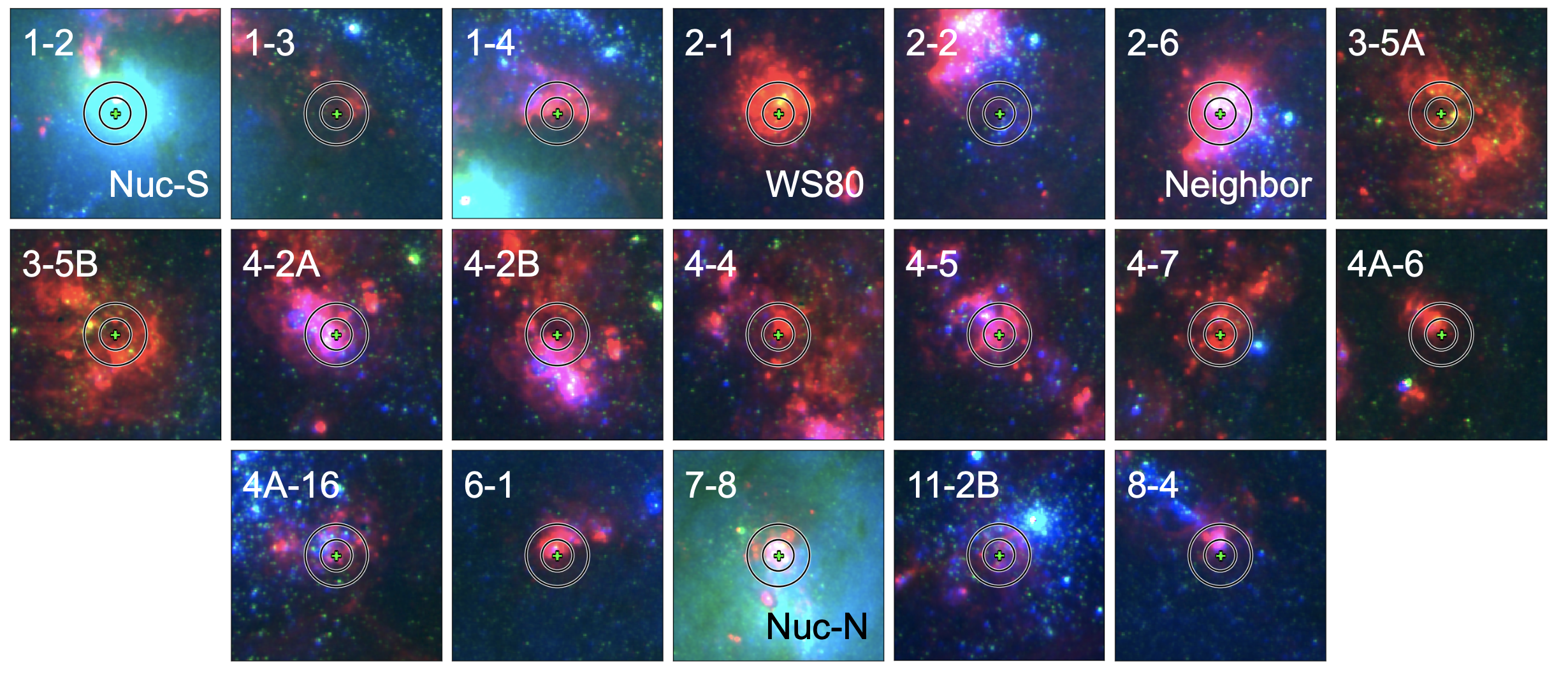}
	\caption{Postage stamps of bright radio sources in V-F150-cont subPa$\alpha$ color image.  The two nuclei (sources \#1 and \#17) and WS80 (source \#4) are identified. The green cross shows the radio source location, which has a positional uncertainty of $\approx0.4\arcsec$, represented by the small circle. Large circle has a radius of 0.8$\arcsec$.
 } \label{fig:brightradio}
\end{figure}
%

Figure~\ref{fig:brightradio} shows $8\times8\arcsec$ regions around each of the bright radio sources, including the southern and northern nuclei (regions 1-2 and 7-8), WS80 (2-1) and its neighbor (2-6).  The color images are created from the $0.55\mu$m image (blue), $1.5\mu$m (green), and continuum subtracted Pa$\alpha$ (red). The green cross shows the source coordinates from \cite{Neff00}.
The circles in each stamp have radii of $0.4\arcsec$ and $0.8\arcsec$, corresponding to the $1\sigma$ and $2\sigma$ radio positional uncertainties, respectively.
The source identifiers correspond to those in Table~\ref{tab:brightradio}. 
Three of these radio sources (2-2, 4-5, and 6-1) have spectral indices $\alpha < -0.4$, which may indicate they are dominated by non-thermal emission, although with only two measurements the published $\alpha$ values should be viewed with some caution.
Eight other sources do not have published spectral indices at all.
Below is our summary of infrared counterparts to each radio source listed in the Table.

\begin{itemize}
\item {\bf Clusters with High Extinction:} In addition to WS80 (2-1), radio sources 1-3 and 4A-6 have 
Pa$\alpha$ emission and high Br$\alpha$/H$\alpha$ and Pa$\alpha$/H$\alpha$ flux ratios, and high A$_V \gea 5$~mag from our SED fitting. 
These three sources are fainter at 1.5$\mu$m than at longer wavelengths, and are included in our embedded cluster catalog.

\item {\bf Clusters with Moderate Extinction:} WS80's neighbor (2-6) and radio source 6-1 has Pa$\alpha$ emission and moderate Pa$\alpha$/H$\alpha$ flux ratios.  Both are included in our embedded cluster catalog, have best-fit extinction A$_V \approx 3$~mag, and are visible at all infrared and optical wavelengths covered by our images.
Despite the low published spectral index of $\alpha=-0.65$, which suggests that source 6-1 might be non-thermal, this radio source is clearly associated with a young star cluster.

\item {\bf Clusters with Little Extinction:} Radio sources 4-2A and 8-4 both have Pa$\alpha$ emission but low Br$\alpha$/H$\alpha$ and Pa$\alpha$/H$\alpha$ flux ratios.  They have counterparts that are extended and bright at optical wavelengths, and were included in the HST-based cluster catalog of \citet{Whitmore10}.
The radio emission is from very young clusters with little extinction.

\item {\bf Non-Thermal:} Source 2-2 has no obvious Pa$\alpha$ emission and source 4-5 shows diffuse Pa$\alpha$ with a shell morphology.  the spectral indices of these sources suggest they are dominated by non-thermal emission. 

\item {\bf No obvious counterpart:} We do not find an obvious peaked infrared source that could be a potential counterpart to radio sources 1-4, 3-5A, 3-5B, 4-2B, 4-4, 4-7, 4A-16, or 11-2B.  Most of these have a partial Pa$\alpha$ shell near the radio position.  Sources 1-4 and 4-4 have a spectral index that suggests they have a thermal component while the rest do not have published spectral indices.

\item {\bf South and North Galaxy Nuclei:} The nuclei of the south and north galaxies are radio sources 1-2 and 7-8, respectively.

\end{itemize}

The bright 6~cm radio sources in the Antennae appear to be somewhat heterogeneous in nature and include very young star clusters with a range of extinction, including some of the most highly extinguished sources in our embedded cluster sample.
Most of the bright 6~cm sources however, are likely non-thermal and not associated with a stellar cluster.

\end{document}

%% file: catalog.tex
\begin{table}
\caption{Catalog of Embedded Star Cluster Candidates}
\label{tab:sample}
\centering
\begin{tabular}{lccccccccc}
\hline\hline
\colhead{Source$^1$} & \colhead{RA} & \colhead{Dec}  & \colhead{Age} & \colhead{A$_V$} & \colhead{Mass} & \colhead{apcorr\tablenotemark{a}} & \colhead{CrossID} \\
\colhead{ID} & \colhead{(J2000)} &  \colhead{(J2000)}  & \colhead{[log ($\tau/\mbox{yr}$)]} & \colhead{(mag)} & \colhead{[log\,($M/M_{\odot}$)]} & \colhead{[mag]} & \colhead{}\\ \hline \hline 
1 & 12:01:52.24 & -18:53:27.96 & 	$5.8^{+0.5}_{-0.5}$ & $3.4^{+0.8}_{-0.6}$ & $4.4\pm0.3$	& -1.2 \\
2 & 12:01:52.96 & -18:53:15.44 & 	$5.7^{+0.6}_{-0.4}$ & $2.4^{+0.6}_{-0.6}$ & $4.3\pm0.3$	& -0.7 \\
3 & 12:01:53.09 & -18:53:09.82 &	$6.4^{+0.7}_{-0.9}$ & $4.8^{+0.9}_{-0.9}$ & $4.8\pm0.7$	& -1.7 \\
4 & 12:01:53.12 & -18:53:09.78 &	$6.6^{+1.6}_{-0.7}$ & $6.6^{+2.4}_{-2.0}$ & $5.1\pm1.2$	& -1.8 & possibly NU1-3\\
5 & 12:01:53.16 & -18:53:08.81 &	$5.8^{+0.5}_{-0.5}$ & $6.6^{+2.4}_{-2.2}$ & $5.0\pm0.8$	& -1.4 \\
6 & 12:01:53.30 & -18:53:08.12 &	$6.4^{+0.6}_{-0.8}$ & $5.2^{+0.4}_{-0.6}$ & $5.0\pm0.7$	& -1.6 \\
7 & 12:01:54.20 & -18:53:08.95 &	$6.0^{+0.6}_{-0.7}$ & $2.2^{+0.6}_{-0.5}$ & $4.5\pm0.2$	& -0.7 \\
8 & 12:01:55.04 & -18:53:11.36 &	$6.4^{+1.8}_{-0.9}$ & $2.2^{+0.8}_{-0.8}$ & $4.3\pm0.3$	& -1.4 \\
9 & 12:01:54.95 & -18:53:05.97 &	$6.3^{+0.6}_{-0.8}$ & $9.2^{+1.4}_{-1.2}$ & $6.5\pm0.3$	& -0.9 & WS80, NU2-1 \\
10 & 12:01:54.83 & -18:53:07.76 &	$5.7^{+0.5}_{-0.4}$ & $3.4^{+1.0}_{-0.8}$ & $5.0\pm0.8$	& -1.5 \\
11 & 12:01:54.77 & -18:53:06.29 &	$6.0^{+1.6}_{-0.6}$ & $3.4^{+4.0}_{-2.6}$ & $4.5\pm0.4$	& -1.3 \\
12 & 12:01:54.68 & -18:53:05.68 &	$5.7^{+0.5}_{-0.4}$ & $3.2^{+0.6}_{-0.6}$ & $4.3\pm0.3$	& -1.0 \\
13 & 12:01:54.59 & -18:53:03.01 &	$6.4^{+0.4}_{-0.1}$ & $3.4^{+0.2}_{-0.2}$ & $5.9\pm0.5$	& -1.3 & Companion, NU2-6 \\
14 & 12:01:55.36 & -18:53:02.65 &	$5.7^{+0.5}_{-0.4}$ & $2.4^{+0.6}_{-0.6}$ & $4.9\pm0.3$	& -1.4 \\
15 & 12:01:55.21 & -18:53:03.19 &	$5.8^{+0.5}_{-0.5}$ & $3.0^{+0.8}_{-1.0}$ & $4.3\pm0.4$	& -1.1 \\
16 & 12:01:55.44 & -18:53:00.74 &	$5.9^{+0.8}_{-0.6}$ & $3.6^{+1.2}_{-1.0}$ & $4.1\pm0.4$	& -0.9 \\
17 & 12:01:55.36 & -18:53:01.86 &	$5.8^{+0.6}_{-0.5}$ & $4.2^{+1.4}_{-1.4}$ & $4.7\pm0.5$	& -1.1 \\
18 & 12:01:55.31 & -18:53:01.50 &	$6.0^{+2.2}_{-0.6}$ & $4.2^{+2.4}_{-2.2}$ & $4.2\pm0.7$	& -1.4 \\
19 & 12:01:55.21 & -18:53:01.25 &	$5.8^{+0.6}_{-0.5}$ & $3.0^{+1.2}_{-1.0}$ & $4.1\pm0.4$	& -1.1 \\
20 & 12:01:54.93 & -18:52:58.59 &	$5.9^{+0.9}_{-0.6}$ & $4.6^{+1.8}_{-1.8}$ & $5.2\pm0.6$	& -1.4 \\
21 & 12:01:54.92 & -18:52:58.19 &	$6.0^{+1.9}_{-0.7}$ & $4.2^{+2.6}_{-2.2}$ & $5.0\pm0.6$	& -1.8 \\
22 & 12:01:55.21 & -18:52:44.58 &	$5.9^{+0.7}_{-0.6}$ & $4.8^{+1.5}_{-1.6}$ & $4.8\pm0.5$	& -1.0 \\
23 & 12:01:54.89 & -18:52:44.98 &	$6.1^{+2.3}_{-0.7}$ & $4.0^{+2.8}_{-2.2}$ & $4.5\pm0.5$	& -0.9 \\
24 & 12:01:54.59 & -18:52:43.97 &	$6.0^{+1.9}_{-0.7}$ & $4.4^{+2.8}_{-2.4}$ & $4.6\pm0.5$	& -1.1 \\
25 & 12:01:55.32 & -18:52:40.01 &	$6.0^{+2.6}_{-0.7}$ & $4.6^{+3.2}_{-2.5}$ & $4.7\pm0.4$	& -1.1 \\
26 & 12:01:55.07 & -18:52:39.50 &	$5.8^{+0.5}_{-0.5}$ & $4.6^{+1.2}_{-1.2}$ & $5.3\pm0.6$	& -1.4 \\
27 & 12:01:55.34 & -18:52:37.96 &	$5.8^{+0.6}_{-0.5}$ & $3.8^{+1.2}_{-1.2}$ & $4.4\pm0.5$	& -0.8 \\
28 & 12:01:55.15 & -18:52:37.96 &	$5.9^{+1.1}_{-0.6}$ & $4.2^{+1.4}_{-1.2}$ & $4.8\pm0.5$	& -1.0 \\
29 & 12:01:56.05 & -18:52:34.97 &	$6.4^{+0.6}_{-0.8}$ & $1.4^{+0.3}_{-0.4}$ & $4.9\pm0.1$	& -1.5 \\
30 & 12:01:54.74 & -18:52:31.37 &	$5.7^{+0.5}_{-0.4}$ & $5.4^{+1.2}_{-1.2}$ & $5.5\pm0.8$	& -1.1 & NU4A-6\\
31 & 12:01:54.94 & -18:52:29.28 &	$5.7^{+0.5}_{-0.4}$ & $3.8^{+0.4}_{-0.4}$ & $4.7\pm0.6$	& -1.7 \\
32 & 12:01:54.96 & -18:52:28.85 &	$5.9^{+1.1}_{-0.6}$ & $6.2^{+1.8}_{-1.6}$ & $4.8\pm0.8$ &	-1.4 \\
33 & 12:01:54.51 & -18:52:28.78 &	$5.7^{+0.6}_{-0.4}$ & $3.2^{+0.8}_{-0.6}$ & $4.0\pm0.6$	& -1.0 \\
34 & 12:01:54.85 & -18:52:25.32 &	$5.9^{+0.9}_{-0.6}$ & $4.8^{+2.2}_{-1.8}$ & $4.8\pm0.7$	& -1.3 \\
35 & 12:01:54.59 & -18:51:56.55 &	$6.4^{+0.7}_{-0.8}$ & $2.8^{+0.6}_{-0.5}$ & $5.6\pm0.3$	& -1.5 & NU6-1 \\
36 & 12:01:54.50 & -18:51:46.19 &	$5.8^{+0.5}_{-0.5}$ & $2.4^{+0.6}_{-0.6}$ & $4.5\pm0.2$	& -0.8 \\
37 & 12:01:55.18 & -18:51:41.08 &	$6.4^{+0.4}_{-1.1}$ & $3.2^{+0.6}_{-0.4}$ & $4.6\pm0.9$	& -1.7 \\
38 & 12:01:55.16 & -18:51:40.43 &	$6.4^{+0.6}_{-1.1}$ & $2.0^{+0.5}_{-0.6}$ & $4.5\pm0.3$	& -1.3 \\
39 & 12:01:53.13 & -18:51:38.23 &	$6.2^{+0.6}_{-0.8}$ & $3.4^{+0.8}_{-0.8}$ & $4.3\pm0.5$	& -1.2 \\
40 & 12:01:52.16 & -18:51:38.81 &	$5.9^{+1.3}_{-0.6}$ & $4.0^{+1.6}_{-1.4}$ & $4.8\pm0.4$	& -1.4 \\
41 & 12:01:52.10 & -18:51:40.36 &	$5.9^{+1.2}_{-0.6}$ & $4.4^{+1.8}_{-1.6}$ & $4.4\pm0.3$ &	-1.1 \\
42 & 12:01:51.97 & -18:51:38.12 &	$6.3^{+0.6}_{-0.8}$ & $3.6^{+0.6}_{-0.6}$ & $5.3\pm0.4$	& -1.4 \\
43 & 12:01:51.05 & -18:51:58.28 &	$5.7^{+0.5}_{-0.4}$ & $4.0^{+0.4}_{-0.4}$ & $4.3\pm0.4$	& -1.3 \\
44 & 12:01:50.80 & -18:52:15.10 &	$6.4^{+0.6}_{-0.8}$ & $4.4^{+0.8}_{-1.0}$ & $5.0\pm0.5$ & 	-1.6 \\
45 & 12:01:50.94 & -18:52:28.09 &	$6.2^{+0.7}_{-0.7}$ & $3.0^{+0.8}_{-0.8}$ & $4.9\pm0.3$	& -1.1 \\ \hline \hline
\end{tabular} 
\tablenotetext{a}{The aperture correction listed for each cluster is the magnitude difference measured between aperture radii of 15 and 5~pixels, $\Delta\mbox{m}_{5-15}~{\rm pix}$, in the NIRCAM F360W filter.}
\tablecomments{The last column cross-lists the ID for radio sources identified by \citet{Neff00}.}
\end{table}

%% file: master.bib
@ARTICLE{Adamo20,
       author = {{Adamo}, A. and {Hollyhead}, K. and {Messa}, M. and {Ryon}, J.~E. and {Bajaj}, V. and {Runnholm}, A. and {Aalto}, S. and {Calzetti}, D. and {Gallagher}, J.~S. and {Hayes}, M.~J. and {Kruijssen}, J.~M.~D. and {K{\"o}nig}, S. and {Larsen}, S.~S. and {Melinder}, J. and {Sabbi}, E. and {Smith}, L.~J. and {{\"O}stlin}, G.},
        title = "{Star cluster formation in the most extreme environments: insights from the HiPEEC survey}",
      journal = {\mnras},
         year = 2020,
        month = dec,
       volume = {499},
       number = {3},
        pages = {3267-3294},
          doi = {10.1093/mnras/staa2380},
archivePrefix = {arXiv},
       eprint = {2008.12794},
 primaryClass = {astro-ph.GA},
       adsurl = {https://ui.adsabs.harvard.edu/abs/2020MNRAS.499.3267A}
}

@ARTICLE{Adamo24,
       author = {{Adamo}, Angela and {Bradley}, Larry D. and {Vanzella}, Eros and {Claeyssens}, Ad{\'e}la{\"\i}de and {Welch}, Brian and {Diego}, Jose M. and {Mahler}, Guillaume and {Oguri}, Masamune and {Sharon}, Keren and {Abdurro'uf} and {Hsiao}, Tiger Yu-Yang and {Xu}, Xinfeng and {Messa}, Matteo and {Lassen}, Augusto E. and {Zackrisson}, Erik and {Brammer}, Gabriel and {Coe}, Dan and {Kokorev}, Vasily and {Ricotti}, Massimo and {Zitrin}, Adi and {Fujimoto}, Seiji and {Inoue}, Akio K. and {Resseguier}, Tom and {Rigby}, Jane R. and {Jim{\'e}nez-Teja}, Yolanda and {Windhorst}, Rogier A. and {Hashimoto}, Takuya and {Tamura}, Yoichi},
        title = "{Bound star clusters observed in a lensed galaxy 460 Myr after the Big Bang}",
      journal = {\nat},
         year = 2024,
        month = aug,
       volume = {632},
       number = {8025},
        pages = {513-516},
          doi = {10.1038/s41586-024-07703-7},
archivePrefix = {arXiv},
       eprint = {2401.03224},
 primaryClass = {astro-ph.GA},
       adsurl = {https://ui.adsabs.harvard.edu/abs/2024Natur.632..513A}
}

@ARTICLE{Aversa11,
       author = {{Aversa}, Alan G. and {Johnson}, Kelsey E. and {Brogan}, Crystal L. and {Goss}, W.~M. and {Pisano}, D.~J.},
        title = "{Very Large Array and ATCA Search for Natal Star Clusters in Nearby Star-forming Galaxies}",
      journal = {\aj},
         year = 2011,
        month = apr,
       volume = {141},
       number = {4},
          eid = {125},
        pages = {125},
          doi = {10.1088/0004-6256/141/4/125},
archivePrefix = {arXiv},
       eprint = {1102.0412},
 primaryClass = {astro-ph.GA},
       adsurl = {https://ui.adsabs.harvard.edu/abs/2011AJ....141..125A}
}

@ARTICLE{Barnes23,
       author = {{Barnes}, Ashley. T. and {Watkins}, Elizabeth J. and {Meidt}, Sharon E. and {Kreckel}, Kathryn and {Sormani}, Mattia C. and {Tre{\ss}}, Robin G. and {Glover}, Simon C.~O. and {Bigiel}, Frank and {Chandar}, Rupali and {Emsellem}, Eric and {Lee}, Janice C. and {Leroy}, Adam K. and {Sandstrom}, Karin M. and {Schinnerer}, Eva and {Rosolowsky}, Erik and {Belfiore}, Francesco and {Blanc}, Guillermo A. and {Boquien}, M{\'e}d{\'e}ric and {Brok}, Jakob den and {Cao}, Yixian and {Chevance}, M{\'e}lanie and {Dale}, Daniel A. and {Egorov}, Oleg V. and {Eibensteiner}, Cosima and {Grasha}, Kathryn and {Groves}, Brent and {Hassani}, Hamid and {Henshaw}, Jonathan D. and {Jeffreson}, Sarah and {Jim{\'e}nez-Donaire}, Mar{\'\i}a J. and {Keller}, Benjamin W. and {Klessen}, Ralf S. and {Koch}, Eric W. and {Kruijssen}, J.~M. Diederik and {Larson}, Kirsten L. and {Li}, Jing and {Liu}, Daizhong and {Lopez}, Laura A. and {Murphy}, Eric J. and {Neumann}, Lukas and {Pety}, J{\'e}r{\^o}me and {Pinna}, Francesca and {Querejeta}, Miguel and {Renaud}, Florent and {Saito}, Toshiki and {Sarbadhicary}, Sumit K. and {Sardone}, Amy and {Smith}, Rowan J. and {Stuber}, Sophia K. and {Sun}, Jiayi and {Thilker}, David A. and {Usero}, Antonio and {Whitmore}, Bradley C. and {Williams}, Thomas G.},
        title = "{PHANGS-JWST First Results: Multiwavelength View of Feedback-driven Bubbles (the Phantom Voids) across NGC 628}",
      journal = {\apjl},
         year = 2023,
        month = feb,
       volume = {944},
       number = {2},
          eid = {L22},
        pages = {L22},
          doi = {10.3847/2041-8213/aca7b9},
archivePrefix = {arXiv},
       eprint = {2212.00812},
 primaryClass = {astro-ph.GA},
       adsurl = {https://ui.adsabs.harvard.edu/abs/2023ApJ...944L..22B}
}

@ARTICLE{Bik03,
    author = {{Bik}, A. and {Lamers}, H.~J.~G.~L.~M. and {Bastian}, N. and {Panagia}, N. and {Romaniello}, M.},
    title = "{Clusters in the inner spiral arms of M 51: The cluster IMF and the formation history}",
    journal = {\aap},
    year = 2003,
    month = jan,
    volume = {397},
    pages = {473-486},
    doi = {10.1051/0004-6361:20021384},
    archivePrefix = {arXiv},
    eprint = {astro-ph/0210594},
    primaryClass = {astro-ph},
    adsurl = {https://ui.adsabs.harvard.edu/abs/2003A&A...397..473B}
}

@ARTICLE{Brandl09,
       author = {{Brandl}, B.~R. and {Snijders}, L. and {den Brok}, M. and {Whelan}, D.~G. and {Groves}, B. and {van der Werf}, P. and {Charmandaris}, V. and {Smith}, J.~D. and {Armus}, L. and {Kennicutt}, Jr., R.~C. and {Houck}, J.~R.},
        title = "{Spitzer-IRS Study of the Antennae Galaxies NGC 4038/39}",
      journal = {\apj},
         year = 2009,
        month = jul,
       volume = {699},
       number = {2},
        pages = {1982-2001},
          doi = {10.1088/0004-637X/699/2/1982},
archivePrefix = {arXiv},
       eprint = {0905.1058},
 primaryClass = {astro-ph.CO},
       adsurl = {https://ui.adsabs.harvard.edu/abs/2009ApJ...699.1982B}
}

@ARTICLE{Brandl12,
       author = {{Brandl}, B.~R. and {Mart{\'\i}n-Hern{\'a}ndez}, N.~L. and {Schaerer}, D. and {Rosenberg}, M. and {van der Werf}, P.~P.},
        title = "{High resolution IR observations of the starburst ring in NGC 7552. One ring to rule them all?}",
      journal = {\aap},
         year = 2012,
        month = jul,
       volume = {543},
          eid = {A61},
        pages = {A61},
          doi = {10.1051/0004-6361/201117568},
archivePrefix = {arXiv},
       eprint = {1205.1922},
 primaryClass = {astro-ph.GA},
       adsurl = {https://ui.adsabs.harvard.edu/abs/2012A&A...543A..61B}
}

@ARTICLE{Calzetti00,
       author = {{Calzetti}, Daniela and {Armus}, Lee and {Bohlin}, Ralph C. and {Kinney}, Anne L. and {Koornneef}, Jan and {Storchi-Bergmann}, Thaisa},
        title = "{The Dust Content and Opacity of Actively Star-forming Galaxies}",
      journal = {\apj},
         year = 2000,
        month = apr,
       volume = {533},
       number = {2},
        pages = {682-695},
          doi = {10.1086/308692},
archivePrefix = {arXiv},
       eprint = {astro-ph/9911459},
 primaryClass = {astro-ph},
       adsurl = {https://ui.adsabs.harvard.edu/abs/2000ApJ...533..682C}
}

@ARTICLE{Calzetti10,
	author = {{Calzetti}, D. and {Wu}, S.-Y. and {Hong}, S. and {Kennicutt}, R.~C. and 
	{Lee}, J.~C. and {Dale}, D.~A. and {Engelbracht}, C.~W. and 
	{van Zee}, L. and {Draine}, B.~T. and {Hao}, C.-N. and {Gordon}, K.~D. and 
	{Moustakas}, J. and {Murphy}, E.~J. and {Regan}, M. and {Begum}, A. and 
	{Block}, M. and {Dalcanton}, J. and {Funes}, J. and {Gil de Paz}, A. and 
	{Johnson}, B. and {Sakai}, S. and {Skillman}, E. and {Walter}, F. and 
	{Weisz}, D. and {Williams}, B. and {Wu}, Y.},
	title = "{The Calibration of Monochromatic Far-Infrared Star Formation Rate Indicators}",
	journal = {\apj},
	archivePrefix = "arXiv",
	eprint = {1003.0961},
	year = 2010,
	month = may,
	volume = 714,
	pages = {1256-1279},
	doi = {10.1088/0004-637X/714/2/1256},
	adsurl = {http://adsabs.harvard.edu/abs/2010ApJ...714.1256C}
}

@ARTICLE{Capriotti96,
       author = {{Capriotti}, Eugene R. and {Hawley}, Suzanne L.},
        title = "{Evaporation, Tidal Disruption, and Orbital Decay of Star Clusters in a Galactic Halo}",
      journal = {\apj},
         year = 1996,
        month = jun,
       volume = {464},
        pages = {765},
          doi = {10.1086/177362},
       adsurl = {https://ui.adsabs.harvard.edu/abs/1996ApJ...464..765C}
}

@ARTICLE{Chandar16,
	author = {{Chandar}, R. and {Whitmore}, B.~C. and {Dinino}, D. and {Kennicutt}, R.~C. and 
	{Chien}, L.-H. and {Schinnerer}, E. and {Meidt}, S.},
	title = "{The Age, Mass, and Size Distributions of Star Clusters in M51}",
	journal = {\apj},
	year = 2016,
	month = jun,
	volume = 824,
	eid = {71},
	pages = {71},
	doi = {10.3847/0004-637X/824/2/71},
	adsurl = {http://adsabs.harvard.edu/abs/2016ApJ...824...71C}
}

@ARTICLE{Chandar23,
       author = {{Chandar}, Rupali and {Caputo}, Miranda and {Mok}, Angus and {Linden}, Sean and {Whitmore}, Bradley C. and {Toscano}, Aimee and {Conyer}, Jaidyn and {Cook}, David O. and {Lee}, Janice C. and {Ubeda}, Leonardo and {White}, Richard},
        title = "{A Tale of Three Dwarfs: No Extreme Cluster Formation in Extreme Star-forming Galaxies}",
      journal = {\apj},
         year = 2023,
        month = jun,
       volume = {949},
       number = {2},
          eid = {116},
        pages = {116},
          doi = {10.3847/1538-4357/acc93b},
archivePrefix = {arXiv},
       eprint = {2304.02450},
 primaryClass = {astro-ph.GA},
       adsurl = {https://ui.adsabs.harvard.edu/abs/2023ApJ...949..116C}
}

@ARTICLE{Claeyssens23,
       author = {{Claeyssens}, Ad{\'e}la{\"\i}de and {Adamo}, Angela and {Richard}, Johan and {Mahler}, Guillaume and {Messa}, Matteo and {Dessauges-Zavadsky}, Miroslava},
        title = "{Star formation at the smallest scales: a JWST study of the clump populations in SMACS0723}",
      journal = {\mnras},
         year = 2023,
        month = apr,
       volume = {520},
       number = {2},
        pages = {2180-2203},
          doi = {10.1093/mnras/stac3791},
archivePrefix = {arXiv},
       eprint = {2208.10450},
 primaryClass = {astro-ph.GA},
       adsurl = {https://ui.adsabs.harvard.edu/abs/2023MNRAS.520.2180C}
}

@ARTICLE{DeLucia24,
       author = {{De Lucia}, Gabriella and {Kruijssen}, J.~M. Diederik and {Trujillo-Gomez}, Sebastian and {Hirschmann}, Michaela and {Xie}, Lizhi},
        title = "{On the origin of globular clusters in a hierarchical universe}",
      journal = {\mnras},
         year = 2024,
        month = may,
       volume = {530},
       number = {3},
        pages = {2760-2777},
          doi = {10.1093/mnras/stae1006},
archivePrefix = {arXiv},
       eprint = {2307.02530},
 primaryClass = {astro-ph.GA},
       adsurl = {https://ui.adsabs.harvard.edu/abs/2024MNRAS.530.2760D}
}

@ARTICLE{Elmegreen10a,
       author = {{Elmegreen}, Bruce G.},
        title = "{The Globular Cluster Mass Function as a Remnant of Violent Birth}",
      journal = {\apjl},
         year = 2010,
        month = apr,
       volume = {712},
       number = {2},
        pages = {L184-L188},
          doi = {10.1088/2041-8205/712/2/L184},
archivePrefix = {arXiv},
       eprint = {1003.0798},
 primaryClass = {astro-ph.CO},
       adsurl = {https://ui.adsabs.harvard.edu/abs/2010ApJ...712L.184E}
}

@ARTICLE{Emig20,
       author = {{Emig}, Kimberly L. and {Bolatto}, Alberto D. and {Leroy}, Adam K. and {Mills}, Elisabeth A.~C. and {Jim{\'e}nez Donaire}, Mar{\'\i}a J. and {Tielens}, Alexander G.~G.~M. and {Ginsburg}, Adam and {Gorski}, Mark and {Krieger}, Nico and {Levy}, Rebecca C. and {Meier}, David S. and {Ott}, J{\"u}rgen and {Rosolowsky}, Erik and {Thompson}, Todd A. and {Veilleux}, Sylvain},
        title = "{Super Star Clusters in the Central Starburst of NGC 4945}",
      journal = {\apj},
         year = 2020,
        month = nov,
       volume = {903},
       number = {1},
          eid = {50},
        pages = {50},
          doi = {10.3847/1538-4357/abb67d},
archivePrefix = {arXiv},
       eprint = {2009.05154},
 primaryClass = {astro-ph.GA},
       adsurl = {https://ui.adsabs.harvard.edu/abs/2020ApJ...903...50E}
}

@ARTICLE{Fall01,
	author = {{Fall}, S. Michael and {Zhang}, Qing},
	title = "{Dynamical Evolution of the Mass Function of Globular Star Clusters}",
	journal = {\apj},
	year = 2001,
	month = Nov,
	volume = {561},
	pages = {751-765},
	doi = {10.1086/323358},
	adsurl = {https://ui.adsabs.harvard.edu/#abs/2001ApJ...561..751F}
}

@ARTICLE{Fall05,
	author = {{Fall}, S.~M. and {Chandar}, R. and {Whitmore}, B.~C.},
	title = "{The Age Distribution of Massive Star Clusters in the Antennae Galaxies}",
	journal = {\apjl},
	eprint = {astro-ph/0509293},
	year = 2005,
	month = oct,
	volume = 631,
	pages = {L133-L136},
	doi = {10.1086/496878},
	adsurl = {http://adsabs.harvard.edu/abs/2005ApJ...631L.133F}
}

@ARTICLE{Fall77,
	author = {{Fall}, S.~M. and {Rees}, M.~J.},
	title = "{Survival and disruption of galactic substructure}",
	journal = {\mnras},
	year = 1977,
	month = nov,
	volume = 181,
	pages = {37P-42P},
	doi = {10.1093/mnras/181.1.37P},
	adsurl = {http://adsabs.harvard.edu/abs/1977MNRAS.181P..37F}
}

@ARTICLE{Finn19,
       author = {{Finn}, Molly K. and {Johnson}, Kelsey E. and {Brogan}, Crystal L. and {Wilson}, Christine D. and {Indebetouw}, Remy and {Harris}, William E. and {Kamenetzky}, Julia and {Bemis}, Ashley},
        title = "{New Insights into the Physical Conditions and Internal Structure of a Candidate Proto-globular Cluster}",
      journal = {\apj},
         year = 2019,
        month = apr,
       volume = {874},
       number = {2},
          eid = {120},
        pages = {120},
          doi = {10.3847/1538-4357/ab0d1e},
archivePrefix = {arXiv},
       eprint = {1903.08669},
 primaryClass = {astro-ph.GA},
       adsurl = {https://ui.adsabs.harvard.edu/abs/2019ApJ...874..120F}
}

@ARTICLE{Gao01,
       author = {{Gao}, Yu and {Lo}, K.~Y. and {Lee}, S. -W. and {Lee}, T. -H.},
        title = "{Molecular Gas and the Modest Star Formation Efficiency in the ``Antennae'' Galaxies: Arp 244=NGC 4038/9}",
      journal = {\apj},
         year = 2001,
        month = feb,
       volume = {548},
       number = {1},
        pages = {172-189},
          doi = {10.1086/318682},
archivePrefix = {arXiv},
       eprint = {astro-ph/0010128},
 primaryClass = {astro-ph},
       adsurl = {https://ui.adsabs.harvard.edu/abs/2001ApJ...548..172G}
}

@ARTICLE{Gilbert00,
       author = {{Gilbert}, Andrea M. and {Graham}, James R. and {McLean}, Ian S. and {Becklin}, E.~E. and {Figer}, Donald F. and {Larkin}, James E. and {Levenson}, N.~A. and {Teplitz}, Harry I. and {Wilcox}, Mavourneen K.},
        title = "{Infrared Spectroscopy of a Massive Obscured Star Cluster in the Antennae Galaxies (NGC 4038/9) with NIRSPEC}",
      journal = {\apjl},
         year = 2000,
        month = apr,
       volume = {533},
       number = {1},
        pages = {L57-L60},
          doi = {10.1086/312599},
archivePrefix = {arXiv},
       eprint = {astro-ph/9912369},
 primaryClass = {astro-ph},
       adsurl = {https://ui.adsabs.harvard.edu/abs/2000ApJ...533L..57G}
}

@ARTICLE{Gilbert07,
       author = {{Gilbert}, Andrea M. and {Graham}, James R.},
        title = "{Feedback in the Antennae Galaxies (NGC 4038/9). I. High-Resolution Infrared Spectroscopy of Winds from Super Star Clusters}",
      journal = {\apj},
         year = 2007,
        month = oct,
       volume = {668},
       number = {1},
        pages = {168-181},
          doi = {10.1086/520910},
archivePrefix = {arXiv},
       eprint = {0706.3935},
 primaryClass = {astro-ph},
       adsurl = {https://ui.adsabs.harvard.edu/abs/2007ApJ...668..168G}
}

@ARTICLE{Gordon2023,
       author = {{Gordon}, Karl D. and {Clayton}, Geoffrey C. and {Decleir}, Marjorie and {Fitzpatrick}, E.~L. and {Massa}, Derck and {Misselt}, Karl A. and {Tollerud}, Erik J.},
        title = "{One Relation for All Wavelengths: The Far-ultraviolet to Mid-infrared Milky Way Spectroscopic R(V)-dependent Dust Extinction Relationship}",
      journal = {\apj},
         year = 2023,
        month = jun,
       volume = {950},
       number = {2},
          eid = {86},
        pages = {86},
          doi = {10.3847/1538-4357/accb59},
archivePrefix = {arXiv},
       eprint = {2304.01991},
 primaryClass = {astro-ph.GA},
       adsurl = {https://ui.adsabs.harvard.edu/abs/2023ApJ...950...86G}
}

@ARTICLE{Goudfrooij03,
       author = {{Goudfrooij}, Paul and {Strader}, Jay and {Brenneman}, Laura and {Kissler-Patig}, Markus and {Minniti}, Dante and {Edwin Huizinga}, J.},
        title = "{Hubble Space Telescope observations of globular cluster systems along the Hubble sequence of spiral galaxies}",
      journal = {\mnras},
         year = 2003,
        month = aug,
       volume = {343},
       number = {2},
        pages = {665-678},
          doi = {10.1046/j.1365-8711.2003.06706.x},
archivePrefix = {arXiv},
       eprint = {astro-ph/0304195},
 primaryClass = {astro-ph},
       adsurl = {https://ui.adsabs.harvard.edu/abs/2003MNRAS.343..665G}
}

@ARTICLE{Fabbiano01,
       author = {{Fabbiano}, G. and {Zezas}, A. and {Murray}, S.~S.},
        title = "{Chandra Observations of ``The Antennae'' Galaxies (NGC 4038/9)}",
      journal = {\apj},
         year = 2001,
        month = jun,
       volume = {554},
       number = {2},
        pages = {1035-1043},
          doi = {10.1086/321397},
archivePrefix = {arXiv},
       eprint = {astro-ph/0102256},
 primaryClass = {astro-ph},
       adsurl = {https://ui.adsabs.harvard.edu/abs/2001ApJ...554.1035F}
}

@ARTICLE{Donnelly25,
       author = {{Donnelly}, Grant P. and {Whitcomb}, Cory M. and {Hands}, Lindsey and {Duval}, Sara E. and {Sandstrom}, Karin and {Smith}, J.-D.~T. and {Carroll}, David and {Dowd}, McKenna and {Hensley}, Brandon S. and {Hunt}, Leslie K. and {Walsh}, Edward and {Watson}, Julie},
        title = "{SIMLA: The Spitzer Infrared Spectrograph Mapping Legacy Archive}",
      journal = {arXiv e-prints},
         year = 2025,
        month = dec,
          eid = {arXiv:2512.12434},
        pages = {arXiv:2512.12434},
          doi = {10.48550/arXiv.2512.12434},
archivePrefix = {arXiv},
       eprint = {2512.12434},
 primaryClass = {astro-ph.IM},
       adsurl = {https://ui.adsabs.harvard.edu/abs/2025arXiv251212434D}
}

@ARTICLE{Whitcomb23,
       author = {{Whitcomb}, C.~M. and {Sandstrom}, K. and {Leroy}, A. and {Smith}, J.-D.~T.},
        title = "{Star Formation and Molecular Gas Diagnostics with Mid- and Far-infrared Emission}",
      journal = {\apj},
         year = 2023,
        month = may,
       volume = {948},
       number = {2},
          eid = {88},
        pages = {88},
          doi = {10.3847/1538-4357/acc316},
archivePrefix = {arXiv},
       eprint = {2212.00180},
 primaryClass = {astro-ph.GA},
       adsurl = {https://ui.adsabs.harvard.edu/abs/2023ApJ...948...88W}
}

@ARTICLE{Kim21,
       author = {{Kim}, Jaeyeon and {Chevance}, M{\'e}lanie and {Kruijssen}, J.~M. Diederik and {Schruba}, Andreas and {Sandstrom}, Karin and {Barnes}, Ashley T. and {Bigiel}, Frank and {Blanc}, Guillermo A. and {Cao}, Yixian and {Dale}, Daniel A. and {Faesi}, Christopher M. and {Glover}, Simon C.~O. and {Grasha}, Kathryn and {Groves}, Brent and {Herrera}, Cinthya and {Klessen}, Ralf S. and {Kreckel}, Kathryn and {Lee}, Janice C. and {Leroy}, Adam K. and {Pety}, J{\'e}r{\^o}me and {Querejeta}, Miguel and {Schinnerer}, Eva and {Sun}, Jiayi and {Usero}, Antonio and {Ward}, Jacob L. and {Williams}, Thomas G.},
        title = "{On the duration of the embedded phase of star formation}",
      journal = {\mnras},
         year = 2021,
        month = jun,
       volume = {504},
       number = {1},
        pages = {487-509},
          doi = {10.1093/mnras/stab878},
archivePrefix = {arXiv},
       eprint = {2012.00019},
 primaryClass = {astro-ph.GA},
       adsurl = {https://ui.adsabs.harvard.edu/abs/2021MNRAS.504..487K}
}

@ARTICLE{Ramambason25,
       author = {{Ramambason}, Lise and {Chevance}, M{\'e}lanie and {Kim}, Jaeyeon and {Belfiore}, Francesco and {Kruijssen}, J.~M. Diederik and {Romanelli}, Andrea and {Amiri}, Amirnezam and {Boquien}, M{\'e}d{\'e}ric and {Chown}, Ryan and {Dale}, Daniel A. and {Dlamini}, Simthembile and {Egorov}, Oleg V. and {Gerasimov}, Ivan and {Glover}, Simon C.~O. and {Grasha}, Kathryn and {Hassani}, Hamid and {Kim}, Hwihyun and {Kreckel}, Kathryn and {Koziol}, Hannah and {Leroy}, Adam K. and {M{\'e}ndez-Delgado}, Jos{\'e} Eduardo and {Neumann}, Justus and {Neumann}, Lukas and {Pan}, Hsi-An and {Pathak}, Debosmita and {Sandstrom}, Karin and {Sarbadhicary}, Sumit K. and {Schinnerer}, Eva and {Sun}, Jiayi and {Sutter}, Jessica and {Thilker}, David A. and {Ubeda}, Leonardo and {Weinbeck}, Tony D. and {Williams}, Thomas G. and {Whitmore}, Bradley C.},
        title = "{Duration and properties of the embedded phase of star formation in 37 nearby galaxies from PHANGS-JWST}",
      journal = {arXiv e-prints},
         year = 2025,
        month = jul,
          eid = {arXiv:2507.01508},
        pages = {arXiv:2507.01508},
          doi = {10.48550/arXiv.2507.01508},
archivePrefix = {arXiv},
       eprint = {2507.01508},
 primaryClass = {astro-ph.GA},
       adsurl = {https://ui.adsabs.harvard.edu/abs/2025arXiv250701508R}
}

@ARTICLE{Howard17,
       author = {{Howard}, Corey and {Pudritz}, Ralph and {Klessen}, Ralf},
        title = "{Ultraviolet Escape Fractions from Giant Molecular Clouds during Early Cluster Formation}",
      journal = {\apj},
         year = 2017,
        month = jan,
       volume = {834},
       number = {1},
          eid = {40},
        pages = {40},
          doi = {10.3847/1538-4357/834/1/40},
archivePrefix = {arXiv},
       eprint = {1611.02708},
 primaryClass = {astro-ph.GA},
       adsurl = {https://ui.adsabs.harvard.edu/abs/2017ApJ...834...40H}
}

@ARTICLE{Graham25,
       author = {{Graham}, Gabrielle B. and {Dale}, Daniel A. and {Smith}, Chase L. and {Brann}, Elisabeth and {Conder}, Kaycee D. and {Crowe}, Samuel and {Dhileepkumar}, Sumitra and {Imming}, Nicole A. and {Mendez}, Emilio and {Pleska}, Zachary and {Sako}, Kelsey and {Amiri}, Amirnezam and {Barnes}, Ashley T. and {Boquien}, M{\'e}d{\'e}ric and {Chandar}, Rupali and {Chown}, Ryan and {Gnedin}, Oleg Y. and {Grasha}, Kathryn and {Hannon}, Stephen and {Hassani}, Hamid and {Indebetouw}, R{\'e}my and {Kim}, Hwihyun and {Kim}, Jaeyeon and {Koziol}, Hannah and {Larson}, Kirsten L. and {Lee}, Janice C. and {Leroy}, Adam K. and {Oakes}, Elias K. and {Jimena Rodr{\'\i}guez}, M. and {Rosolowsky}, Erik and {Sandstrom}, Karin and {Schinnerer}, Eva and {Sutter}, Jessica and {Thilker}, David A. and {Ubeda}, Leonardo and {Whitmore}, Bradley C. and {Weinbeck}, Tony D. and {Williams}, Thomas G. and {Wofford}, Aida and {M{\'e}ndez-Delgado}, J. Eduardo and {Tian}, Qiushi Chris and {the PHANGS Collaboration}},
        title = "{PAH Marks the Spot: Digging for Buried Clusters in Nearby Star-forming Galaxies}",
      journal = {arXiv e-prints},
         year = 2025,
        month = nov,
          eid = {arXiv:2511.11920},
        pages = {arXiv:2511.11920},
          doi = {10.48550/arXiv.2511.11920},
archivePrefix = {arXiv},
       eprint = {2511.11920},
 primaryClass = {astro-ph.GA},
       adsurl = {https://ui.adsabs.harvard.edu/abs/2025arXiv251111920G}
}

@ARTICLE{Gregg24,
       author = {{Gregg}, Benjamin and {Calzetti}, Daniela and {Adamo}, Angela and {Bajaj}, Varun and {Ryon}, Jenna E. and {Linden}, Sean T. and {Correnti}, Matteo and {Cignoni}, Michele and {Messa}, Matteo and {Sabbi}, Elena and {Gallagher}, John S. and {Grasha}, Kathryn and {Pedrini}, Alex and {Gutermuth}, Robert A. and {Melinder}, Jens and {Kotulla}, Ralf and {P{\'e}rez}, Gustavo and {Krumholz}, Mark R. and {Bik}, Arjan and {{\"O}stlin}, G{\"o}ran and {Johnson}, Kelsey E. and {Bortolini}, Giacomo and {Smith}, Linda J. and {Tosi}, Monica and {Maji}, Subhransu and {Faustino Vieira}, Helena},
        title = "{Feedback in Emerging Extragalactic Star Clusters, FEAST: The Relation between 3.3 {\ensuremath{\mu}}m Polycyclic Aromatic Hydrocarbon Emission and Star Formation Rate Traced by Ionized Gas in NGC 628}",
      journal = {\apj},
         year = 2024,
        month = aug,
       volume = {971},
       number = {1},
          eid = {115},
        pages = {115},
          doi = {10.3847/1538-4357/ad54b4},
archivePrefix = {arXiv},
       eprint = {2405.09667},
 primaryClass = {astro-ph.GA},
       adsurl = {https://ui.adsabs.harvard.edu/abs/2024ApJ...971..115G}
}

@ARTICLE{Guillard12,
       author = {{Guillard}, P. and {Boulanger}, F. and {Pineau des For{\^e}ts}, G. and {Falgarone}, E. and {Gusdorf}, A. and {Cluver}, M.~E. and {Appleton}, P.~N. and {Lisenfeld}, U. and {Duc}, P. -A. and {Ogle}, P.~M. and {Xu}, C.~K.},
        title = "{Turbulent Molecular Gas and Star Formation in the Shocked Intergalactic Medium of Stephan's Quintet}",
      journal = {\apj},
         year = 2012,
        month = apr,
       volume = {749},
       number = {2},
          eid = {158},
        pages = {158},
          doi = {10.1088/0004-637X/749/2/158},
archivePrefix = {arXiv},
       eprint = {1202.2862},
 primaryClass = {astro-ph.CO},
       adsurl = {https://ui.adsabs.harvard.edu/abs/2012ApJ...749..158G}
}

@ARTICLE{Haas05,
       author = {{Haas}, M. and {Chini}, R. and {Klaas}, U.},
        title = "{Exceptional H$_{2}$ emission in the Antennae galaxies: Pre-starburst shocks from the galaxy collision}",
      journal = {\aap},
         year = 2005,
        month = apr,
       volume = {433},
       number = {2},
        pages = {L17-L20},
          doi = {10.1051/0004-6361:200500010},
       adsurl = {https://ui.adsabs.harvard.edu/abs/2005A&A...433L..17H}
}

@ARTICLE{He22,
       author = {{He}, Hao and {Wilson}, Christine and {Brunetti}, Nathan and {Finn}, Molly and {Bemis}, Ashley and {Johnson}, Kelsey},
        title = "{Embedded Young Massive Star Clusters in the Antennae Merger}",
      journal = {\apj},
         year = 2022,
        month = mar,
       volume = {928},
       number = {1},
          eid = {57},
        pages = {57},
          doi = {10.3847/1538-4357/ac5628},
archivePrefix = {arXiv},
       eprint = {2202.08077},
 primaryClass = {astro-ph.GA},
       adsurl = {https://ui.adsabs.harvard.edu/abs/2022ApJ...928...57H}
}

@ARTICLE{Herrera17,
       author = {{Herrera}, C.~N. and {Boulanger}, F.},
        title = "{The impact of a massive star cluster on its surrounding matter in the Antennae overlap region}",
      journal = {\aap},
         year = 2017,
        month = apr,
       volume = {600},
          eid = {A139},
        pages = {A139},
          doi = {10.1051/0004-6361/201628454},
archivePrefix = {arXiv},
       eprint = {1701.00835},
 primaryClass = {astro-ph.GA},
       adsurl = {https://ui.adsabs.harvard.edu/abs/2017A&A...600A.139H}
}

@ARTICLE{Hummer87,
       author = {{Hummer}, D.~G. and {Storey}, P.~J.},
        title = "{Recombination-line intensities for hydrogenic ions - I. Case B calculations for H I and He II.}",
      journal = {\mnras},
         year = 1987,
        month = feb,
       volume = {224},
        pages = {801-820},
          doi = {10.1093/mnras/224.3.801},
       adsurl = {https://ui.adsabs.harvard.edu/abs/1987MNRAS.224..801H}
}

@ARTICLE{Jog92,
       author = {{Jog}, Chanda J. and {Solomon}, P.~M.},
        title = "{A Triggering Mechanism for Enhanced Star Formation in Colliding Galaxies}",
      journal = {\apj},
         year = 1992,
        month = mar,
       volume = {387},
        pages = {152},
          doi = {10.1086/171067},
       adsurl = {https://ui.adsabs.harvard.edu/abs/1992ApJ...387..152J}
}

@ARTICLE{Johnson01,
       author = {{Johnson}, Kelsey E. and {Kobulnicky}, Henry A. and {Massey}, Philip and {Conti}, Peter S.},
        title = "{A Sample of Clusters of Extragalactic Ultracompact H II Regions}",
      journal = {\apj},
         year = 2001,
        month = oct,
       volume = {559},
       number = {2},
        pages = {864-877},
          doi = {10.1086/322335},
archivePrefix = {arXiv},
       eprint = {astro-ph/0107181},
 primaryClass = {astro-ph},
       adsurl = {https://ui.adsabs.harvard.edu/abs/2001ApJ...559..864J}
}

@ARTICLE{Johnson03,
       author = {{Johnson}, Kelsey E. and {Kobulnicky}, Henry A.},
        title = "{The Spectral Energy Distributions of Infant Super-Star Clusters in Henize 2-10 from 7 Millimeters to 6 Centimeters}",
      journal = {\apj},
         year = 2003,
        month = nov,
       volume = {597},
       number = {2},
        pages = {923-928},
          doi = {10.1086/378585},
archivePrefix = {arXiv},
       eprint = {astro-ph/0308303},
 primaryClass = {astro-ph},
       adsurl = {https://ui.adsabs.harvard.edu/abs/2003ApJ...597..923J}
}

@ARTICLE{Johnson04,
       author = {{Johnson}, Kelsey E. and {Indebetouw}, R{\'e}my and {Watson}, Christer and {Kobulnicky}, Henry A.},
        title = "{Revealing the Young Starburst in Haro 3 with Radio and Infrared Imaging}",
      journal = {\aj},
         year = 2004,
        month = aug,
       volume = {128},
       number = {2},
        pages = {610-616},
          doi = {10.1086/422017},
archivePrefix = {arXiv},
       eprint = {astro-ph/0406493},
 primaryClass = {astro-ph},
       adsurl = {https://ui.adsabs.harvard.edu/abs/2004AJ....128..610J}
}

@ARTICLE{Johnson09,
       author = {{Johnson}, Kelsey E. and {Hunt}, Leslie K. and {Reines}, Amy E.},
        title = "{Probing Star Formation at Low Metallicity: The Radio Emission of Super Star Clusters in SBS 0335-052}",
      journal = {\aj},
         year = 2009,
        month = apr,
       volume = {137},
       number = {4},
        pages = {3788-3799},
          doi = {10.1088/0004-6256/137/4/3788},
archivePrefix = {arXiv},
       eprint = {0902.0796},
 primaryClass = {astro-ph.CO},
       adsurl = {https://ui.adsabs.harvard.edu/abs/2009AJ....137.3788J}
}

@ARTICLE{Johnson15,
       author = {{Johnson}, K.~E. and {Leroy}, A.~K. and {Indebetouw}, R. and {Brogan}, C.~L. and {Whitmore}, B.~C. and {Hibbard}, J. and {Sheth}, K. and {Evans}, A.~S.},
        title = "{The Physical Conditions in a Pre-super Star Cluster Molecular Cloud in the Antennae Galaxies}",
      journal = {\apj},
         year = 2015,
        month = jun,
       volume = {806},
       number = {1},
          eid = {35},
        pages = {35},
          doi = {10.1088/0004-637X/806/1/35},
archivePrefix = {arXiv},
       eprint = {1503.06477},
 primaryClass = {astro-ph.GA},
       adsurl = {https://ui.adsabs.harvard.edu/abs/2015ApJ...806...35J}
}

@ARTICLE{Kavelaars97,
       author = {{Kavelaars}, J.~J. and {Hanes}, D.~A.},
        title = "{A comparison of the globular cluster luminosity functions of the inner and outer halo of the Milky Way and M31}",
      journal = {\mnras},
         year = 1997,
        month = mar,
       volume = {285},
       number = {4},
        pages = {L31-L34},
          doi = {10.1093/mnras/285.4.L31},
archivePrefix = {arXiv},
       eprint = {astro-ph/9608191},
 primaryClass = {astro-ph},
       adsurl = {https://ui.adsabs.harvard.edu/abs/1997MNRAS.285L..31K}
}

@ARTICLE{Kepley14,
       author = {{Kepley}, Amanda A. and {Reines}, Amy E. and {Johnson}, Kelsey E. and {Walker}, Lisa May},
        title = "{High Resolution Radio and Optical Observations of the Central Starburst in the Low-metallicity Dwarf Galaxy II Zw 40}",
      journal = {\aj},
         year = 2014,
        month = feb,
       volume = {147},
       number = {2},
          eid = {43},
        pages = {43},
          doi = {10.1088/0004-6256/147/2/43},
archivePrefix = {arXiv},
       eprint = {1312.4988},
 primaryClass = {astro-ph.GA},
       adsurl = {https://ui.adsabs.harvard.edu/abs/2014AJ....147...43K}
}

@ARTICLE{Klaas10,
       author = {{Klaas}, U. and {Nielbock}, M. and {Haas}, M. and {Krause}, O. and {Schreiber}, J.},
        title = "{Tracing the sites of obscured star formation in the Antennae galaxies with Herschel-PACS}",
      journal = {\aap},
         year = 2010,
        month = jul,
       volume = {518},
          eid = {L44},
        pages = {L44},
          doi = {10.1051/0004-6361/201014670},
archivePrefix = {arXiv},
       eprint = {1005.2290},
 primaryClass = {astro-ph.CO},
       adsurl = {https://ui.adsabs.harvard.edu/abs/2010A&A...518L..44K}
}

@ARTICLE{Knutas25,
       author = {{Knutas}, Alice and {Adamo}, Angela and {Pedrini}, Alex and {Linden}, Sean T. and {Bajaj}, Varun and {Ryon}, Jenna E. and {Gregg}, Benjamin and {Ali}, Ahmad A. and {Andersson}, Eric P. and {Bik}, Arjan and {Bortolini}, Giacomo and {Buckner}, Anne S.~M. and {Calzetti}, Daniela and {Duarte-Cabral}, Ana and {Elmegreen}, Bruce G. and {Faustino Vieira}, Helena and {Gallagher}, John S. and {Grasha}, Kathryn and {Johnson}, Kelsey and {Lai}, Thomas S. -Y. and {Lapeer}, Drew and {Messa}, Matteo and {{\"O}stlin}, G{\"o}ran and {Sabbi}, Elena and {Smith}, Linda J. and {Tosi}, Monica},
        title = "{FEAST: JWST uncovers the emerging timescales of young star clusters in M83}",
      journal = {arXiv e-prints},
         year = 2025,
        month = may,
          eid = {arXiv:2505.08874},
        pages = {arXiv:2505.08874},
          doi = {10.48550/arXiv.2505.08874},
archivePrefix = {arXiv},
       eprint = {2505.08874},
 primaryClass = {astro-ph.GA},
       adsurl = {https://ui.adsabs.harvard.edu/abs/2025arXiv250508874K}
}

@ARTICLE{Kobulnicky99,
       author = {{Kobulnicky}, Henry A. and {Johnson}, Kelsey E.},
        title = "{Signatures of the Youngest Starbursts: Optically Thick Thermal Bremsstrahlung Radio Sources in Henize 2-10}",
      journal = {\apj},
         year = 1999,
        month = dec,
       volume = {527},
       number = {1},
        pages = {154-166},
          doi = {10.1086/308075},
archivePrefix = {arXiv},
       eprint = {astro-ph/9907233},
 primaryClass = {astro-ph},
       adsurl = {https://ui.adsabs.harvard.edu/abs/1999ApJ...527..154K}
}

@ARTICLE{Kunze96,
       author = {{Kunze}, D. and {Rigopoulou}, D. and {Lutz}, D. and {Egami}, E. and {Feuchtgruber}, H. and {Genzel}, R. and {Spoon}, H.~W.~W. and {Sturm}, E. and {Sternberg}, A. and {Moorwood}, A.~F.~M. and {de Graauw}, T.},
        title = "{SWS spectroscopy of the colliding galaxies NGC4038/39.}",
      journal = {\aap},
         year = 1996,
        month = nov,
       volume = {315},
        pages = {L101-L104},
       adsurl = {https://ui.adsabs.harvard.edu/abs/1996A&A...315L.101K}
}

@ARTICLE{Lai23,
       author = {{Lai}, Thomas S. -Y. and {Armus}, Lee and {Bianchin}, Marina and {D{\'\i}az-Santos}, Tanio and {Linden}, Sean T. and {Privon}, George C. and {Inami}, Hanae and {U}, Vivian and {Bohn}, Thomas and {Evans}, Aaron S. and {Larson}, Kirsten L. and {Hensley}, Brandon S. and {Smith}, J. -D.~T. and {Malkan}, Matthew A. and {Song}, Yiqing and {Stierwalt}, Sabrina and {van der Werf}, Paul P. and {McKinney}, Jed and {Aalto}, Susanne and {Buiten}, Victorine A. and {Rich}, Jeff and {Charmandaris}, Vassilis and {Appleton}, Philip and {Barcos-Mu{\~n}oz}, Loreto and {B{\"o}ker}, Torsten and {Finnerty}, Luke and {Kader}, Justin A. and {Law}, David R. and {Medling}, Anne M. and {Brown}, Michael J.~I. and {Hayward}, Christopher C. and {Howell}, Justin and {Iwasawa}, Kazushi and {Kemper}, Francisca and {Marshall}, Jason and {Mazzarella}, Joseph M. and {M{\"u}ller-S{\'a}nchez}, Francisco and {Murphy}, Eric J. and {Sanders}, David and {Surace}, Jason},
        title = "{GOALS-JWST: Small Neutral Grains and Enhanced 3.3 {\ensuremath{\mu}}m PAH Emission in the Seyfert Galaxy NGC 7469}",
      journal = {\apjl},
         year = 2023,
        month = nov,
       volume = {957},
       number = {2},
          eid = {L26},
        pages = {L26},
          doi = {10.3847/2041-8213/ad0387},
archivePrefix = {arXiv},
       eprint = {2307.15169},
 primaryClass = {astro-ph.GA},
       adsurl = {https://ui.adsabs.harvard.edu/abs/2023ApJ...957L..26L}
}

@ARTICLE{Li17,
	author = {{Li}, H. and {Gnedin}, O.~Y. and {Gnedin}, N.~Y. and {Meng}, X. and 
	{Semenov}, V.~A. and {Kravtsov}, A.~V.},
	title = "{Star Cluster Formation in Cosmological Simulations. I. Properties of Young Clusters}",
	journal = {\apj},
	archivePrefix = "arXiv",
	eprint = {1608.03244},
	year = 2017,
	month = jan,
	volume = 834,
	eid = {69},
	pages = {69},
	doi = {10.3847/1538-4357/834/1/69},
	adsurl = {http://adsabs.harvard.edu/abs/2017ApJ...834...69L}
}

@ARTICLE{Li20,
       author = {{Li}, Hui and {Vogelsberger}, Mark and {Marinacci}, Federico and {Sales}, Laura V. and {Torrey}, Paul},
        title = "{The effects of subgrid models on the properties of giant molecular clouds in galaxy formation simulations}",
      journal = {\mnras},
         year = 2020,
        month = dec,
       volume = {499},
       number = {4},
        pages = {5862-5872},
          doi = {10.1093/mnras/staa3122},
archivePrefix = {arXiv},
       eprint = {2001.07214},
 primaryClass = {astro-ph.GA},
       adsurl = {https://ui.adsabs.harvard.edu/abs/2020MNRAS.499.5862L}
}

@ARTICLE{Linden23,
       author = {{Linden}, Sean T. and {Evans}, Aaron S. and {Armus}, Lee and {Rich}, Jeffrey A. and {Larson}, Kirsten L. and {Lai}, Thomas and {Privon}, George C. and {U}, Vivian and {Inami}, Hanae and {Bohn}, Thomas and {Song}, Yiqing and {Barcos-Mu{\~n}oz}, Loreto and {Charmandaris}, Vassilis and {Medling}, Anne M. and {Stierwalt}, Sabrina and {Diaz-Santos}, Tanio and {B{\"o}ker}, Torsten and {van der Werf}, Paul and {Aalto}, Susanne and {Appleton}, Philip and {Brown}, Michael J.~I. and {Hayward}, Christopher C. and {Howell}, Justin H. and {Iwasawa}, Kazushi and {Kemper}, Francisca and {Frayer}, David T. and {Law}, David and {Malkan}, Matthew A. and {Marshall}, Jason and {Mazzarella}, Joseph M. and {Murphy}, Eric J. and {Sanders}, David and {Surace}, Jason},
        title = "{GOALS-JWST: Revealing the Buried Star Clusters in the Luminous Infrared Galaxy VV 114}",
      journal = {\apjl},
         year = 2023,
        month = feb,
       volume = {944},
       number = {2},
          eid = {L55},
        pages = {L55},
          doi = {10.3847/2041-8213/acb335},
archivePrefix = {arXiv},
       eprint = {2210.05763},
 primaryClass = {astro-ph.GA},
       adsurl = {https://ui.adsabs.harvard.edu/abs/2023ApJ...944L..55L}
}

@ARTICLE{Maji17,
       author = {{Maji}, Moupiya and {Zhu}, Qirong and {Li}, Yuexing and {Charlton}, Jane and {Hernquist}, Lars and {Knebe}, Alexander},
        title = "{The Formation and Evolution of Star Clusters in Interacting Galaxies}",
      journal = {\apj},
         year = 2017,
        month = aug,
       volume = {844},
       number = {2},
          eid = {108},
        pages = {108},
          doi = {10.3847/1538-4357/aa7aa1},
archivePrefix = {arXiv},
       eprint = {1606.07091},
 primaryClass = {astro-ph.GA},
       adsurl = {https://ui.adsabs.harvard.edu/abs/2017ApJ...844..108M}
}

@ARTICLE{McCrady07,
       author = {{McCrady}, Nate and {Graham}, James R.},
        title = "{Super Star Cluster Velocity Dispersions and Virial Masses in the M82 Nuclear Starburst}",
      journal = {\apj},
         year = 2007,
        month = jul,
       volume = {663},
       number = {2},
        pages = {844-856},
          doi = {10.1086/518357},
archivePrefix = {arXiv},
       eprint = {0704.0478},
 primaryClass = {astro-ph},
       adsurl = {https://ui.adsabs.harvard.edu/abs/2007ApJ...663..844M}
}

@ARTICLE{Messa21,
       author = {{Messa}, Matteo and {Calzetti}, Daniela and {Adamo}, Angela and {Grasha}, Kathryn and {Johnson}, Kelsey E. and {Sabbi}, Elena and {Smith}, Linda J. and {Bajaj}, Varun and {Finn}, Molly K. and {Lin}, Zesen},
        title = "{Looking for Obscured Young Star Clusters in NGC 1313}",
      journal = {\apj},
         year = 2021,
        month = mar,
       volume = {909},
       number = {2},
          eid = {121},
        pages = {121},
          doi = {10.3847/1538-4357/abe0b5},
archivePrefix = {arXiv},
       eprint = {2011.09392},
 primaryClass = {astro-ph.GA},
       adsurl = {https://ui.adsabs.harvard.edu/abs/2021ApJ...909..121M}
}

@ARTICLE{Messa24,
       author = {{Messa}, Matteo and {Dessauges-Zavadsky}, Miroslava and {Adamo}, Angela and {Richard}, Johan and {Claeyssens}, Ad{\'e}la{\"\i}de},
        title = "{Properties of the brightest young stellar clumps in extremely lensed galaxies at redshifts 4 to 5}",
      journal = {\mnras},
         year = 2024,
        month = apr,
       volume = {529},
       number = {3},
        pages = {2162-2179},
          doi = {10.1093/mnras/stae565},
archivePrefix = {arXiv},
       eprint = {2402.14920},
 primaryClass = {astro-ph.GA},
       adsurl = {https://ui.adsabs.harvard.edu/abs/2024MNRAS.529.2162M}
}

@ARTICLE{Mirabel98,
       author = {{Mirabel}, I.~F. and {Vigroux}, L. and {Charmandaris}, V. and {Sauvage}, M. and {Gallais}, P. and {Tran}, D. and {Cesarsky}, C. and {Madden}, S.~C. and {Duc}, P. -A.},
        title = "{The dark side of star formation in the Antennae galaxies}",
      journal = {\aap},
         year = 1998,
        month = may,
       volume = {333},
        pages = {L1-L4},
          doi = {10.48550/arXiv.astro-ph/9802176},
archivePrefix = {arXiv},
       eprint = {astro-ph/9802176},
 primaryClass = {astro-ph},
       adsurl = {https://ui.adsabs.harvard.edu/abs/1998A&A...333L...1M}
}

@ARTICLE{Miller97,
       author = {{Miller}, Bryan W. and {Whitmore}, Bradley C. and {Schweizer}, Francois and {Fall}, S. Michael},
        title = "{The Star Cluster System of the Merger Remnant NGC 7252}",
      journal = {\aj},
         year = 1997,
        month = dec,
       volume = {114},
        pages = {2381},
          doi = {10.1086/118655},
       adsurl = {https://ui.adsabs.harvard.edu/abs/1997AJ....114.2381M}
}

@ARTICLE{Mok19,
	author = {{Mok}, Angus and {Chandar}, Rupali and {Fall}, S. Michael},
	title = "{Constraints on Upper Cutoffs in the Mass Functions of Young Star Clusters}",
	journal = {\apj},
	year = "2019",
	month = "Feb",
	volume = {872},
	number = {1},
	eid = {93},
	pages = {93},
	doi = {10.3847/1538-4357/aaf6ea},
	archivePrefix = {arXiv},
	eprint = {1806.11192},
	primaryClass = {astro-ph.GA},
	adsurl = {https://ui.adsabs.harvard.edu/abs/2019ApJ...872...93M}
}

@ARTICLE{Mowla22,
       author = {{Mowla}, Lamiya and {Iyer}, Kartheik G. and {Desprez}, Guillaume and {Estrada-Carpenter}, Vicente and {Martis}, Nicholas S. and {Noirot}, Ga{\"e}l and {Sarrouh}, Ghassan T. and {Strait}, Victoria and {Asada}, Yoshihisa and {Abraham}, Roberto G. and {Brammer}, Gabriel and {Sawicki}, Marcin and {Willott}, Chris J. and {Bradac}, Marusa and {Doyon}, Ren{\'e} and {Muzzin}, Adam and {Pacifici}, Camilla and {Ravindranath}, Swara and {Zabl}, Johannes},
        title = "{The Sparkler: Evolved High-redshift Globular Cluster Candidates Captured by JWST}",
      journal = {\apjl},
         year = 2022,
        month = oct,
       volume = {937},
       number = {2},
          eid = {L35},
        pages = {L35},
          doi = {10.3847/2041-8213/ac90ca},
archivePrefix = {arXiv},
       eprint = {2208.02233},
 primaryClass = {astro-ph.GA},
       adsurl = {https://ui.adsabs.harvard.edu/abs/2022ApJ...937L..35M}
}

@ARTICLE{Murphy11,
       author = {{Murphy}, E.~J. and {Condon}, J.~J. and {Schinnerer}, E. and {Kennicutt}, R.~C. and {Calzetti}, D. and {Armus}, L. and {Helou}, G. and {Turner}, J.~L. and {Aniano}, G. and {Beir{\~a}o}, P. and {Bolatto}, A.~D. and {Brandl}, B.~R. and {Croxall}, K.~V. and {Dale}, D.~A. and {Donovan Meyer}, J.~L. and {Draine}, B.~T. and {Engelbracht}, C. and {Hunt}, L.~K. and {Hao}, C. -N. and {Koda}, J. and {Roussel}, H. and {Skibba}, R. and {Smith}, J. -D.~T.},
        title = "{Calibrating Extinction-free Star Formation Rate Diagnostics with 33 GHz Free-free Emission in NGC 6946}",
      journal = {\apj},
         year = 2011,
        month = aug,
       volume = {737},
       number = {2},
          eid = {67},
        pages = {67},
          doi = {10.1088/0004-637X/737/2/67},
archivePrefix = {arXiv},
       eprint = {1105.4877},
 primaryClass = {astro-ph.CO},
       adsurl = {https://ui.adsabs.harvard.edu/abs/2011ApJ...737...67M}
}

@ARTICLE{Neff00,
       author = {{Neff}, Susan G. and {Ulvestad}, James S.},
        title = "{VLA Observations of the Nearby Merger NGC 4038/4039: H II Regions and Supernova Remnants in the ``Antennae''}",
      journal = {\aj},
         year = 2000,
        month = aug,
       volume = {120},
       number = {2},
        pages = {670-696},
          doi = {10.1086/301503},
       adsurl = {https://ui.adsabs.harvard.edu/abs/2000AJ....120..670N}
}

@ARTICLE{Ogle10,
       author = {{Ogle}, Patrick and {Boulanger}, Francois and {Guillard}, Pierre and {Evans}, Daniel A. and {Antonucci}, Robert and {Appleton}, P.~N. and {Nesvadba}, Nicole and {Leipski}, Christian},
        title = "{Jet-powered Molecular Hydrogen Emission from Radio Galaxies}",
      journal = {\apj},
         year = 2010,
        month = dec,
       volume = {724},
       number = {2},
        pages = {1193-1217},
          doi = {10.1088/0004-637X/724/2/1193},
archivePrefix = {arXiv},
       eprint = {1009.4533},
 primaryClass = {astro-ph.CO},
       adsurl = {https://ui.adsabs.harvard.edu/abs/2010ApJ...724.1193O}
}

@ARTICLE{Peng10,
	author = {{Peng}, Y.-j. and {Lilly}, S.~J. and {Kova{\v c}}, K. and {Bolzonella}, M. and 
	{Pozzetti}, L. and {Renzini}, A. and {Zamorani}, G. and {Ilbert}, O. and 
	{Knobel}, C. and {Iovino}, A. and {Maier}, C. and {Cucciati}, O. and 
	{Tasca}, L. and {Carollo}, C.~M. and {Silverman}, J. and {Kampczyk}, P. and 
	{de Ravel}, L. and {Sanders}, D. and {Scoville}, N. and {Contini}, T. and 
	{Mainieri}, V. and {Scodeggio}, M. and {Kneib}, J.-P. and {Le F{\`e}vre}, O. and 
	{Bardelli}, S. and {Bongiorno}, A. and {Caputi}, K. and {Coppa}, G. and 
	{de la Torre}, S. and {Franzetti}, P. and {Garilli}, B. and 
	{Lamareille}, F. and {Le Borgne}, J.-F. and {Le Brun}, V. and 
	{Mignoli}, M. and {Perez Montero}, E. and {Pello}, R. and {Ricciardelli}, E. and 
	{Tanaka}, M. and {Tresse}, L. and {Vergani}, D. and {Welikala}, N. and 
	{Zucca}, E. and {Oesch}, P. and {Abbas}, U. and {Barnes}, L. and 
	{Bordoloi}, R. and {Bottini}, D. and {Cappi}, A. and {Cassata}, P. and 
	{Cimatti}, A. and {Fumana}, M. and {Hasinger}, G. and {Koekemoer}, A. and 
	{Leauthaud}, A. and {Maccagni}, D. and {Marinoni}, C. and {McCracken}, H. and 
	{Memeo}, P. and {Meneux}, B. and {Nair}, P. and {Porciani}, C. and 
	{Presotto}, V. and {Scaramella}, R.},
	title = "{Mass and Environment as Drivers of Galaxy Evolution in SDSS and zCOSMOS and the Origin of the Schechter Function}",
	journal = {\apj},
	archivePrefix = "arXiv",
	eprint = {1003.4747},
	primaryClass = "astro-ph.CO",
	year = 2010,
	month = sep,
	volume = 721,
	pages = {193-221},
	doi = {10.1088/0004-637X/721/1/193},
	adsurl = {http://adsabs.harvard.edu/abs/2010ApJ...721..193P}
}

@ARTICLE{Renaud15,
       author = {{Renaud}, Florent and {Bournaud}, Fr{\'e}d{\'e}ric and {Duc}, Pierre-Alain},
        title = "{A parsec-resolution simulation of the Antennae galaxies: formation of star clusters during the merger}",
      journal = {\mnras},
         year = 2015,
        month = jan,
       volume = {446},
       number = {2},
        pages = {2038-2054},
          doi = {10.1093/mnras/stu2208},
archivePrefix = {arXiv},
       eprint = {1410.5754},
 primaryClass = {astro-ph.GA},
       adsurl = {https://ui.adsabs.harvard.edu/abs/2015MNRAS.446.2038R}
}

@ARTICLE{Rodriguez23,
       author = {{Rodr{\'\i}guez}, M. Jimena and {Lee}, Janice C. and {Whitmore}, B.~C. and {Thilker}, David A. and {Maschmann}, Daniel and {Chandar}, Rupali and {Deger}, Sinan and {Boquien}, M{\'e}d{\'e}ric and {Dale}, Daniel A. and {Larson}, Kirsten L. and {Williams}, Thomas G. and {Kim}, Hwihyun and {Schinnerer}, Eva and {Rosolowsky}, Erik and {Leroy}, Adam K. and {Emsellem}, Eric and {Sandstrom}, Karin M. and {Kruijssen}, J.~M. Diederik and {Grasha}, Kathryn and {Watkins}, Elizabeth J. and {Barnes}, Ashley. T. and {Sormani}, Mattia C. and {Kim}, Jaeyeon and {Anand}, Gagandeep S. and {Chevance}, M{\'e}lanie and {Bigiel}, F. and {Klessen}, Ralf S. and {Hassani}, Hamid and {Liu}, Daizhong and {Faesi}, Christopher M. and {Cao}, Yixian and {Belfiore}, Francesco and {Pessa}, Ismael and {Kreckel}, Kathryn and {Groves}, Brent and {Pety}, J{\'e}r{\^o}me and {Indebetouw}, R{\'e}my and {Egorov}, Oleg V. and {Blanc}, Guillermo A. and {Saito}, Toshiki and {Hughes}, Annie},
        title = "{PHANGS-JWST First Results: Dust-embedded Star Clusters in NGC 7496 Selected via 3.3 {\ensuremath{\mu}}m PAH Emission}",
      journal = {\apjl},
         year = 2023,
        month = feb,
       volume = {944},
       number = {2},
          eid = {L26},
        pages = {L26},
          doi = {10.3847/2041-8213/aca653},
       adsurl = {https://ui.adsabs.harvard.edu/abs/2023ApJ...944L..26R}
}

@ARTICLE{Rodriguez25,
       author = {{Rodr{\'\i}guez}, M. Jimena and {Lee}, Janice C. and {Indebetouw}, Remy and {Whitmore}, B.~C. and {Maschmann}, Daniel and {Williams}, Thomas G. and {Chandar}, Rupali and {Barnes}, A.~T. and {Gnedin}, Oleg Y. and {Sandstrom}, Karin M. and {Rosolowsky}, Erik and {Leroy}, Adam K. and {Thilker}, David A. and {Kim}, Hwihyun and {Sun}, Jiayi and {Klessen}, Ralf S. and {Groves}, Brent and {Wofford}, Aida and {Boquien}, M{\'e}d{\'e}ric and {Dale}, Daniel A. and {{\'U}beda}, Leonardo and {Larson}, Kirsten L. and {Grasha}, Kathryn and {Johnson}, Kelsey E. and {Levy}, Rebecca C. and {Bigiel}, Frank and {Hassani}, Hamid and {Sarbadhicary}, Sumit K.},
        title = "{Tracing the Earliest Stages of Star and Cluster Formation in 19 Nearby Galaxies with PHANGS-JWST and HST: Compact 3.3 {\ensuremath{\mu}}m Polycyclic Aromatic Hydrocarbon Emitters and Their Relation to the Optical Census of Star Clusters}",
      journal = {\apj},
         year = 2025,
        month = apr,
       volume = {983},
       number = {2},
          eid = {137},
        pages = {137},
          doi = {10.3847/1538-4357/adbb69},
archivePrefix = {arXiv},
       eprint = {2412.07862},
 primaryClass = {astro-ph.GA},
       adsurl = {https://ui.adsabs.harvard.edu/abs/2025ApJ...983..137R}
}

@ARTICLE{Rosolowsky05,
	author = {{Rosolowsky}, E.},
	title = "{The Mass Spectra of Giant Molecular Clouds in the Local Group}",
	journal = {\pasp},
	eprint = {astro-ph/0508679},
	year = 2005,
	month = dec,
	volume = 117,
	pages = {1403-1410},
	doi = {10.1086/497582},
	adsurl = {http://adsabs.harvard.edu/abs/2005PASP..117.1403R}
}

@ARTICLE{Rubin70,
       author = {{Rubin}, Vera C. and {Ford}, W. Kent, Jr. and {D'Odorico}, Sandro},
        title = "{Emission-Line Intensities and Radial Velocities in the Interacting Galaxies NGC 4038-4039}",
      journal = {\apj},
         year = 1970,
        month = jun,
       volume = {160},
        pages = {801},
          doi = {10.1086/150473},
       adsurl = {https://ui.adsabs.harvard.edu/abs/1970ApJ...160..801R}
}

@ARTICLE{Senchyna24,
       author = {{Senchyna}, Peter and {Plat}, Adele and {Stark}, Daniel P. and {Rudie}, Gwen C. and {Berg}, Danielle and {Charlot}, St{\'e}phane and {James}, Bethan L. and {Mingozzi}, Matilde},
        title = "{GN-z11 in Context: Possible Signatures of Globular Cluster Precursors at Redshift 10}",
      journal = {\apj},
         year = 2024,
        month = may,
       volume = {966},
       number = {1},
          eid = {92},
        pages = {92},
          doi = {10.3847/1538-4357/ad235e},
archivePrefix = {arXiv},
       eprint = {2303.04179},
 primaryClass = {astro-ph.GA},
       adsurl = {https://ui.adsabs.harvard.edu/abs/2024ApJ...966...92S}
}

@ARTICLE{Smith07,
       author = {{Smith}, J.~D.~T. and {Draine}, B.~T. and {Dale}, D.~A. and {Moustakas}, J. and {Kennicutt}, Jr., R.~C. and {Helou}, G. and {Armus}, L. and {Roussel}, H. and {Sheth}, K. and {Bendo}, G.~J. and {Buckalew}, B.~A. and {Calzetti}, D. and {Engelbracht}, C.~W. and {Gordon}, K.~D. and {Hollenbach}, D.~J. and {Li}, A. and {Malhotra}, S. and {Murphy}, E.~J. and {Walter}, F.},
        title = "{The Mid-Infrared Spectrum of Star-forming Galaxies: Global Properties of Polycyclic Aromatic Hydrocarbon Emission}",
      journal = {\apj},
         year = 2007,
        month = feb,
       volume = {656},
       number = {2},
        pages = {770-791},
          doi = {10.1086/510549},
archivePrefix = {arXiv},
       eprint = {astro-ph/0610913},
 primaryClass = {astro-ph},
       adsurl = {https://ui.adsabs.harvard.edu/abs/2007ApJ...656..770S}
}

@ARTICLE{Stanford90,
       author = {{Stanford}, S.~A. and {Sargent}, A.~I. and {Sanders}, D.~B. and {Scoville}, N.~Z.},
        title = "{CO Aperture Synthesis of NGC 4038/39 (Arp 244)}",
      journal = {\apj},
         year = 1990,
        month = feb,
       volume = {349},
        pages = {492},
          doi = {10.1086/168334},
       adsurl = {https://ui.adsabs.harvard.edu/abs/1990ApJ...349..492S}
}

@ARTICLE{Thilker22,
       author = {{Thilker}, David A. and {Whitmore}, Bradley C. and {Lee}, Janice C. and {Deger}, Sinan and {Chandar}, Rupali and {Larson}, Kirsten L. and {Hannon}, Stephen and {Ubeda}, Leonardo and {Dale}, Daniel A. and {Glover}, Simon C.~O. and {Grasha}, Kathryn and {Klessen}, Ralf S. and {Kruijssen}, J.~M. Diederik and {Rosolowsky}, Erik and {Schruba}, Andreas and {White}, Richard L. and {Williams}, Thomas G.},
        title = "{PHANGS-HST: new methods for star cluster identification in nearby galaxies}",
      journal = {\mnras},
         year = 2022,
        month = jan,
       volume = {509},
       number = {3},
        pages = {4094-4127},
          doi = {10.1093/mnras/stab3183},
archivePrefix = {arXiv},
       eprint = {2106.13366},
 primaryClass = {astro-ph.GA},
       adsurl = {https://ui.adsabs.harvard.edu/abs/2022MNRAS.509.4094T}
}

@ARTICLE{Tsuge21b,
       author = {{Tsuge}, Kisetsu and {Tachihara}, Kengo and {Fukui}, Yasuo and {Sano}, Hidetoshi and {Tokuda}, Kazuki and {Ueda}, Junko and {Iono}, Daisuke},
        title = "{The formation of the young massive cluster B1 in the Antennae Galaxies (NGC 4038/NGC 4039) triggered by cloud-cloud collision}",
      journal = {\pasj},
         year = 2021,
        month = apr,
       volume = {73},
       number = {2},
        pages = {417-430},
          doi = {10.1093/pasj/psab008},
archivePrefix = {arXiv},
       eprint = {2005.04075},
 primaryClass = {astro-ph.GA},
       adsurl = {https://ui.adsabs.harvard.edu/abs/2021PASJ...73..417T}
}

@ARTICLE{Vanzella22,
       author = {{Vanzella}, E. and {Castellano}, M. and {Bergamini}, P. and {Meneghetti}, M. and {Zanella}, A. and {Calura}, F. and {Caminha}, G.~B. and {Rosati}, P. and {Cupani}, G. and {Me{\v{s}}tri{\'c}}, U. and {Brammer}, G. and {Tozzi}, P. and {Mercurio}, A. and {Grillo}, C. and {Sani}, E. and {Cristiani}, S. and {Nonino}, M. and {Merlin}, E. and {Pignataro}, G.~V.},
        title = "{High star cluster formation efficiency in the strongly lensed Sunburst Lyman-continuum galaxy at z = 2.37}",
      journal = {\aap},
         year = 2022,
        month = mar,
       volume = {659},
          eid = {A2},
        pages = {A2},
          doi = {10.1051/0004-6361/202141590},
archivePrefix = {arXiv},
       eprint = {2106.10280},
 primaryClass = {astro-ph.GA},
       adsurl = {https://ui.adsabs.harvard.edu/abs/2022A&A...659A...2V}
}

@ARTICLE{Vanzella23,
       author = {{Vanzella}, E. and {Loiacono}, F. and {Bergamini}, P. and {Me{\v{s}}tri{\'c}}, U. and {Castellano}, M. and {Rosati}, P. and {Meneghetti}, M. and {Grillo}, C. and {Calura}, F. and {Mignoli}, M. and {Brada{\v{c}}}, M. and {Adamo}, A. and {Rihtar{\v{s}}i{\v{c}}}, G. and {Dickinson}, M. and {Gronke}, M. and {Zanella}, A. and {Annibali}, F. and {Willott}, C. and {Messa}, M. and {Sani}, E. and {Acebron}, A. and {Bolamperti}, A. and {Comastri}, A. and {Gilli}, R. and {Caputi}, K.~I. and {Ricotti}, M. and {Gruppioni}, C. and {Ravindranath}, S. and {Mercurio}, A. and {Strait}, V. and {Martis}, N. and {Pascale}, R. and {Caminha}, G.~B. and {Annunziatella}, M. and {Nonino}, M.},
        title = "{An extremely metal-poor star complex in the reionization era: Approaching Population III stars with JWST}",
      journal = {\aap},
         year = 2023,
        month = oct,
       volume = {678},
          eid = {A173},
        pages = {A173},
          doi = {10.1051/0004-6361/202346981},
archivePrefix = {arXiv},
       eprint = {2305.14413},
 primaryClass = {astro-ph.GA},
       adsurl = {https://ui.adsabs.harvard.edu/abs/2023A&A...678A.173V}
}

@ARTICLE{Whelan11,
       author = {{Whelan}, David G. and {Johnson}, Kelsey E. and {Whitney}, Barbara A. and {Indebetouw}, R{\'e}my and {Wood}, Kenneth},
        title = "{The Infrared Properties of Embedded Super Star Clusters: Predictions from Three-dimensional Radiative Transfer Models}",
      journal = {\apj},
         year = 2011,
        month = mar,
       volume = {729},
       number = {2},
          eid = {111},
        pages = {111},
          doi = {10.1088/0004-637X/729/2/111},
archivePrefix = {arXiv},
       eprint = {1101.3324},
 primaryClass = {astro-ph.CO},
       adsurl = {https://ui.adsabs.harvard.edu/abs/2011ApJ...729..111W}
}

@ARTICLE{Whitmore95,
       author = {{Whitmore}, Bradley C. and {Schweizer}, Francois},
        title = "{Hubble Space Telescope Observations of Young Star Clusters in NGC 4038/4039, ``The Antennae'' Galaxies}",
      journal = {\aj},
         year = 1995,
        month = mar,
       volume = {109},
        pages = {960},
          doi = {10.1086/117334},
       adsurl = {https://ui.adsabs.harvard.edu/abs/1995AJ....109..960W}
}

@ARTICLE{Whitaker25,
       author = {{Whitaker}, Katherine E. and {Cutler}, Sam E. and {Chandar}, Rupali and {Pan}, Richard and {Setton}, David J. and {Furtak}, Lukas J. and {Bezanson}, Rachel and {Labb{\'e}}, Ivo and {Leja}, Joel and {Suess}, Katherine A. and {Wang}, Bingjie and {Weaver}, John R. and {Atek}, Hakim and {Brammer}, Gabriel B. and {Feldmann}, Robert and {F{\"o}rster Schreiber}, Natascha M. and {Glazebrook}, Karl and {de Graaff}, Anna and {Greene}, Jenny E. and {Khullar}, Gourav and {Marchesini}, Danilo and {Maseda}, Michael V. and {Miller}, Tim B. and {Mo}, Houjun and {Mowla}, Lamiya A. and {Nanayakkara}, Themiya and {Nelson}, Erica J. and {Price}, Sedona H. and {Rizzo}, Francesca and {van Dokkum}, Pieter and {Williams}, Christina C. and {Zhang}, Yanzhe and {Zhang}, Yunchong and {Zitrin}, Adi},
        title = "{Discovery of Ancient Globular Cluster Candidates in The Relic, a Quiescent Galaxy at z=2.5}",
      journal = {arXiv e-prints},
         year = 2025,
        month = jan,
          eid = {arXiv:2501.07627},
        pages = {arXiv:2501.07627},
          doi = {10.48550/arXiv.2501.07627},
archivePrefix = {arXiv},
       eprint = {2501.07627},
 primaryClass = {astro-ph.GA},
       adsurl = {https://ui.adsabs.harvard.edu/abs/2025arXiv250107627W}
}

@ARTICLE{Whitmore99,
       author = {{Whitmore}, Bradley C. and {Zhang}, Qing and {Leitherer}, Claus and {Fall}, S. Michael and {Schweizer}, Fran{\c{c}}ois and {Miller}, Bryan W.},
        title = "{The Luminosity Function of Young Star Clusters in ``the Antennae'' Galaxies (NGC 4038-4039)}",
      journal = {\aj},
         year = 1999,
        month = oct,
       volume = {118},
       number = {4},
        pages = {1551-1576},
          doi = {10.1086/301041},
archivePrefix = {arXiv},
       eprint = {astro-ph/9907430},
 primaryClass = {astro-ph},
       adsurl = {https://ui.adsabs.harvard.edu/abs/1999AJ....118.1551W}
}

@ARTICLE{Whitmore02,
       author = {{Whitmore}, Bradley C. and {Zhang}, Qing},
        title = "{What Fraction of the Young Clusters in the Antennae Galaxies Are ``Missing''?}",
      journal = {\aj},
         year = 2002,
        month = sep,
       volume = {124},
       number = {3},
        pages = {1418-1434},
          doi = {10.1086/341822},
archivePrefix = {arXiv},
       eprint = {astro-ph/0207100},
 primaryClass = {astro-ph},
       adsurl = {https://ui.adsabs.harvard.edu/abs/2002AJ....124.1418W}
}

@ARTICLE{Whitmore10,
	author = {{Whitmore}, B.~C. and {Chandar}, R. and {Schweizer}, F. and 
	{Rothberg}, B. and {Leitherer}, C. and {Rieke}, M. and {Rieke}, G. and 
	{Blair}, W.~P. and {Mengel}, S. and {Alonso-Herrero}, A.},
	title = "{The Antennae Galaxies (NGC 4038/4039) Revisited: Advanced Camera for Surveys and NICMOS Observations of a Prototypical Merger}",
	journal = {\aj},
	archivePrefix = "arXiv",
	eprint = {1005.0629},
	primaryClass = "astro-ph.EP",
	year = 2010,
	month = jul,
	volume = 140,
	pages = {75-109},
	doi = {10.1088/0004-6256/140/1/75},
	adsurl = {http://adsabs.harvard.edu/abs/2010AJ....140...75W}
}

@ARTICLE{Whitmore14,
       author = {{Whitmore}, Bradley C. and {Brogan}, Crystal and {Chandar}, Rupali and {Evans}, Aaron and {Hibbard}, John and {Johnson}, Kelsey and {Leroy}, Adam and {Privon}, George and {Remijan}, Anthony and {Sheth}, Kartik},
        title = "{ALMA Observations of the Antennae Galaxies. I. A New Window on a Prototypical Merger}",
      journal = {\apj},
         year = 2014,
        month = nov,
       volume = {795},
       number = {2},
          eid = {156},
        pages = {156},
          doi = {10.1088/0004-637X/795/2/156},
archivePrefix = {arXiv},
       eprint = {1410.4473},
 primaryClass = {astro-ph.GA},
       adsurl = {https://ui.adsabs.harvard.edu/abs/2014ApJ...795..156W}
}

@ARTICLE{Wilson03,
       author = {{Wilson}, Christine D. and {Scoville}, Nicholas and {Madden}, Suzanne C. and {Charmandaris}, Vassilis},
        title = "{The Mass Function of Supergiant Molecular Complexes and Implications for Forming Young Massive Star Clusters in the Antennae (NGC 4038/4039)}",
      journal = {\apj},
         year = 2003,
        month = dec,
       volume = {599},
       number = {2},
        pages = {1049-1066},
          doi = {10.1086/379344},
archivePrefix = {arXiv},
       eprint = {astro-ph/0308545},
 primaryClass = {astro-ph},
       adsurl = {https://ui.adsabs.harvard.edu/abs/2003ApJ...599.1049W}
}

@ARTICLE{Zhang99,
	author = {{Zhang}, Q. and {Fall}, S.~M.},
	title = "{The Mass Function of Young Star Clusters in the ``Antennae'' Galaxies}",
	journal = {\apjl},
	eprint = {astro-ph/9911229},
	year = 1999,
	month = dec,
	volume = 527,
	pages = {L81-L84},
	doi = {10.1086/312412},
	adsurl = {http://adsabs.harvard.edu/abs/1999ApJ...527L..81Z}
}

@ARTICLE{Schweizer07,
       author = {{Schweizer}, Fran{\c{c}}ois and {Seitzer}, Patrick},
        title = "{Remnant of a ``Wet'' Merger: NGC 34 and Its Young Massive Clusters, Young Stellar Disk, and Strong Gaseous Outflow}",
      journal = {\aj},
         year = 2007,
        month = may,
       volume = {133},
       number = {5},
        pages = {2132-2155},
          doi = {10.1086/513317},
archivePrefix = {arXiv},
       eprint = {astro-ph/0702645},
 primaryClass = {astro-ph},
       adsurl = {https://ui.adsabs.harvard.edu/abs/2007AJ....133.2132S}
}

@software{photutils,
  author       = {Larry Bradley and
                  Brigitta Sip{\H o}cz and
                  Thomas Robitaille and
                  Erik Tollerud and
                  Z\`e Vin{\'{\i}}cius and
                  Christoph Deil and
                  Kyle Barbary and
                  Tom J Wilson and
                  Ivo Busko and
                  Axel Donath and
                  Hans Moritz G{\"u}nther and
                  Mihai Cara and
                  P. L. Lim and
                  Sebastian Me{\ss}linger and
                  Simon Conseil and
                  Zach Burnett and
                  Azalee Bostroem and
                  Michael Droettboom and
                  E. M. Bray and
                  Lars Andersen Bratholm and
                  Adam Ginsburg and
                  William Jamieson and
                  Geert Barentsen and
                  Matt Craig and
                  Brett M. Morris and
                  Marshall Perrin and
                  Shivangee Rathi and
                  Sergio Pascual and
                  Iskren Y. Georgiev},
  title        = {astropy/photutils: 2.0.2},
  month        = oct,
  year         = 2024,
  publisher    = {Zenodo},
  version      = {2.0.2},
  doi          = {10.5281/zenodo.13989456},
  url          = {https://doi.org/10.5281/zenodo.13989456},
}

@ARTICLE{Riess+11,
       author = {{Riess}, Adam G. and {Macri}, Lucas and {Casertano}, Stefano and {Lampeitl}, Hubert and {Ferguson}, Henry C. and {Filippenko}, Alexei V. and {Jha}, Saurabh W. and {Li}, Weidong and {Chornock}, Ryan},
        title = "{A 3\% Solution: Determination of the Hubble Constant with the Hubble Space Telescope and Wide Field Camera 3}",
      journal = {\apj},
         year = 2011,
        month = apr,
       volume = {730},
       number = {2},
          eid = {119},
        pages = {119},
          doi = {10.1088/0004-637X/730/2/119},
archivePrefix = {arXiv},
       eprint = {1103.2976},
 primaryClass = {astro-ph.CO},
       adsurl = {https://ui.adsabs.harvard.edu/abs/2011ApJ...730..119R}
}

@ARTICLE{Conroy+09,
       author = {{Conroy}, Charlie and {Gunn}, James E. and {White}, Martin},
        title = "{The Propagation of Uncertainties in Stellar Population Synthesis Modeling. I. The Relevance of Uncertain Aspects of Stellar Evolution and the Initial Mass Function to the Derived Physical Properties of Galaxies}",
      journal = {\apj},
         year = 2009,
        month = jul,
       volume = {699},
       number = {1},
        pages = {486-506},
          doi = {10.1088/0004-637X/699/1/486},
archivePrefix = {arXiv},
       eprint = {0809.4261},
 primaryClass = {astro-ph},
       adsurl = {https://ui.adsabs.harvard.edu/abs/2009ApJ...699..486C}
}

@ARTICLE{Conroy+10,
       author = {{Conroy}, Charlie and {White}, Martin and {Gunn}, James E.},
        title = "{The Propagation of Uncertainties in Stellar Population Synthesis Modeling. II. The Challenge of Comparing Galaxy Evolution Models to Observations}",
      journal = {\apj},
         year = 2010,
        month = jan,
       volume = {708},
       number = {1},
        pages = {58-70},
          doi = {10.1088/0004-637X/708/1/58},
archivePrefix = {arXiv},
       eprint = {0904.0002},
 primaryClass = {astro-ph.CO},
       adsurl = {https://ui.adsabs.harvard.edu/abs/2010ApJ...708...58C}
}

@software{python-fsps,
  author       = {Ben Johnson and
                  Dan Foreman-Mackey and
                  Jonathan Sick and
                  Joel Leja and
                  Mike Walmsley and
                  Erik Tollerud and
                  Henry Leung and
                  Spencer Scott and
                  Minjung Park},
  title        = {dfm/python-fsps: v0.4.7},
  month        = jun,
  year         = 2024,
  publisher    = {Zenodo},
  version      = {v0.4.7},
  doi          = {10.5281/zenodo.12447779},
  url          = {https://doi.org/10.5281/zenodo.12447779},
}

@ARTICLE{Kroupa01,
       author = {{Kroupa}, Pavel},
        title = "{On the variation of the initial mass function}",
      journal = {\mnras},
         year = 2001,
        month = apr,
       volume = {322},
       number = {2},
        pages = {231-246},
          doi = {10.1046/j.1365-8711.2001.04022.x},
archivePrefix = {arXiv},
       eprint = {astro-ph/0009005},
 primaryClass = {astro-ph},
       adsurl = {https://ui.adsabs.harvard.edu/abs/2001MNRAS.322..231K}
}

@ARTICLE{Bressan+12,
       author = {{Bressan}, Alessandro and {Marigo}, Paola and {Girardi}, L{\'e}o. and {Salasnich}, Bernardo and {Dal Cero}, Claudia and {Rubele}, Stefano and {Nanni}, Ambra},
        title = "{PARSEC: stellar tracks and isochrones with the PAdova and TRieste Stellar Evolution Code}",
      journal = {\mnras},
         year = 2012,
        month = nov,
       volume = {427},
       number = {1},
        pages = {127-145},
          doi = {10.1111/j.1365-2966.2012.21948.x},
archivePrefix = {arXiv},
       eprint = {1208.4498},
 primaryClass = {astro-ph.SR},
       adsurl = {https://ui.adsabs.harvard.edu/abs/2012MNRAS.427..127B}
}

@ARTICLE{Nguyen+22,
       author = {{Nguyen}, C.~T. and {Costa}, G. and {Girardi}, L. and {Volpato}, G. and {Bressan}, A. and {Chen}, Y. and {Marigo}, P. and {Fu}, X. and {Goudfrooij}, P.},
        title = "{PARSEC V2.0: Stellar tracks and isochrones of low- and intermediate-mass stars with rotation}",
      journal = {\aap},
         year = 2022,
        month = sep,
       volume = {665},
          eid = {A126},
        pages = {A126},
          doi = {10.1051/0004-6361/202244166},
archivePrefix = {arXiv},
       eprint = {2207.08642},
 primaryClass = {astro-ph.SR},
       adsurl = {https://ui.adsabs.harvard.edu/abs/2022A&A...665A.126N}
}

@ARTICLE{MILES1,
       author = {{S{\'a}nchez-Bl{\'a}zquez}, P. and {Gorgas}, J. and {Cardiel}, N. and {Gonz{\'a}lez}, J.~J.},
        title = "{Stellar populations of early-type galaxies in different environments. II. Ages and metallicities}",
      journal = {\aap},
         year = 2006,
        month = oct,
       volume = {457},
       number = {3},
        pages = {809-821},
          doi = {10.1051/0004-6361:20064845},
archivePrefix = {arXiv},
       eprint = {astro-ph/0604568},
 primaryClass = {astro-ph},
       adsurl = {https://ui.adsabs.harvard.edu/abs/2006A&A...457..809S}
}

@ARTICLE{Vazdekis+10,
       author = {{Vazdekis}, A. and {S{\'a}nchez-Bl{\'a}zquez}, P. and {Falc{\'o}n-Barroso}, J. and {Cenarro}, A.~J. and {Beasley}, M.~A. and {Cardiel}, N. and {Gorgas}, J. and {Peletier}, R.~F.},
        title = "{Evolutionary stellar population synthesis with MILES - I. The base models and a new line index system}",
      journal = {\mnras},
         year = 2010,
        month = jun,
       volume = {404},
       number = {4},
        pages = {1639-1671},
          doi = {10.1111/j.1365-2966.2010.16407.x},
archivePrefix = {arXiv},
       eprint = {1004.4439},
 primaryClass = {astro-ph.CO},
       adsurl = {https://ui.adsabs.harvard.edu/abs/2010MNRAS.404.1639V}
}

@ARTICLE{Vazdekis+16,
       author = {{Vazdekis}, A. and {Koleva}, M. and {Ricciardelli}, E. and {R{\"o}ck}, B. and {Falc{\'o}n-Barroso}, J.},
        title = "{UV-extended E-MILES stellar population models: young components in massive early-type galaxies}",
      journal = {\mnras},
         year = 2016,
        month = dec,
       volume = {463},
       number = {4},
        pages = {3409-3436},
          doi = {10.1093/mnras/stw2231},
archivePrefix = {arXiv},
       eprint = {1612.01187},
 primaryClass = {astro-ph.GA},
       adsurl = {https://ui.adsabs.harvard.edu/abs/2016MNRAS.463.3409V}
}

@ARTICLE{Goudfrooij16,
       author = {{Goudfrooij}, Paul and {Fall}, S. Michael},
        title = "{Evolution of the Mass and Luminosity Functions of Globular Star Clusters}",
      journal = {\apj},
         year = 2016,
        month = dec,
       volume = {833},
       number = {1},
          eid = {8},
        pages = {8},
          doi = {10.3847/0004-637X/833/1/8},
archivePrefix = {arXiv},
       eprint = {1611.09876},
 primaryClass = {astro-ph.SR},
       adsurl = {https://ui.adsabs.harvard.edu/abs/2016ApJ...833....8G}
}

@software{Robitaille2012,
       author = {{Robitaille}, Thomas and {Bressert}, Eli},
        title = "{APLpy: Astronomical Plotting Library in Python}",
 howpublished = {Astrophysics Source Code Library, record ascl:1208.017},
         year = 2012,
        month = aug,
          eid = {ascl:1208.017},
archivePrefix = {ascl},
       eprint = {1208.017},
       adsurl = {https://ui.adsabs.harvard.edu/abs/2012ascl.soft08017R}
}
